# Reliable Radiologic Skeletal Muscle Area Assessment – A Biomarker for Cancer Cachexia Diagnosis

**Sabeen Ahmed [1,7,*], Nathan Parker [2], Margaret Park [3,6], Daniel Jeong [4], Lauren Peres [5], Evan W. Davis [3,5], Jennifer B. Permuth [3,5], Erin Siegel [8], Matthew B. Schabath [5], Yasin Yilmaz [7], and Ghulam Rasool [1,7]**

[1] Department of Machine Learning, H. Lee Moffitt Cancer Center and Research Institute, Tampa, FL; sabeen.ahmed@moffitt.org, ghulam.rasool@moffitt.org
[2] Department of Health Outcomes and Behavior, H. Lee Moffitt Cancer Center and Research Institute, Tampa, FL; nathan.parker@moffitt.org
[3] Department of GI Oncology, H. Lee Moffitt Cancer Center and Research Institute, Tampa, FL; margaret.park@moffitt.org, evan.davis@moffitt.org, jenny.permuth@moffitt.org
[4] Diagnostic Imaging and Interventional Radiology, H. Lee Moffitt Cancer Center and Research Institute, Tampa, FL; daniel.jeong@moffitt.org
[5] Department of Cancer Epidemiology, H. Lee Moffitt Cancer Center and Research Institute, Tampa, FL; lauren.peres@moffitt.org, evan.davis@moffitt.org, jenny.permuth@moffitt.org, matthew.schabath@moffitt.org
[6] Department of Biostatistics and Bioinformatics, H. Lee Moffitt Cancer Center and Research Institute, Tampa, FL; margaret.park@moffitt.org
[7] Department of Electrical Engineering, University of South Florida, Tampa, FL; yasiny@usf.edu, sabeen.ahmed@moffitt.org, ghulam.rasool@moffitt.org
[8] Epidemiology and Genomics Research Program, National Cancer Institute, NIH; erin.siegel@nih.gov

* Corresponding author: sabeen.ahmed@moffitt.org

**Abstract:** Cancer cachexia is a common metabolic disorder characterized by severe muscle atrophy which is associated with poor prognosis and quality of life. Monitoring skeletal muscle area (SMA) longitudinally through computed tomography (CT) scans, an imaging modality routinely acquired in cancer care, is an effective way to identify and track this condition. However, existing tools often lack full automation and exhibit inconsistent accuracy, limiting their potential for integration into clinical workflows. To address these challenges, we developed SMAART-AI (Skeletal Muscle Assessment-Automated and Reliable Tool-based on AI), an end-to-end automated pipeline powered by deep learning models (nnU-Net 2D) trained on mid-third lumbar level CT images with 5-fold cross-validation, ensuring generalizability and robustness. SMAART-AI incorporates an uncertainty-based mechanism to flag high-error SMA predictions for expert review, enhancing reliability. We combined the SMA, skeletal muscle index, BMI, and clinical data to train a multi-layer perceptron (MLP) model designed to predict cachexia at the time of cancer diagnosis. Tested on the gastroesophageal cancer dataset, SMAART-AI achieved a Dice score of 97.80% ± 0.93%, with SMA estimated across all four datasets in this study at a median absolute error of 2.48% compared to manual annotations with SliceOmatic. Uncertainty metrics—variance, entropy, and coefficient of variation—strongly correlated with SMA prediction errors (0.83, 0.76, and 0.73 respectively). The MLP model predicts cachexia with 79% precision, providing clinicians with a reliable tool for early diagnosis and intervention. By combining automation, accuracy, and uncertainty awareness, SMAART-AI bridges the gap between research and clinical application, offering a transformative approach to managing cancer cachexia.

## 1. Introduction

Cancer cachexia, a multifactorial syndrome characterized by involuntary weight loss, skeletal muscle atrophy, and fatigue, presents a significant challenge in cancer management [1]. Affecting approximately 80% of cancer patients and contributing to 20–30% of cancer-related deaths, cachexia has a profound impact on patient outcomes, reducing quality of life and complicating disease treatment [2-4]. Cancer cachexia is more prevalent in the gastroesophageal, pancreatic, colorectal, lung, and hematological cancers [2, 5, 6]. Since cachexia is irreversible at later stages, early detection and monitoring are essential for timely interventions that help maintain muscle mass, improve treatment tolerance, and enhance survival rates [3, 7-9].

Current clinical assessments of cachexia often use anthropometric measurements like weight, body mass index (BMI), waist circumference, and bioelectrical impedance analysis (BIA) because they are easy to collect in clinical or research studies [2, 3, 10]. However, these measures have limitations; a patient's weight might remain relatively stable despite skeletal muscle loss, and BIA is affected by factors like hydration levels, recent exercise, and limited tissue specificity [3, 10]. When available, computed tomography (CT) imaging offers a more precise alternative, with skeletal muscle area (SMA) in single images from thoracic and abdominal scans providing a reliable estimate of overall muscle mass [11-13]. Despite its efficacy, manual extraction and annotation of single slices are time-consuming, limiting its practical use in routine clinical settings.

Conventional machine learning methods have been used to automate skeletal muscle segmentation in CT scans using various atlas-based techniques [14-17]. However, these methods depend on handcrafted features, requiring significant domain expertise and manual input. Deep learning (DL) approaches, especially convolutional neural networks (CNNs), have yielded superior performance in body composition analysis by learning features directly from imaging data. This advancement has streamlined tissue segmentation into a two-step process: identifying the mid-L3 slice and then segmenting skeletal muscle [18-25].

Despite their promise, existing DL models face several challenges when considered for real-world clinical use [26, 27]. Many models trained on large datasets perform well on benchmark datasets but struggle with real-world data, which may be out-of-distribution or noisy [28]. Out-of-distribution data refers to cases where the real-world data differs significantly from the data used for training, such as CT scans from different populations or imaging protocols. Noisy data includes inconsistencies that obscure patterns the model is designed to learn, such as artifacts like metal implants, poor image quality, or motion artifacts. Owing to their design, these DL models can fail without issuing any warning to the users [29-32]. The lack of availability of model development source code and pre-trained weights severely hamper study reproducibility. This makes it difficult for researchers to replicate results or build on prior work, ultimately limiting further research and clinical adoption [33].

Several open-source and proprietary tools are available for skeletal muscle segmentation, including SliceOmatic (by TomoVision) [34], ABACS (Automatic Body Composition Analyzer using Computed Tomography Image Segmentation by Voronoi Analytics) [35], DAFS (Data Analysis Facilitation Suite by Voronoi Analytics) [36], AW Server (Advanced Workstation Server by General Electric) [37], and TotalSegmentator [38]. Among these, SliceOmatic and AW Server offer manual segmentation guided by Hounsfield windowing, while ABACS (as a plug-in used in conjunction with SliceOmatic), DAFS, and TotalSegmentator provide the automated AI-based segmentation option. Although these automated tools have demonstrated an advantage over purely manual segmentation in terms of the time taken for the task [22, 24, 25]; they still lack the full automation needed for clinical integration. Furthermore, the accuracy of these tools often decreases when processing real-world noisy or out-of-distribution data without warning the user, and for some, the segmented masks cannot be retrieved for further evaluation or correction. This reduces their reliability and potential for integration into clinical workflows.

To address these shortcomings, we propose SMAART-AI (**S**keletal **M**uscle **A**ssessment-**A**utomated and **R**eliable **T**ool based on **AI**), an end-to-end image processing and machine learning pipeline that processes CT scans to identify the required image slices, accurately segments skeletal muscle pixels, and performs uncertainty estimation to assess model confidence [39, 40]. The schematic layout of the proposed tool and study design is depicted in Figure 1. We have made the code available for reproducibility and further research at [https://github.com/Beemd/SM_Segmentation](https://github.com/Beemd/SM_Segmentation).

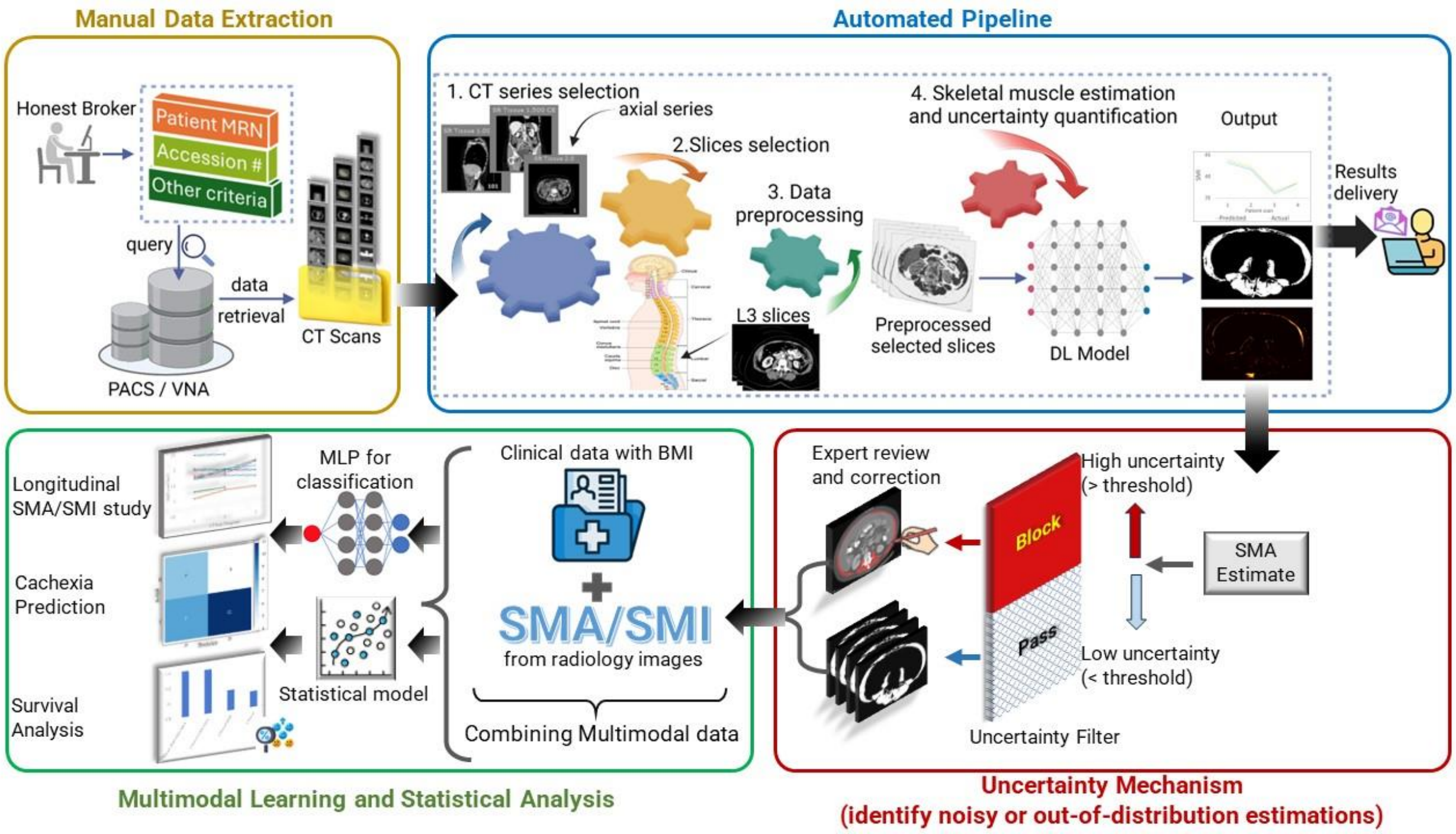

**Figure 1. Overview of the proposed framework for S̲keletal M̲uscle A̲ssessment – A̲utomated and R̲eliable T̲ool - based on A̲I̲ (SMAART-AI).** This tool can easily be integrated into clinical workflows to assess and monitor skeletal muscle area (SMA) changes as a biomarker for cancer cachexia diagnosis. The data extraction is a manual

process presented in **Manual Data Extraction**, where patient CTs are retrieved from PACS/VNA systems based on specific criteria. This data is then processed through the **Automated Pipeline** of **SMAART-AI**, starting with CT series selection (axial series) and slice selection at the third lumbar (L3) level. The selected slices undergo data preprocessing to prepare the images for being passed on to the trained deep learning (DL) model for inference. The DL model segments the skeletal muscle to estimate the SMA, generates the corresponding uncertainty map, and calculates the uncertainty metrics. An **Uncertainty Filtering Mechanism** then applies thresholding to identify high and low uncertainty cases, blocking noisy and out-of-distribution images with a high probability of degraded performance by the DL model. The blocked images and corresponding segmentations are passed on for expert review and correction, ensuring reliable SMA and skeletal muscle index (SMI) estimations. These estimations are combined with clinical data (for example, age, height, gender, weight, BMI, race, ethnicity, cancer stage) to form a multimodal dataset, which is then used for **Multimodal Learning and Statistical Analysis**. The SMA and SMI are monitored longitudinally for patients identified to be both cachectic and non-cachectic at the time of cancer diagnosis. A multi-layer perceptron (MLP) model is trained for cachexia and recurrence classification, and the survival analysis shows better performance with multimodal data compared to unimodal clinical data alone.

This study makes several key contributions:

1. **Automated Measurement and Monitoring of Skeletal Muscle**: We provide an end-to-end data processing/machine learning pipeline for automated SMA and SMI tracking across patient scans, which can be easily integrated into clinical workflows. The pipeline provides the generated mask as well as the uncertainty map in DICOM and PNG formats.
2. **Comparative Analysis**: We benchmarked our model's performance against various propriety and open-source tools, including SliceOmatic (manual), ABACS (SliceOmatic AI-based module), DAFS, AW Server, and TotalSegmentator. To our knowledge, this is the first study comparing the performance of licensed tools like ABACS, DAFS, and AW Server, which are used within hospital research centers for estimating SMA.
3. **Uncertainty Estimation**: We evaluated uncertainty estimation techniques, including dropout, ensemble, and post hoc calibration, to assess model reliability. Multiple metrics, such as variance, entropy, and the coefficient of variation, were compared to quantify estimated uncertainty and detect degraded model performance using a threshold-based mechanism. Our results indicate that certain metrics consistently exhibit strong correlations between estimated uncertainty and SMA across datasets. Based on these findings, we identified a single metric that performs reliably across datasets, enabling a standardized mechanism for detecting degraded model performance in clinical settings.
4. **Uncertainty Maps**: We generated visual maps using the uncertainty information to enhance interpretability and guide clinical decision-making. These maps are especially useful when the input image is noisy.
5. **Survival Analysis**: We conducted a survival analysis and evaluated the concordance index using different data combinations: (i) clinical data and BMI, (ii) clinical data, SMA, and SMI, and (iii) a multimodal approach integrating clinical data, BMI, SMA, and SMI. Our results demonstrate that incorporating multimodal data improves the concordance index.

6. **Cachexia and Recurrence Prediction:** We developed machine learning models to predict cachexia and disease recurrence at the time of cancer diagnosis. The models leverage multimodal data, including SMA and SMI extracted from radiology images, combined with clinical data and BMI. To our knowledge, this is the first study applying machine learning for cachexia and recurrence prediction, with the potential for integration into clinical practice.

## 2. Materials and Methods

### *2.1. Datasets*

This study utilized patient data and CT images acquired from cancer patients treated at the H. Lee Moffitt Cancer Center and Research Institute (Tampa, Florida). Four cohorts of patients included in this study were 24 patients with gastroesophageal cancer, 60 patients with colorectal cancer, 153 patients with pancreatic cancer and cysts, and 324 patients with ovarian cancer (Table 1). This study has been reviewed by an Institutional Review Board to ensure that the study is carried out in an ethical manner. The pancreatic cohort also included patients from the Florida Pancreas Collaborative study [41]. All CT scans in this study were in DICOM format.

**Table 1.** Datasets used in this study.

| Cancer site | No. of patients | No. of CT scans |
|---|---|---|
| Gastroesophageal | 24 | 71 |
| Colorectal | 60 | 60 |
| Pancreatic | 153 | 222 |
| Ovarian | 324 | 324 |

The gastroesophageal dataset included 24 patients, each with multiple scans, one at each time point, totaling 71 scans. Specifically, 5 patients had 2 scans, 16 had 3 scans, 2 had 4 scans, and 1 patient had 5 scans. Only the mid-L3 slices from the non-contrast axial series, along with the corresponding skeletal muscle masks and SMA, were available.

The colorectal dataset included CT scans from 60 patients, each with one scan taken at diagnosis or before treatment/surgery. As summarized in Table 2, all 60 patients were included for statistical analysis based on the availability of clinical data. All CT scans were multi-slice and multi-series.

**Table 2**. Colorectal cancer cohort

| | Total patient count = 60 |
|---|---|
| Age at diagnosis, mean (SD) | 61.93 ± 12.50 |
| Sex, N | |
| Female | 28 |
| Male | 32 |
| Ethnicity, N | |
| Non-Hispanic/Non-Latinx | 53 |
| Hispanic/Latinx | 7 |
| Race, N | |
| White | 56 |
| Other | 4 |
| AJCC-7 Stage, N | |
| I | 8 |
| II | 17 |
| III | 6 |
| IV | 1 |
| NA | 2 |
| Grade/ Differentiation, N | |
| Well | 3 |
| Moderate | 41 |
| Poor | 5 |
| Undifferentiated | 6 |
| NA | 5 |

The pancreatic dataset included 153 patients, each diagnosed with a pancreatic mass that was either malignant or benign [41, 42]. The cohort consists of patients diagnosed with pre-malignant lesions, intraductal papillary mucinous neoplasms (IPMNs), and those diagnosed with cancer, pancreatic ductal adenocarcinoma (PDAC), and pancreatic neuroendocrine tumors (PNET). A subset of patients with baseline scans also had scans at the first follow-up, and a further subset of those with first follow-up scans had scans at the second follow-up. Each time point included one CT scan per patient. Baseline scans were taken around the time of cancer diagnosis, with subsequent scans approximately six months apart, resulting in a total of 222 scans from 153 unique patients. All CT scans were multi-slice and multi-series. Based on the availability of clinical data at diagnosis, a cohort of 130 patients was selected for statistical analysis, as summarized in Table 3. Out of these 130 patients, 89 were PDAC patients.

**Table 3**. Pancreatic cancer cohort.

| | **Total patient count = 130** |
|---|---|
| Age at diagnosis, mean (SD) | 67.81 ± 10.80 |
| Sex, N | |
| Female | 58 |
| Male | 72 |
| Race and Ethnicity, N | |
| Non-Hispanic White | 107 |
| Hispanic/Latinx | 13 |
| Non-Hispanic Black | 10 |
| TNM Stage (Pathological), N | |
| 1: 0 (T0/Tis, N0, M0) | 8 |
| 2: IA (T1, N0, M0) | 17 |
| 3: IB (T2, N0, M0) | 15 |
| 4: IIA (T3, N0, M0 | 20 |
| 5: IIA (T1, N1, M0) | 1 |
| 6: IIA (T2, N1, M0) | 8 |
| 7: IIB (T3, N1, M0) | 4 |
| 8: III (T4, Any N, M0) | 19 |
| 9: IV (Any T, Any N, M1) | 25 |
| 99: NA | 13 |

The ovarian cancer cohort consisted of 324 scans, with each patient having one scan around the time of cancer diagnosis. All scans were processed through SMAART-AI, and a subset of 175 patients, summarized in Table 2, was randomly selected for statistical analysis due to the time-consuming nature of manual marking required for validation. The dataset included multi-slice and multi-series CT scans with all or some combinations of views (axial, sagittal, coronal), maximum intensity projections, contrast, and non-contrast, and within contrast, arterial, venous, and delayed phases. Axial CT series in the post contrast venous phase, which is commonly present, was used for processing when available.

**Table 4**. Ovarian cancer cohort.

| | **Total patient count = 175** |
|---|---|
| Age at diagnosis, mean (SD) | 64.29 ± 10.58 |
| Ethnicity, N | |
| Non-Hispanic | 165 |
| Hispanic | 10 |
| Race, N | |
| Black | 6 |
| White | 163 |
| Other | 6 |
| FIGO Stage, N | |
| I | 9 |
| II | 14 |
| III | 116 |
| IV | 36 |
| Grade/ Differentiation, N | |
| Moderate | 6 |
| Poor | 45 |
| Undifferentiated | 83 |
| NA | 41 |
| Tumor Sequence number, N | |
| 00 | 140 |
| 01 | 5 |
| 02 | 23 |

*2.2. Data Processing*

2.2.1. Annotations for DL Model Training and Testing

**Gastroesophageal cancer dataset**: Segmentation masks and the segmented area for the mid-L3 slice were created using the SliceOmatic tool (version 6, TomoVision). Experts manually generated these masks using the Hounsfield unit (HU) window and made corrections using the 'region growing' mode.

**Pancreatic dataset**: A radiologist evaluated SMA for the manually selected end of the L3 or start of the L4 slice (referred to as end-L3 in this study) from the axial series using the AW Server tool. Evaluation of SMA using the AW Server was based on HU windowing for muscle, and no manual correction was made. Only the SMA and end-L3 CT slice information were available, but not the skeletal muscle mask. For comparison, SMA for the same end-L3 slices was estimated using the manual SliceOmatic tool and SMAART-AI. SMAART-AI was used for all available scans, whereas SliceOmatic, being a manual tool, was used for 109 randomly selected scans only. For the pancreatic cancer dataset, the cachexia status was available and determined by an independent team based on one of the following two criteria:

1. The two-stage system defined by Fearon et al. [43] categorizes patients as either cachectic or non-cachectic. Cachexia was diagnosed if there was >5% weight loss over the past six months when a participant had ≥ 20 BMI, or >2% weight loss for patients with a BMI < 20.
2. The four-stage system defined by Vigano et al. [8] classifies patients into four categories: pre-cachectic, cachectic, refractory, and non-cachectic. This classification was based on five criteria: (i) Biochemical markers from lab reports (elevated C-reactive protein or leukocytes, hypoalbuminemia, or anemia), (ii) Food intake (normal or decreased), (iii) Moderate weight loss over the past six months (≤ 5%), (iv) Significant weight loss over the past six months (> 5%), (v) Performance status (Eastern Cooperative Oncology Group Performance Status ≥ 3). A patient was categorized as non-cachectic if none of these criteria were met.

If insufficient information was available to apply the four-stage system, the two-stage system was used instead.

**Colorectal cancer dataset**: The SMA for the mid-L3 slices was estimated using DAFS, an AI-based tool by Voronoi Health Analytics. This tool automatically selected the mid-L3 slice based on its internal algorithm. Only the SMA values were available, without the skeletal muscle mask or information about the specific mid-L3 slice images identified by DAFS. For comparison, SMA was estimated at the manually determined mid-L3 level for 53 randomly selected CT images using SliceOmatic (the manual tool). SMAART-AI estimated the SMA for all available scans at the same manually identified mid-L3 level and at the automatically identified mid-L3 slice using the internal mechanism explained in section 2.2.2.

**Ovarian cancer dataset**: SMA values for the mid-L3 slices identified manually, were evaluated using ABACS by a radiologist. ABACS, an AI-based tool is a plug-in available within SliceOmatic (TomoVision) by Voronoi Health Analytics. The ABACS tool automatically generates skeletal muscle masks for manually selected slices. For this dataset, while SMA values and mid-L3 CT slice information were available, the skeletal muscle mask itself was not. Manual estimation of SMA using SliceOmatic was performed for 154 randomly selected patient scans at the same mid-L3 slice for comparison with ABACS. SMAART-AI estimated the SMA for all available scans at the same manually identified mid-L3 level and at the automatically identified mid-L3 slice for comparison with ABACS (the automated tool available as a plug-in within SliceOmatic) and SliceOmatic (manual tool).

In this study, SliceOmatic refers to the manual tool, whereas ABACS refers to the automatic tool available as a plug-in within SliceOmatic.

2.2.2. DICOM Data Processing

**Preprocessing.** The DICOM CT images in the training, validation, and test datasets were adjusted to the skeletal muscle HU range of -29 to 150, followed by conversion to PNG format. The PNG images were normalized by subtracting the mean and dividing by the standard deviation, resulting in pixel values converted to a standard normal distribution with a mean of 0 and a standard deviation of 1.

**Postprocessing**. The output segmentation masks from the DL model are converted to PNG and DICOM formats. The DICOM format enables the user to view the generated mask in a DICOM viewer and make corrections if deemed necessary.

The L3 slices identified by the automated pipeline in each patient scan were processed to determine the mid-L3 slice and derive the corresponding skeletal muscle area. The mid-L3 SMA was estimated in two ways:

1. Identifying a single mid-L3 slice based on the following rules:
    - Mid-slice-index = (slice-count/2) – 1, for mid-L3 slice-count ≤ 12
    - Mid-slice-index = (slice-count/2) – 2, for mid-L3 slice-count ≤ 32
    - Mid-slice-index = (slice-count/2) – 3, for mid-L3 slice-count > 32
2. Calculating the average area of mid-L3 slices as an alternative to the single mid-L3 SMA:
    - For mid-L3 slice count > 12: The average area is calculated from five slices, two slices above and two slices below the identified mid-L3 slice, including the mid-L3 slice itself.
    - For mid-L3 slice count ≤ 12: The average area is calculated from three slices, one above and one below the identified mid-L3 slice, including the mid-L3 slice itself.

### *2.3. Automated Pipeline Framework (SMAART-AI)*

An end-to-end pipeline was developed to monitor patients' SMA and SMI. The pipeline began by identifying the axial series in CT scans and locating slices corresponding to the third lumbar vertebral level (L3). These L3 slices were converted to PNG format and processed by the DL model, which identified skeletal muscle pixels in each slice. The DL model then generated the pixel-level skeletal muscle segmentations along with uncertainty maps as its output. In the case of multiple scans at different time points for the same patient, a plot for longitudinal monitoring of SMA/SMI was included in the output [44]. Additionally, the model generated a file containing patient IDs, scan dates, CT series number, slice number, corresponding SMA, quantified uncertainty value, and study/series descriptions to distinguish contrast and venous-phase axial series. A user-defined threshold on uncertainty was used to segregate expected high-error SMA predictions by the DL model in case of out-of-distribution or noisy images. CT images with high SMA errors were manually annotated using SliceOmatic. The SMA values estimated by the DL model, along with the manually generated SMA for high-uncertainty cases, were used alongside clinical data for survival analysis, cachexia, and recurrence prediction.

#### 2.3.1. Automated Selection of Both Axial Series and Lumbar Level Slice Selection

SMAART-AI identified all axial series within the complete CT scan in DICOM format using patient orientation data from the DICOM header attribute 'ImageOrientationPatient.' Each axial series could contain multiple groups, which were identified (if present). The image slices were sorted using DICOM header attributes such as study and series instance, slice thickness, spacing between slices, frame of reference, image position, CT series, and acquisition number.

L3 slices were then identified using the open-source tool TotalSegmentator [38] (available on GitHub: https://github.com/wasserth/TotalSegmentator). TotalSegmentator segments 117 anatomical structures in CT images and saves each segmented axial series in NIfTI format. Each anatomical structure has been assigned an index number, with the L3 vertebra assigned index number 29. The identified axial series were processed by TotalSegmentator to produce segmented NIfTI files, and index number 29 was used to identify the L3 slices.

#### 2.3.2. nnU-Net for Segmentation

**Deep Learning Model Architecture:**

The nnU-Net framework was used to train and validate a DL model for identifying skeletal muscle [45]. The U-Net architecture has been chosen for its exceptional performance in medical imaging segmentation tasks [46]. Since the training dataset consisted of single L3 slices, we used 2D nnU-Net architectures. Two model options with different architectures were available: 'PlainConvUNet' and 'ResidualEncoderUNet.'

The architecture of PlainConvUNet included 7 encoder stages, each with 2 convolutional blocks, followed by 2 convolutional blocks in the bottleneck stage. The decoder also had 7 stages, each containing 2 convolutional blocks. Skip connections link the output of each encoder stage to the corresponding decoder stage. Each convolutional block in both encoder and decoder has a convolutional layer, followed by an instance normalization layer and a Leaky ReLU activation function.

The 'ResidualEncoderUNet' consisted of 7 stages in both the encoder and decoder, connected by skip connections. The residual blocks were distributed as follows:

Stage 1: 1 residual block

Stage 2: 3 residual blocks

Stage 3: 4 residual blocks

Stages 4 to 7: 6 residual blocks each

Bottleneck stage: 6 residual blocks.

Each residual block contained two convolutional blocks, followed by two sets of convolutional layers, instance normalization layers, and a Leaky ReLU activation layer.

**DL Model Training:** The two DL models were trained using the nnU-Net framework with 5-fold cross-validation. For each fold, the training started with randomly initialized weights and ran for 1000 epochs with a learning rate of 10e-2. The training dataset consisted of 45 mid-L3 slice images from the gastroesophageal dataset and 15 end-L3 slice images from the pancreatic cancer dataset. The test set included 25 images from the gastroesophageal dataset, and the performance of the DL model was evaluated using the average Dice score and Jacquard index. The results of this test set have been reported. Additionally, the trained model was used to run inference on the pancreatic, colorectal, and ovarian datasets.

The pixel-wise average probability from the output of the models across the 5 folds was calculated, and all pixels with an average probability greater than 0.5 were marked as skeletal muscle. There are cases termed as false positives and false negatives. False positive pixels were not part of the skeletal muscle in the manually marked mask; however, the DL model identified the pixels as part of it. False negative pixels were part of the skeletal muscle, but the model did not mark them as part of the skeletal muscle mask it produced.

2.3.3. Uncertainty Estimation Methods and Metrics

DL has two main types of uncertainty: aleatoric and epistemic. Aleatoric uncertainty arises from the noise inherent in the data for which the DL model is trained and is irreducible. Epistemic uncertainty is related to the DL model's parameters and can be reduced by training with a larger, more diverse dataset. To estimate uncertainty in the DL model predictions for SMA, we applied three different techniques, quantifying the total uncertainty (aleatoric + epistemic), epistemic, and aleatoric uncertainty.

1. **Calibration**: The 'netcal' Python library [47], specifically the 'LogisticCalibration' method (also known as Platt scaling), was used. The calibration model was trained using the DL model outputs and corresponding labels. During inference, the DL model output was passed through the calibration model to calibrate it.
2. **Monte Carlo Dropout**: A dropout layer with a 20% probability (p=0.2) was added after each convolutional layer in the 'ResidualEncoderUNet' architecture. The model was trained with 5-fold cross-validation and inference for each image was repeated 20 times per fold. The average of the 5-fold ensemble predictions at each iteration was taken. The final dropout prediction was the pixel-wise mean of these twenty ensemble averages. The uncertainty estimate was calculated as the mean of the pixel-wise variance across these ensemble averages.

3. **Model Ensemble**: Ten models were used—five with the 'PlainConvUNet' architecture and five with 'ResidualEncoderUNet'. Each set of five models was part of the 5-fold cross-validation. The final prediction was derived by taking the pixel-wise mean of these ten models, and the uncertainty estimate was calculated from the mean pixel-wise variance across these models.

The following metrics were used for quantifying the uncertainty estimated using the different techniques [48]:

1. Average Probability: Calculated by taking the average of the output probabilities of the predicted class at each pixel in a single image. This metric captures the total uncertainty.
2. Average probability-SM: This is the average output probability of pixels marked as skeletal muscle (SM) only. This metric captures the total uncertainty.
3. Average Calibrated Probability: Average of the calibrated output probabilities of the predicted class at each pixel in a single image. This metric captures the total uncertainty.
4. Coefficient of Variation (pixel-wise): The average of the pixel-wise coefficient of variation, calculated from the ensemble or dropout outputs as the ratio of the standard deviation to the mean, multiplied by 100. This metric captures the epistemic uncertainty.
5. Coefficient of Variation (SMA): Calculated using the standard deviation and mean of the SMA estimated by each model in the ensemble or multiple inferences in case of the dropout method. This metric captures the epistemic uncertainty.
6. Average Variance: It is calculated as the average of the variance computed for each pixel. The pixel-wise variance is calculated using the output probabilities from the ensemble models or multiple inferences using the dropout method. This metric captures the epistemic uncertainty.
7. Average Variance-SM: Average of the variance for pixels identified as being part of the skeletal muscle (SM) only. This metric captures the epistemic uncertainty.
8. Average Entropy: Estimates the total uncertainty by calculating the binary entropy at each pixel based on the average output probabilities across pixels in either an ensemble of models or multiple inferences with dropout. The average entropy of all pixels across the image is reported.
9. Expected Entropy of the Ensemble: Estimates aleatoric uncertainty by calculating the binary entropy at each pixel for all the models in the ensemble. The average entropy is computed for each pixel across all models, and the final reported value is the mean of these pixel-wise average entropies across the entire image.

The Monte Carlo dropout technique was applied only to the gastroesophageal dataset.

#### 2.3.4. Statistical Tests for Uncertainty Methods and Metrics

The Pearson correlation coefficient (r) was calculated between each uncertainty method/metric and the difference between the SMA estimated by SMAART-AI and that estimated by SliceOmatic. The interpretation of r values is as follows: $|r| = 0$ indicates no relationship, $0 < |r| \leq 0.3$ indicates a weak relationship, $0.3 < |r| \leq 0.5$ indicates a moderate relationship, $0.5 < |r| \leq 0.7$ indicates a strong relationship, $|r| > 0.7$ indicates a very strong relationship, and $|r| = 1$ represents a perfect relationship. The statistical significance of these correlations was assessed using the Student's t-test, with a significance level set at 95% ($p < 0.05$ indicating statistical significance). This analysis established the degree of association between each uncertainty method/metric and the error in SMA estimated by SMAART-AI, providing insights into how well each method can identify cases with potentially high estimation errors.

#### 2.3.5. Mechanism for Identifying High Error SMA Predictions by SMAART-AI

The ensemble technique, combined with the uncertainty metric of average variance, was used to identify cases with a high probability of having considerable errors in SMA estimation by SMAART-AI across colorectal, pancreatic, and ovarian datasets. These errors included both underestimation and overestimation of SMA by SMAART-AI. A unique threshold was manually selected for each dataset based on the estimated uncertainty values. Cases exceeding this threshold were flagged as potentially high error, indicating a significant deviation from SMA measurements obtained using SliceOmatic. These expected high-error cases were forwarded for expert review.

### *2.4. Statistical Analysis and Predictions*

#### 2.4.1 Survival Analysis

For survival analysis, we used the 'CoxPHFitter' tool from the 'lifelines' Python library on three datasets: pancreatic, colorectal, and ovarian cancer [49]. The input data included variables such as age, gender, race, ethnicity, weight, height, cancer stage, BMI, SMI, SMA, time to event (TTE), and vital status. The analysis was performed using unimodal clinical data around the time of cancer diagnosis and multimodal data integrating clinical data with SMA/SMI derived from radiology images [50-55]. Various penalizer values with the 'CoxPHFitter' tool were used to determine optimal performance for combinations of SMA, SMI, and BMI.

#### 2.4.2 Predictions

The pancreatic cancer dataset included additional information on cachexia status, while the ovarian cancer dataset included recurrence data corresponding to each patient in their respective cohorts.

**Cachexia**: Cachexia prediction was a binary classification task for which we trained an MLP model with three linear layers, each followed by dropout and an output layer with sigmoid activation. Dropout probabilities were 0.2 after the first two layers and 0.5 after the third. The linear layers contained 256, 128, and 32 nodes, respectively. The model was trained for 50 epochs with a learning rate of 5e-5. A dataset of 100 pancreatic cancer patients was split into training and validation sets (85:15 ratio), and the trained MLP was evaluated on a test set of 30 PDAC patients only. The complete dataset of 130 patients had 70 identified as being cachectic and 60 as non-cachectic.

**Recurrence**: Recurrence prediction was a binary classification task for which we trained an MLP model with three linear layers, each followed by a dropout and sigmoid output layer. Dropout probabilities were 0.75 after the first layer, 0.5 after the second, and 0.65 after the third. The linear layers contained 64, 32, and 16 nodes, respectively. The model was trained for 200 epochs with a learning rate of 5e-4. A dataset of 125 ovarian cancer patients was split into training and validation sets (85:15 ratio), and the trained MLP was evaluated on a test set of 50 patients. SMOTE (Synthetic Minority Over-sampling Technique) was used to address the class imbalance in the training data [56]. The complete dataset of 175 patients has 116 who recurred and 59 did not.

Different hyperparameters for the MLP models were used for predicting cachexia and recurrence to optimize the respective tasks.

## 3. Results

### *3.1. Comparison of the Predicted SMA between SMAART-AI, TotalSegmentator, DAFS, ABACS, AW Server, and SliceOmatic*

#### 3.1.1. Gastroesophageal Cancer

Table 5 compares the SMA estimated by SMAART-AI using the ensemble technique, with the SMA estimated manually by experts using SliceOmatic for the held-out test set of the gastroesophageal cancer dataset. The table presents the pixel count marked as part of the skeletal muscle, which is used as a proxy for SMA. The mean and median absolute differences between the model's estimation and SliceOmatic are 2.44% and 0.81%, respectively. Additionally, the mean and median Jacquard scores are 94.21% and 94.84%, while the Dice scores are 96.96% and 97.35%, respectively.

Table 6 compares the SMA estimated by SMAART-AI using the dropout technique, with the SMA estimated manually by experts using SliceOmatic for the held-out test set of the gastroesophageal cancer dataset. The mean and median absolute differences between SMAART-AI's estimation and SliceOmatic are 2.72% and 1.06%, respectively. Additionally, the mean and median Jacquard scores are 93.95% and 94.93%, while the Dice scores are 96.82% and 97.40%, respectively.

The false positive count represents the number of pixels marked by SMAART-AI as part of the skeletal muscle but not included in the skeletal muscle mask generated manually by SliceOmatic. Similarly, the false negative count refers to the number of pixels that were not marked by SMAART-AI as skeletal muscle but were included in the mask generated manually by SliceOmatic. Cases highlighted in red indicate noisy or out-of-distribution CT images, leading to degraded performance by SMAART-AI on these images.

**Table 5.** Model Performance using the Ensemble Technique for Skeletal Muscle Area Estimation on the Gastroesophageal Cancer Dataset.

| Patient ID.scan | Model's Pixel Count | SliceOmatic Pixel Count | Difference in Area (%) | Jacquard Score (%) | Dice Score (%) | False Positive | False Negative |
|---|---|---|---|---|---|---|---|
| 2.2 | 21691 | 21499 | 0.89% | 97.39 | 98.68 | 381 | 189 |
| 2.4 | 18029 | 18058 | -0.16% | 96.98 | 98.47 | 262 | 291 |
| 3.1 | 20611 | 20449 | 0.79% | 94.14 | 96.98 | 701 | 539 |
| 4.1 | 20012 | 19979 | 0.17% | 98.98 | 99.49 | 119 | 86 |
| 4.2 | 21854 | 21496 | 1.67% | 97.04 | 98.50 | 505 | 147 |
| 5.1 | 13159 | 13309 | -1.13% | 94.52 | 97.18 | 298 | 448 |
| 5.2 | 12704 | 12801 | -0.76% | 95.10 | 97.49 | 272 | 369 |
| 5.3 | 15321 | 15436 | -0.75% | 94.54 | 97.19 | 374 | 489 |
| 5.4 | 16502 | 16563 | -0.37% | 94.59 | 97.22 | 429 | 490 |
| 7.3 | 30976 | 29453 | 5.17% | 86.69 | 92.87 | 2916 | 1393 |
| 9.1 | 18829 | 18955 | -0.66% | 97.28 | 98.62 | 197 | 323 |
| 9.2 | 19162 | 18795 | 1.95% | 95.01 | 97.44 | 669 | 302 |
| 9.3 | 25130 | 22612 | 11.14% | 86.11 | 92.53 | 3041 | 523 |
| 15.1 | 7483 | 7068 | 5.87% | 92.58 | 96.14 | 488 | 73 |
| 15.2 | 20140 | 19974 | 0.83% | 98.26 | 99.12 | 259 | 93 |
| 15.3 | 18132 | 18497 | -1.97% | 96.90 | 98.42 | 106 | 471 |
| 15.4 | 21075 | 20532 | 2.64% | 92.34 | 96.02 | 1100 | 557 |
| 15.5 | 21586 | 21762 | -0.81% | 92.98 | 96.36 | 701 | 877 |
| 16.1 | 18104 | 18137 | -0.18% | 97.94 | 98.96 | 172 | 205 |
| 16.2 | 19361 | 19068 | 1.54% | 94.12 | 96.97 | 729 | 436 |
| 16.3 | 15484 | 15609 | -0.80% | 96.21 | 98.07 | 238 | 363 |
| 21.1 | 6739 | 6704 | 0.52% | 96.85 | 98.40 | 125 | 90 |
| 21.3 | 19017 | 19063 | -0.24% | 94.84 | 97.35 | 481 | 527 |
| 21.5 | 19436 | 19455 | -0.10% | 93.99 | 96.90 | 612 | 612 |
| 23.2 | 22404 | 18694 | 19.85% | 79.75 | 88.73 | 4170 | 460 |

The results in red represent noisy or out-of-distribution images.

**Table 6.** Model Performance using the Dropout Technique for Skeletal Muscle Area Estimation on the Gastroesophageal Cancer Dataset.

| Patient ID.scan | Model's Pixel Count | SliceOmatic Pixel Count | Difference in Area (%) | Jacquard Score (%) | Dice Score (%) | False Positive | False Negative |
|---|---|---|---|---|---|---|---|
| 2.2 | 21625 | 21499 | 0.586 | 97.64 | 98.81 | 320 | 194 |
| 2.4 | 18484 | 18058 | 2.360 | 95.44 | 97.67 | 639 | 213 |
| 3.1 | 20233 | 20449 | -1.056 | 91.45 | 95.54 | 800 | 1016 |
| 4.1 | 20027 | 19979 | 0.240 | 98.96 | 99.48 | 129 | 81 |
| 4.2 | 21958 | 21496 | 2.149 | 96.86 | 98.41 | 577 | 115 |
| 5.1 | 13139 | 13309 | -1.277 | 94.61 | 97.23 | 281 | 451 |
| 5.2 | 12696 | 12801 | -0.820 | 95.72 | 97.82 | 226 | 331 |
| 5.3 | 15428 | 15436 | -0.052 | 94.21 | 97.02 | 456 | 464 |
| 5.4 | 16608 | 16563 | 0.272 | 93.08 | 96.42 | 617 | 572 |
| 7.3 | 31294 | 29453 | 6.251 | 86.56 | 92.80 | 3108 | 1267 |
| 9.1 | 18680 | 18955 | -1.451 | 97.41 | 98.69 | 109 | 384 |
| 9.2 | 19225 | 18795 | 2.288 | 94.93 | 97.40 | 709 | 279 |
| 9.3 | 25422 | 22612 | 12.427 | 84.90 | 91.83 | 3366 | 556 |
| 15.1 | 7457 | 7068 | 5.504 | 93.33 | 96.55 | 445 | 56 |
| 15.2 | 20073 | 19974 | 0.496 | 98.33 | 99.16 | 218 | 119 |
| 15.3 | 18149 | 18497 | -1.881 | 96.73 | 98.34 | 131 | 479 |
| 15.4 | 21098 | 20532 | 2.757 | 92.86 | 96.30 | 1054 | 488 |
| 15.5 | 21892 | 21762 | 0.597 | 93.36 | 96.56 | 815 | 685 |
| 16.1 | 18079 | 18137 | -0.320 | 97.85 | 98.91 | 168 | 226 |
| 16.2 | 19411 | 19068 | 1.799 | 93.88 | 96.84 | 779 | 436 |
| 16.3 | 15541 | 15609 | -0.436 | 96.22 | 98.07 | 266 | 334 |
| 21.1 | 6726 | 6704 | 0.328 | 97.27 | 98.62 | 104 | 82 |
| 21.3 | 19030 | 19063 | -0.173 | 94.93 | 97.40 | 479 | 512 |
| 21.5 | 19458 | 19455 | 0.015 | 94.21 | 97.02 | 582 | 579 |
| 23.2 | 22911 | 18694 | 22.56 | 78.11 | 87.71 | 4665 | 448 |

The results in red represent noisy or out-of-distribution images.

#### 3.1.2. Colorectal Cancer

Figure 2 and Figure A1 show the SMA estimated using the mid-L3 slice from 90 scans of 60 patients having SMA from more than one axial series per scan in a subset of the patient scans. DAFS and SMAART-AI determine the mid-slice using their respective mechanisms, while TotalSegmentator uses the mid-slice determined by SMAART-AI. A comparison of the SMA estimates from SMAART-AI, DAFS, and TotalSegmentator reveals that both DAFS and TotalSegmentator consistently underestimate the SMA compared to SMAART-AI. The median and mean of the absolute differences between DAFS and SMAART-AI are 19.67% and 21.38%, respectively, while for TotalSegmentator and the SMAART-AI, the median and mean absolute differences are 15.73% and 15.58%, respectively. The mean/median estimated SMA is 121.1/123.03 cm² for DAFS, 126.89/124.69 cm² for TotalSegmentator, and 146.44/144.11 cm² for SMAART-AI.

Figure 3 and Figure A2 present the area estimates for 53 patient scans at the mid-L3 level (determined by SMAART-AI), the average area of slices above, below, and including mid-L3 (determined by SMAART-AI), SMAART-AI at the manually selected mid-L3, and SliceOmatic. In most cases, there is little difference between the mid-L3 SMA and the average SMA around mid-L3, with the average SMA having a mean/median of 143.22/141.81 cm², which closely matches SMAART-AI's estimate for the mid-slice 143.42/139.82. The absolute difference between the average and single mid-L3 SMA has a mean of 0.66% and a median of 0.53%. For the manually selected mid-L3 slice, the absolute difference between the SMA estimated by SMAART-AI and SliceOmatic shows a mean of 2.21% and a median of 1.38%. The mean/median areas for the manually selected mid-L3 are 138.71/137.40cm² from SliceOmatic and 141.01/139.75 cm² from SMAART-AI. The absolute difference between the estimations by DAFS versus SliceOmatic has a mean of 14.48% and a median of 14.27%, showing that DAFS underestimates the SMA for all the scans.

SMAART-AI performed well, with less than a 2.5% difference in 68% of cases when the mid-L3 slice was manually selected, though this dropped to 42% when SMAART-AI automatically selected the slice (Figure 3). Most of the considerable differences in the estimated SMA by SMAART-AI occur in CT images that are out-of-distribution or have varying levels of noise as shown in Figure 8.

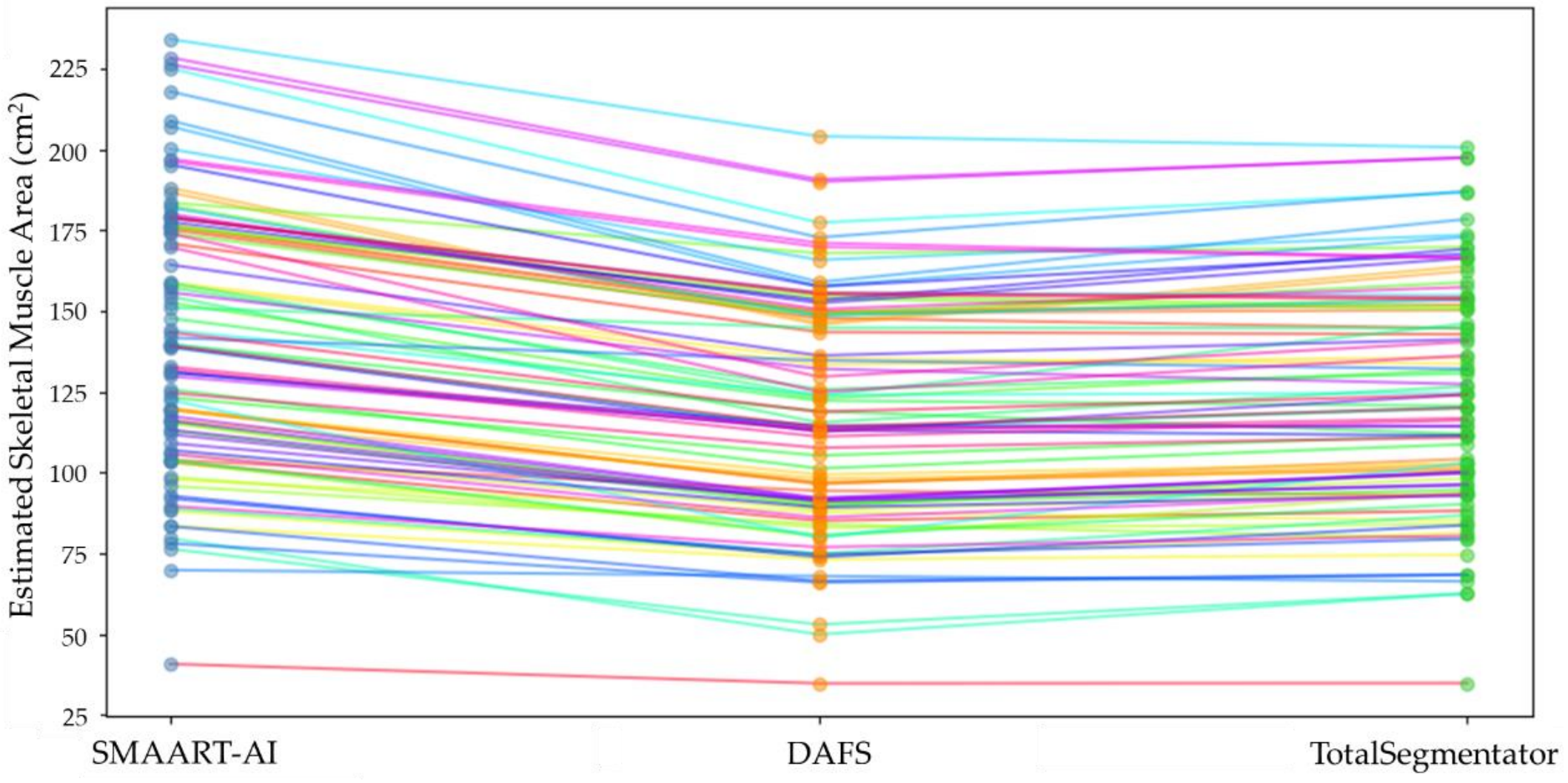

(**a**)

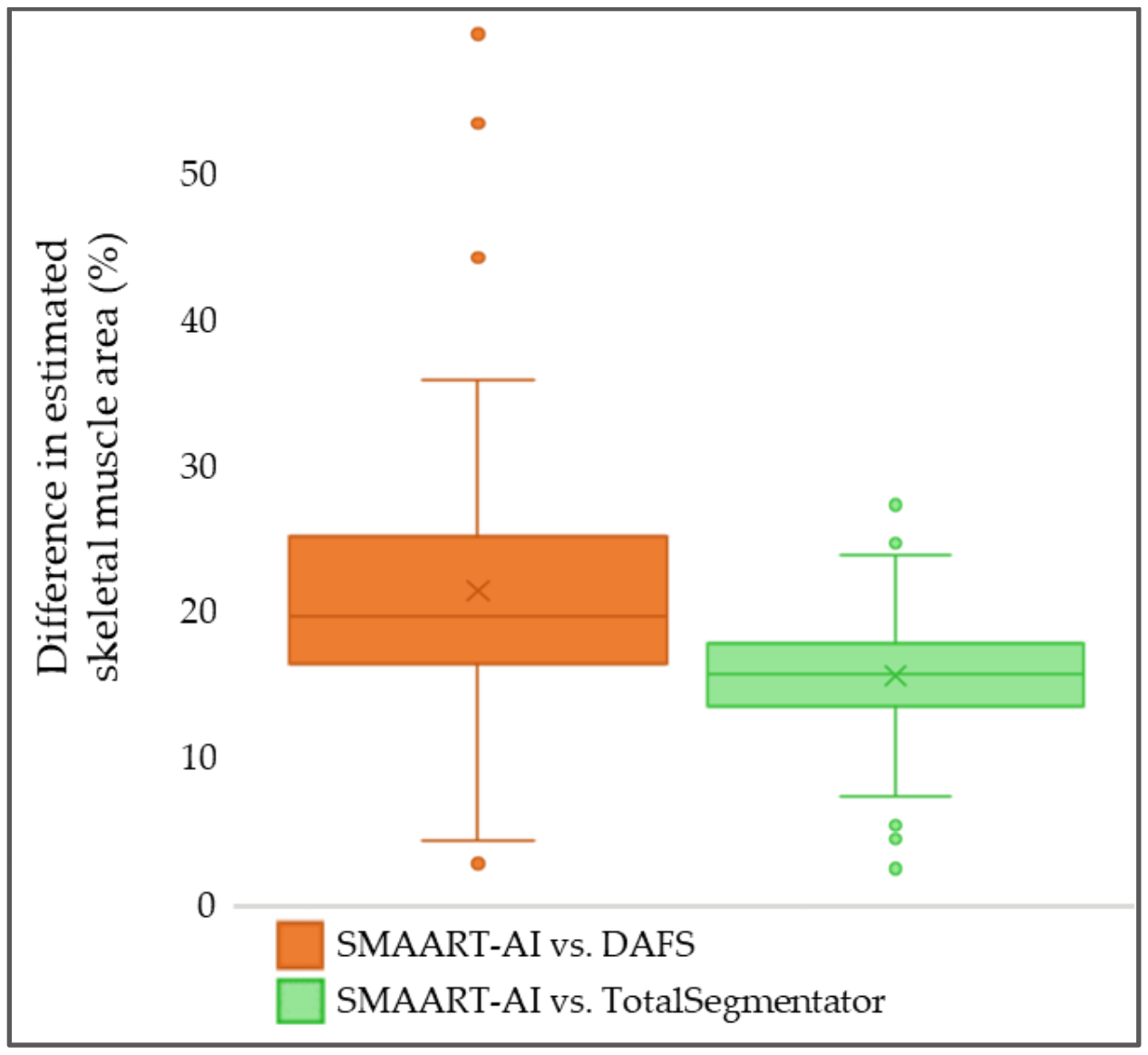

(**b**)

**Figure 2. Comparative analysis of SMA estimation using different tools for the colorectal cancer dataset.** (**a**) Comparison of SMA estimated for 60 patients (90 scans, including multiple axial series per patient) at the mid-L3 level by SMAART-AI, DAFS, and TotalSegmentator. Both DAFS and TotalSegmentator consistently estimate lower SMA values compared to SMAART-AI, with DAFS generally estimating lower values than TotalSegmentator. The mid-L3 slice used by SMAART-AI and DAFS is determined automatically by their respective pipelines, while TotalSegmentator used the mid-L3 slice determined by our proposed pipeline. (**b**) The box plot of the distribution of differences between SMA estimated by SMAART-AI and DAFS indicates a large discrepancy where DAFS underestimates the SMA, and the absolute difference has a median of 19.67%. This large difference in SMA estimations between SMAART-AI and DAFS can be potentially due to variation in the selected mid-L3 slice or poor DAFS performance on this dataset. Box plot of the distribution of differences between SMA predictions by SMAART-AI and TotalSegmentator suggests that TotalSegmentator consistently underestimates SMA in most cases with a median absolute difference of 15.73%.

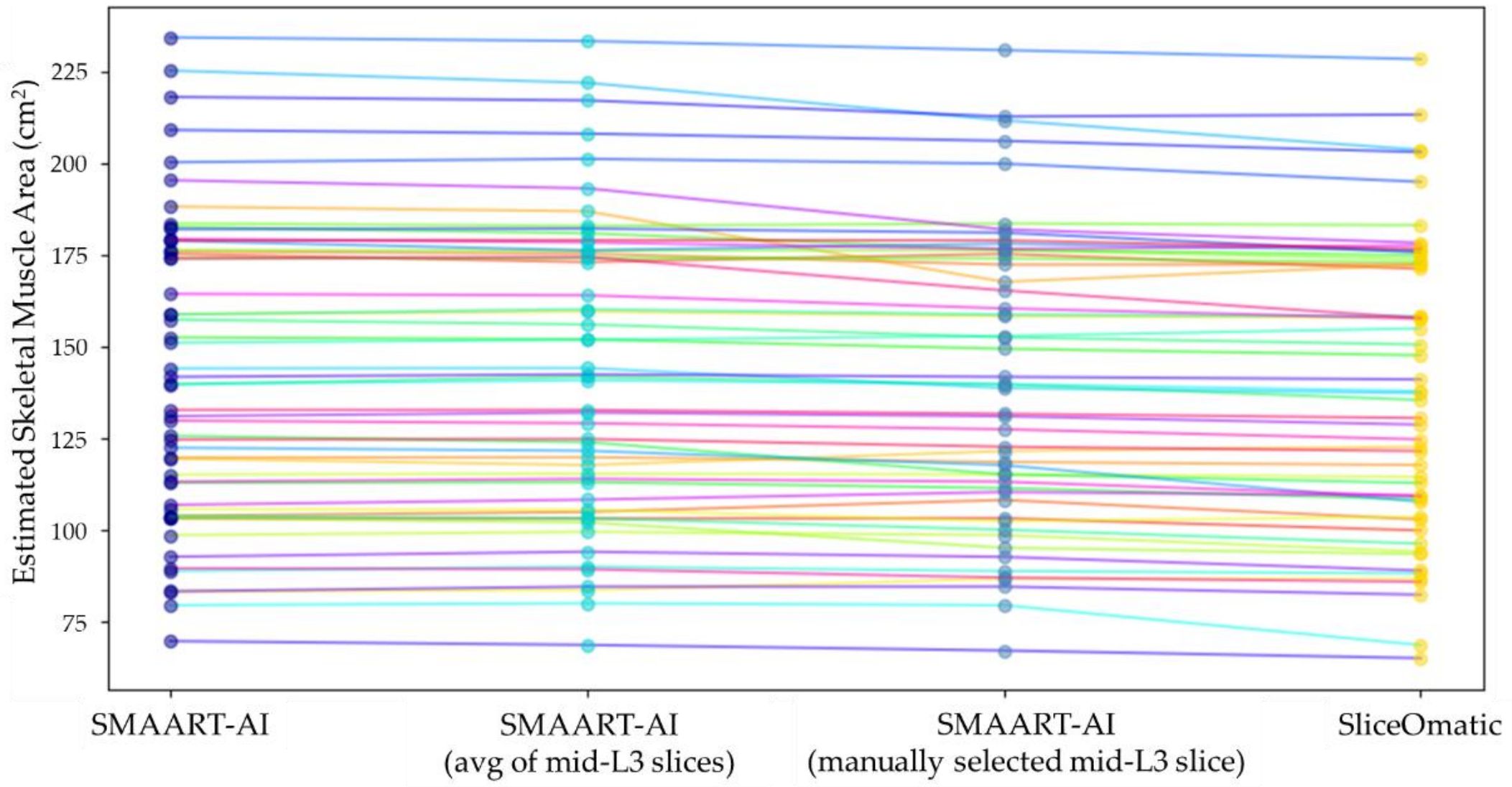

(**a**)

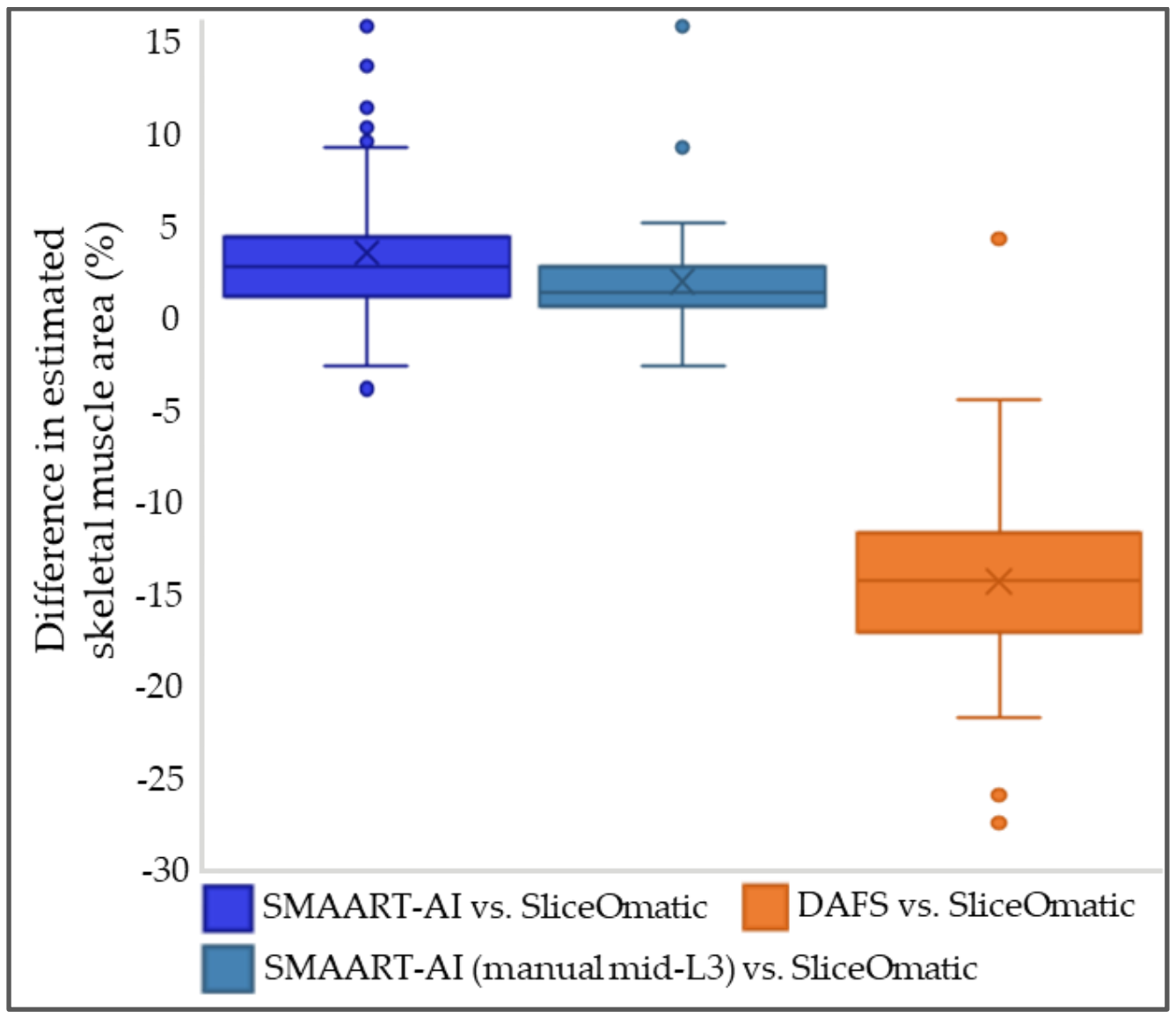

(**b**)

**Figure 3. Benchmarking SMA estimation by SMAART-AI and DAFS versus SliceOmatic for the colorectal cancer dataset.** (**a**) Comparison of SMA estimated for 53 patients by different methods: SMAART-AI, SMAART-AI using the average of slices around and including the pipeline determined mid-L3, SMAART-AI at the manually determined mid-L3 slice, and SliceOmatic (at the same manually

determined mid-L3 slice as used with SMAART-AI). The SMA values estimated by SMAART-AI at the automatically selected mid-L3 slices and the average of slices surrounding and including automatically selected mid-L3 are nearly identical in most cases but slightly differ from estimations made by SMAART-AI using manually selected mid-L3. However, estimations made by SMAART-AI and SliceOmatic at the same manually selected mid-L3 closely align with each other in most cases. (**b**) The box plot of the distribution of differences between SMAART-AI's SMA estimates, and SliceOmatic shows good agreement in many cases, with a median difference close to 0% and a small interquartile range compared to DAFS versus SliceOmatic. Differences greater than 2.5% are mainly observed when the automated and manually selected mid-L3 slices differ significantly, or when the CT image quality is poor with a lot of noise, or when the image is out-of-distribution, The box plot for the distribution of the difference between SMAART-AI's estimation and SliceOmatic at the manually determined mid-L3 indicates strong model performance overall with a median difference of 1.33% and a very small interquartile range. Larger discrepancies are mainly observed in low-quality noisy images, while differences up to 2% may be attributed to the fact that the model may include connective tissues as part of skeletal muscle. The comparison of the average area of slices around and including mid-L3 with the single mid-L3 slice has a median difference of 0.53% and a mean difference of 0.66%, indicating that adjacent slices provide similar area estimates. The box plot for the distribution of the difference between DAFS and SliceOmatic shows that DAFS underestimates the SMA with a median difference of -14.27% and most of the difference values being high beyond -5%. Overall, the model outperforms both DAFS and TotalSegmentator.

#### 3.1.3. Pancreatic Cancer

Figure 4 and Figure A3 compare the SMA estimations made by SMAART-AI, AW Server, and SliceOmatic for 153 patients and 222 patient scans at the manually selected end-L3 slice. The results indicate that TotalSegmentator consistently underestimates the SMA, while SMAART-AI and AW Server provide more closely aligned values. The mean and median estimated areas are 118.81 $cm^2$ and 116.97 $cm^2$ for TotalSegmentator, 135.02 $cm^2$ and 131.64 $cm^2$ for SMAART-AI, and 131.00 $cm^2$ and 127.70 $cm^2$ for the AW Server. The mean and median absolute differences between SMAART-AI and AW Server are 4.37% and 3.04%, respectively, while the absolute differences between SMAART-AI and TotalSegmentator are 14.38% and 14.82%.

Figure 5 and Figure A4 present SMA estimates by SMAART-AI at the manually and automatically determined end-L3 slices, with a mean and median absolute difference of 2.30% and 2.94%, respectively. The mean and median estimated areas at the manually selected end-L3 slice for SMAART-AI, AW Server, and SliceOmatic are 133.45 $cm^2$ and 131.95 $cm^2$, 130.89 $cm^2$ and 133.00 $cm^2$, and 129.39$cm^2$ and 131.60 $cm^2$, respectively. The mean and median areas estimated by SMAART-AI using the automatically identified end-L3 slice are 134.12 $cm^2$ and 131.95 $cm^2$. The absolute differences between SliceOmatic and AW Server, as well as between SliceOmatic and SMAART-AI (all SMA estimations made at the same manually selected end-L3 slice) have a mean and median of 4.74% and 3.03%, and 2.41% and 1.68%, respectively. SMAART-AI and AW Server show close agreement with SliceOmatic in approximately 87% and 61% of the cases, respectively. The data includes 79 patients and 109 patient scans.

SMAART-AI consistently matched manual SliceOmatic segmentation within a 2.5% difference in 67% of cases when the end-L3 slice was identified manually and in 49% of cases when the end-L3 slice was identified automatically by SMAART-AI, outperforming AW Server (43%).

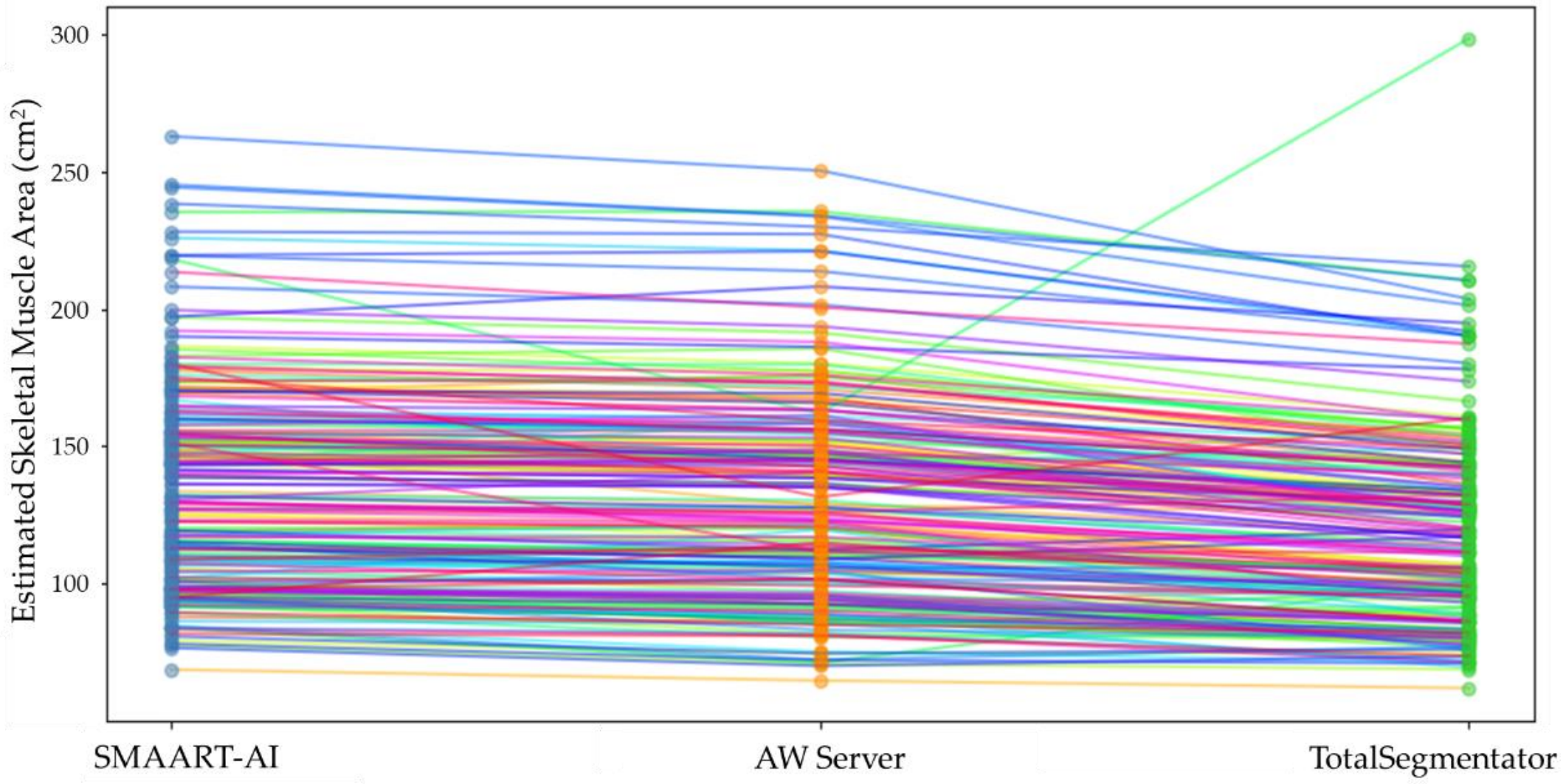

(**a**)

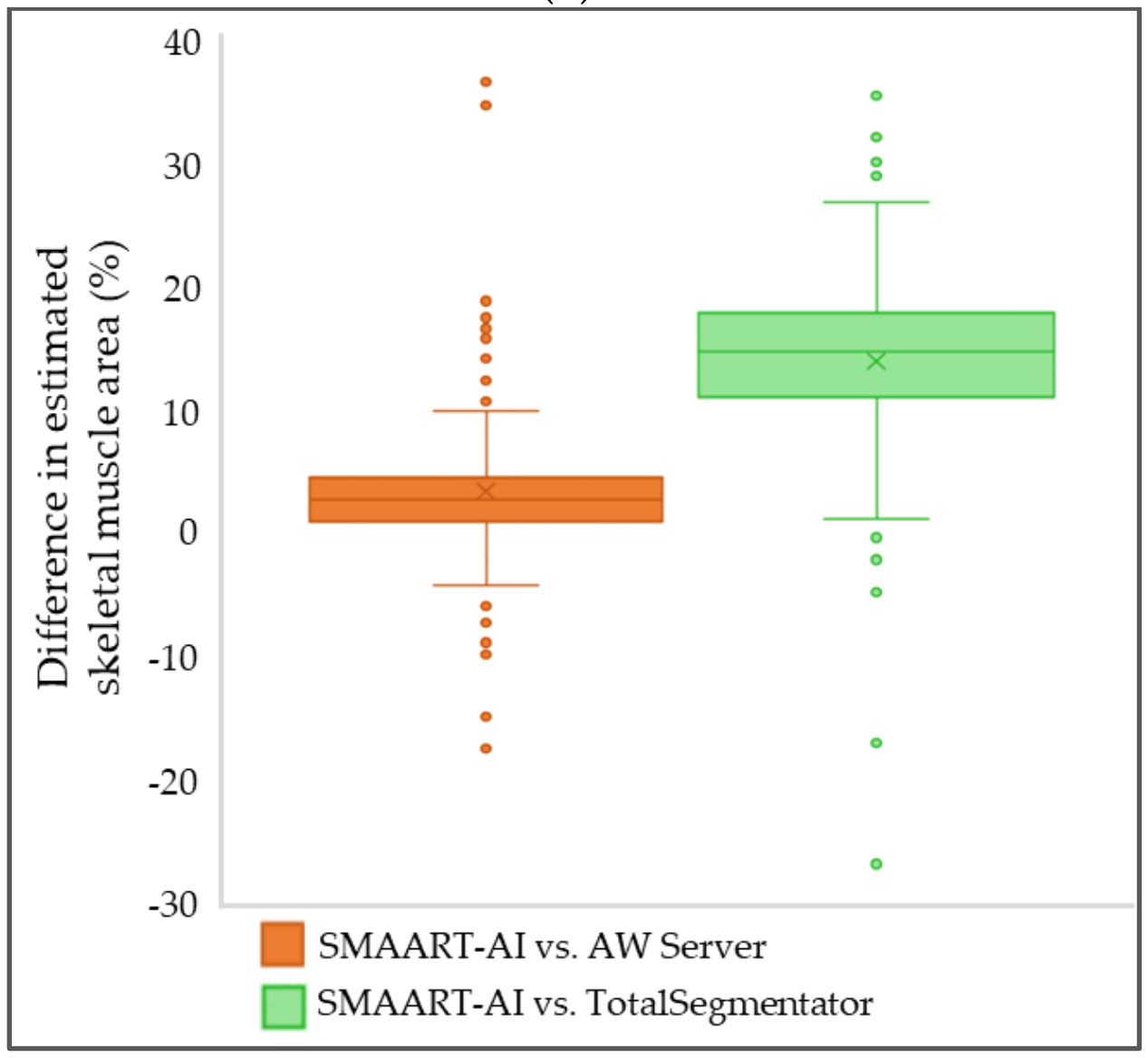

(**b**)

**Figure 4. Comparative analysis of SMA estimation using different tools for the pancreatic dataset**. (**a**) Comparison of skeletal muscle area (SMA) estimates from 222 patient scans at the manually determined end-L3 slice by SMAART-AI, AW Server, and TotalSegmentator shows that TotalSegmentator consistently estimates lower values in most cases compared to both SMAART-AI and AW Server. (**b**) The box plot for the distribution of the differences between the SMA estimated by SMAART-AI and AW Server shows the median to be close to 0 and has a small interquartile range compared to the differences between the SMA estimations by SMAART-AI and TotalSegmentator. Overall, SMAART-AI tends to underestimate in some cases but generally overestimates compared to AW Server. The box plot for the distribution of the differences in estimates between SMAART-AI and TotalSegmentator indicates

significant disagreement in most cases, with a median of 14.77% and 81% of the absolute differences being greater than 10%.

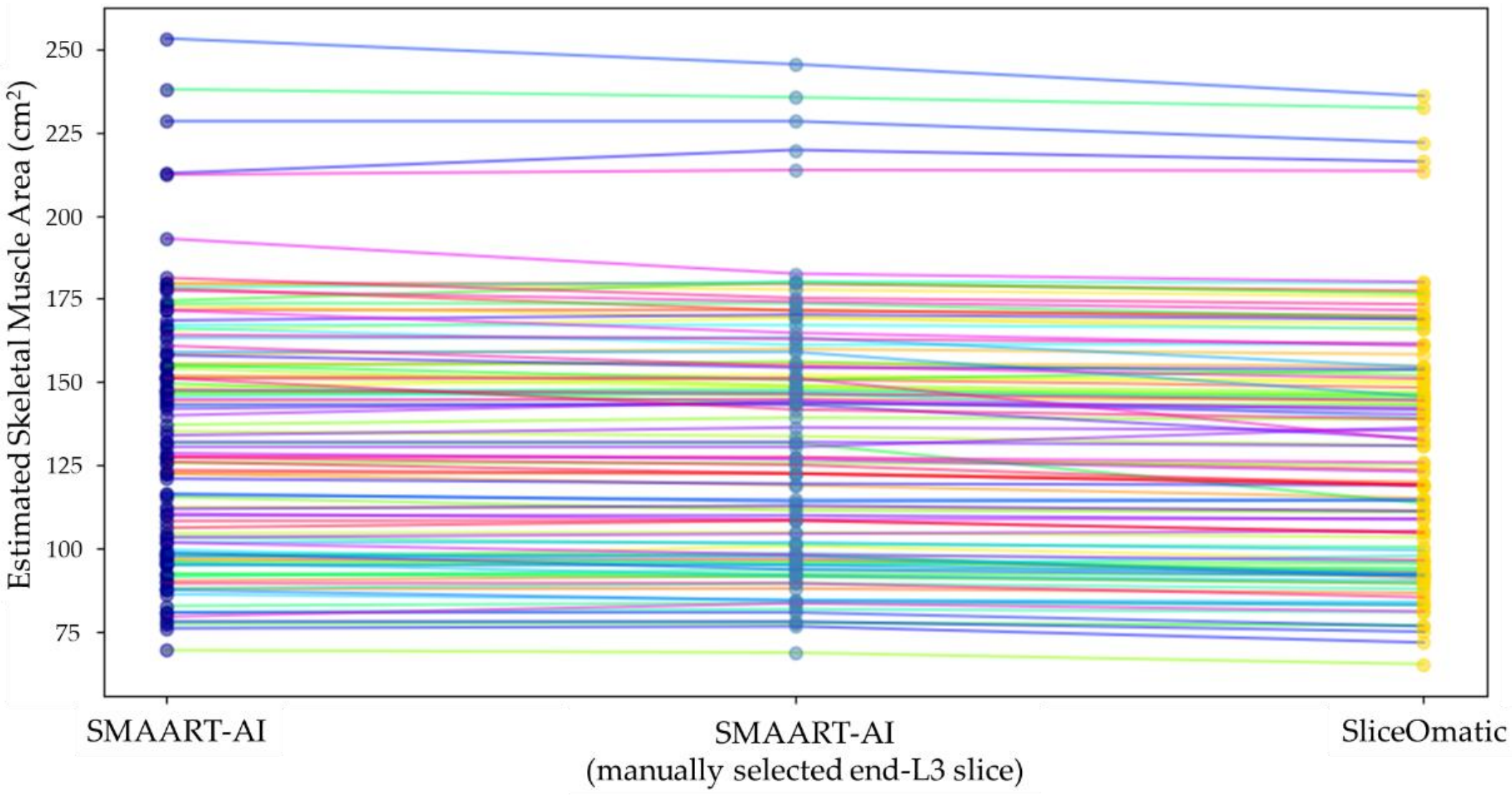

(**a**)

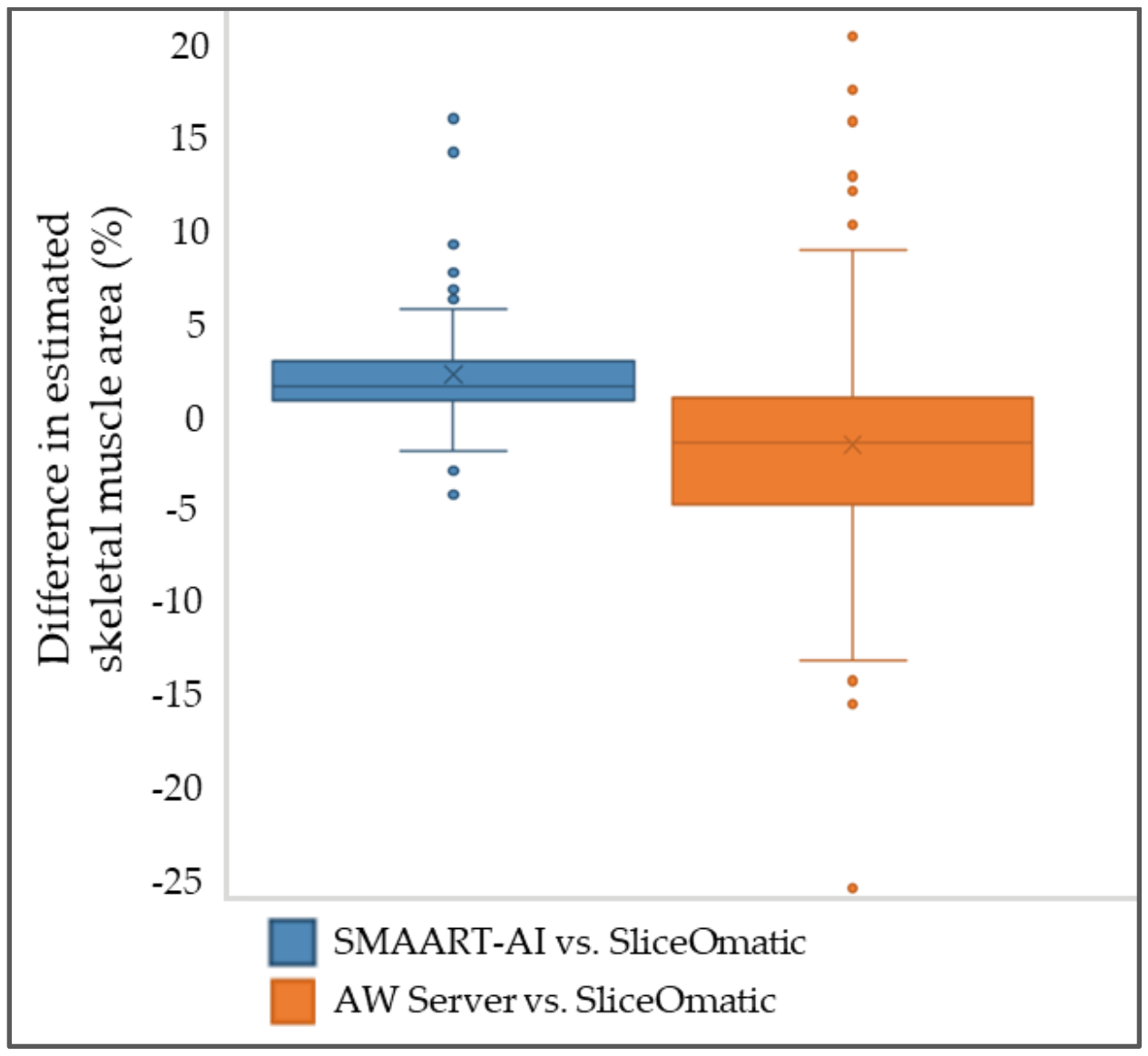

(**b**)

**Figure 5. Benchmarking SMA estimation by SMAART-AI and AW server versus SliceOmatic for the pancreatic dataset**. (**a**) Comparison of SMA estimates for 109 patient scans (each represented by a separate line), using SMAART-AI at both automatically and manually identified end-L3 slices, and SliceOmatic (at the same manually identified end-L3 slices). The SMA values from both automated and manually selected slices are nearly identical in most cases, closely matching the reference end-L3 SMA determined by SliceOmatic. (**b**) Box plot of the distribution of differences in SMA between estimates from SMAART-AI and SliceOmatic at the manually determined end-L3 slice shows a small interquartile range and median close to 0 compared to the box plot of the difference between AW Server versus SliceOmatic which has a larger interquartile range but median closer to 0 than SMAART-AI versus

SliceOmatic. However, the distribution of the differences between AW Server versus SliceOmatic is skewed towards underestimation, with the median at -1.45%. SMAART-AI tends to overestimate the SMA in some cases, whereas AW Server shows a slight bias towards underestimation. Overall, SMAART-AI performs better than AW Server when benchmarked against SliceOmatic.

#### 3.1.4. Ovarian Cancer

Figure 6 and Figure A5 present the SMA estimates from SMAART-AI, ABACS, and TotalSegmentator for the manually determined mid-L3 slice from 324 patient scans. The results reveal that TotalSegmentator consistently underestimates the SMA compared to ABACS and SMAART-AI. The mean and median estimated areas are 120.85 $cm^2$ and 116.89 $cm^2$ for SMAART-AI, 118.14 $cm^2$ and 114.2 $cm^2$ for ABACS, and 100.90 $cm^2$ and 97.54 $cm^2$ for TotalSegmentator. The median and mean absolute differences between SMAART-AI and ABACS are 4.96% and 3.23%, respectively, while the median and mean absolute differences between SMAART-AI and TotalSegmentator are 20.66% and 17.31%, respectively. In 45% of the estimations, the absolute difference between SMAART-AI versus ABACS is less than or equal to 3%.

In Figure 7 and Figure A6, the comparison of SMA estimates from 154 patient scans is presented for several methods: SMAART-AI at the automatically determined mid-L3, the average SMA of slices around and including the automatically identified mid-L3, the SMA at the manually determined mid-L3, ABACS, and SliceOmatic. The mean and median SMA values are 125.78 $cm^2$ and 123.55 $cm^2$ for SMAART-AI at automatically determined mid-L3, 125.46 $cm^2$ and 123.42 $cm^2$ for the average of slices around automatic mid-L3, 123.98 $cm^2$ and 121.49 $cm^2$ for the manually selected mid-L3, 120.98 $cm^2$ and 118.25 $cm^2$ for ABACS, and 112.01 $cm^2$ and 108.85 $cm^2$ for SliceOmatic. Additionally, the mean and median of the absolute difference in SMA between ABACS and SliceOmatic at the manually determined mid-L3 are 10.32% and 6.21%, respectively, while the mean and median absolute differences between SMAART-AI and SliceOmatic are 11.09% and 7.08%.

ABACS performed better in more cases overall, 33% within a difference of 2.5% from SliceOmatic, compared to 26% by SMAART-AI when the mid-L3 slice was identified manually and 18% when the mid-L3 slices were identified automatically by SMAART-AI. Nevertheless, SMAART-AI made accurate estimations in comparison with ABACS in certain images as shown in Figure 9.

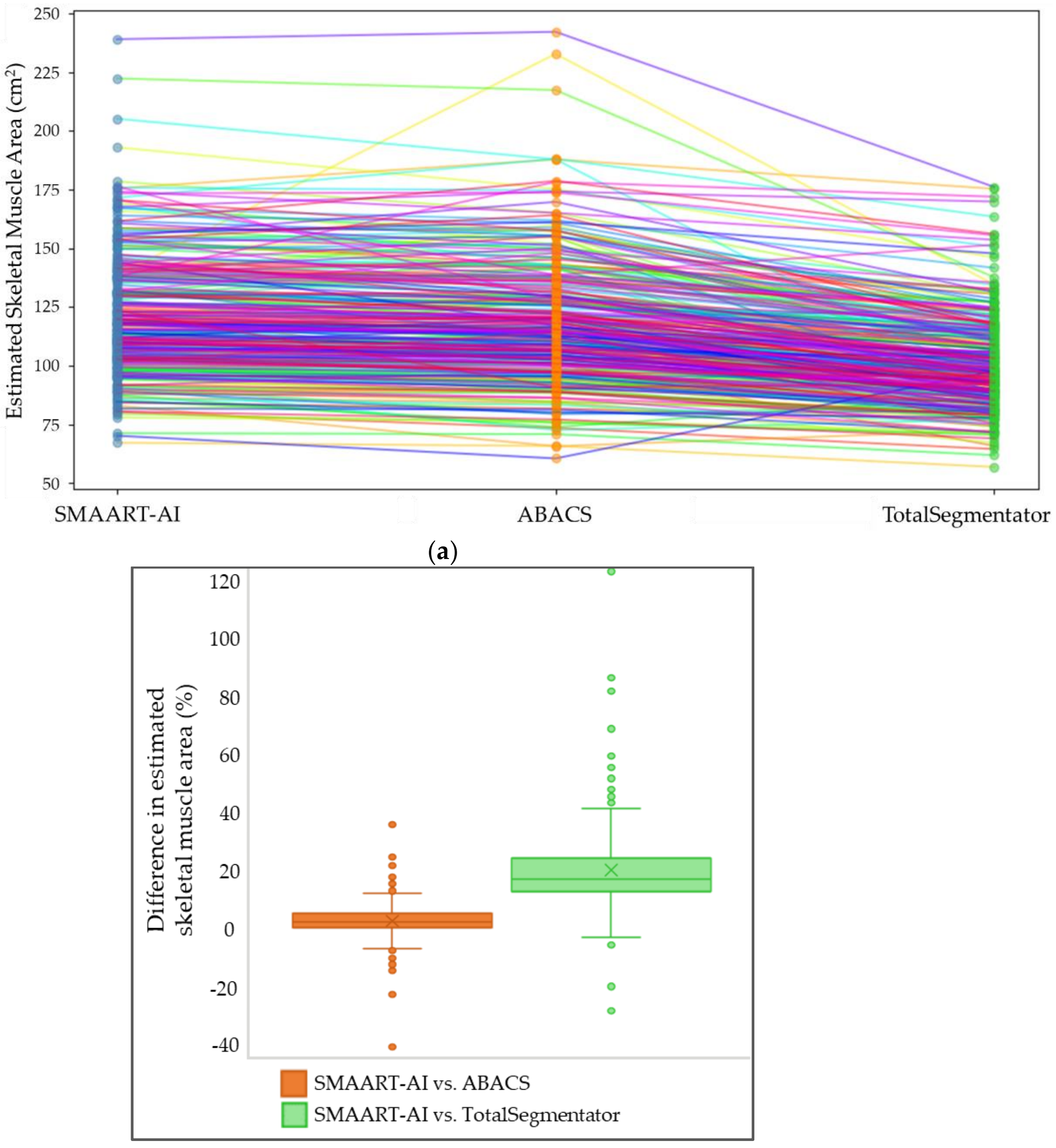

**Figure 6. Comparative analysis of SMA estimation using different tools for the ovarian cancer dataset**. (**a**) A comparison of skeletal muscle area (SMA) from 324 patient scans estimated by SMAART-AI, ABACS, and TotalSegmentator, at the same manually determined mid-L3 slice, shows that TotalSegmentator estimates lower values compared to SMAART-AI and ABACS with some exceptions. There are a few odd cases where ABACS estimates very large values for SMA compared to SMAART-AI and TotalSegmentator. (**b**) The box plot of the distribution of the difference between the SMA estimated by SMAART-AI and ABACS has a median close to zero and the interquartile range much smaller than for the distribution of differences between SMAART-AI versus TotalSegmentator. The box plot of the distribution of the difference between SMAART-AI and TotalSegmentator shows high disagreement in most cases, with a median difference of 17.17%.

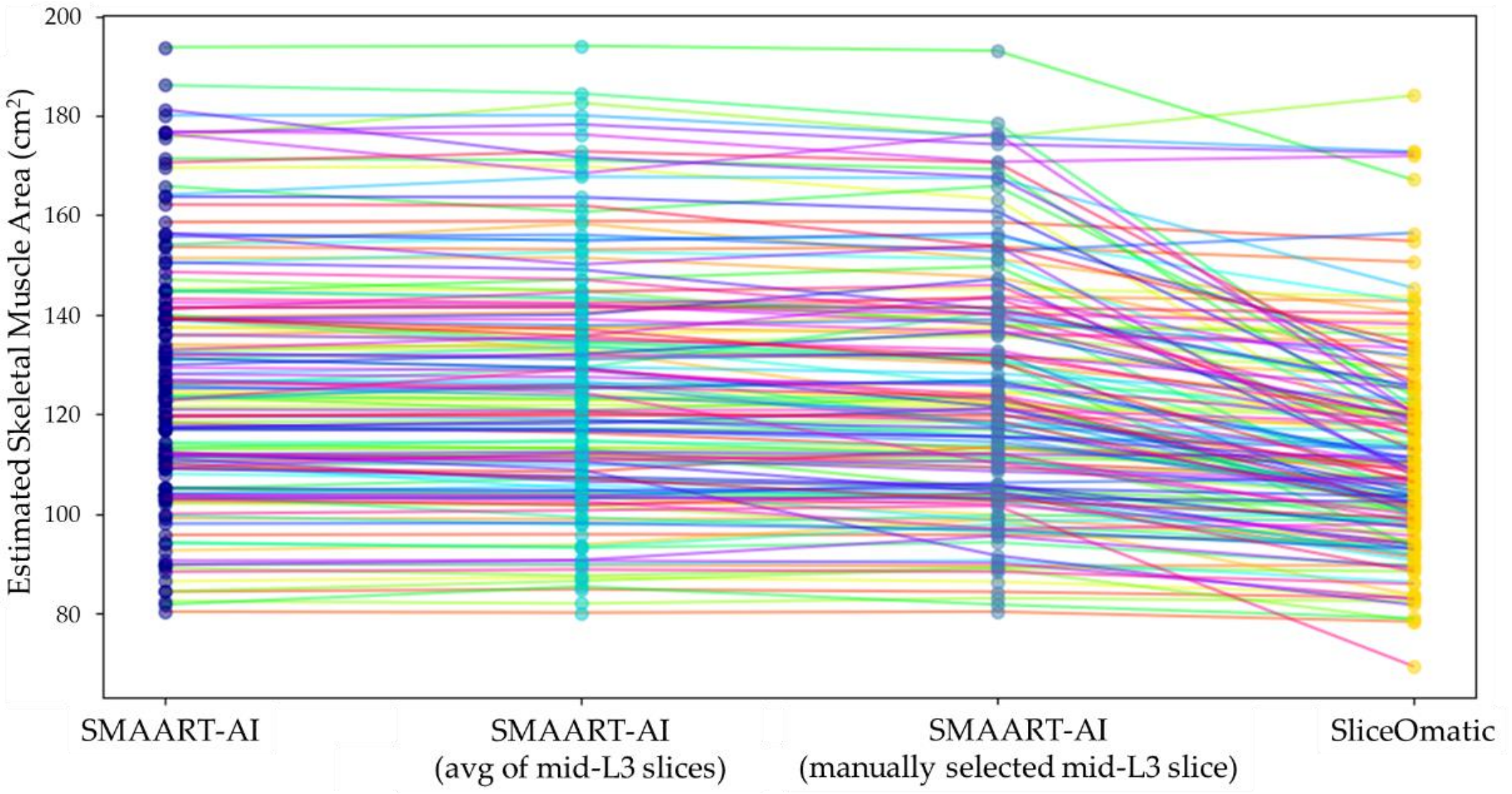

(**a**)

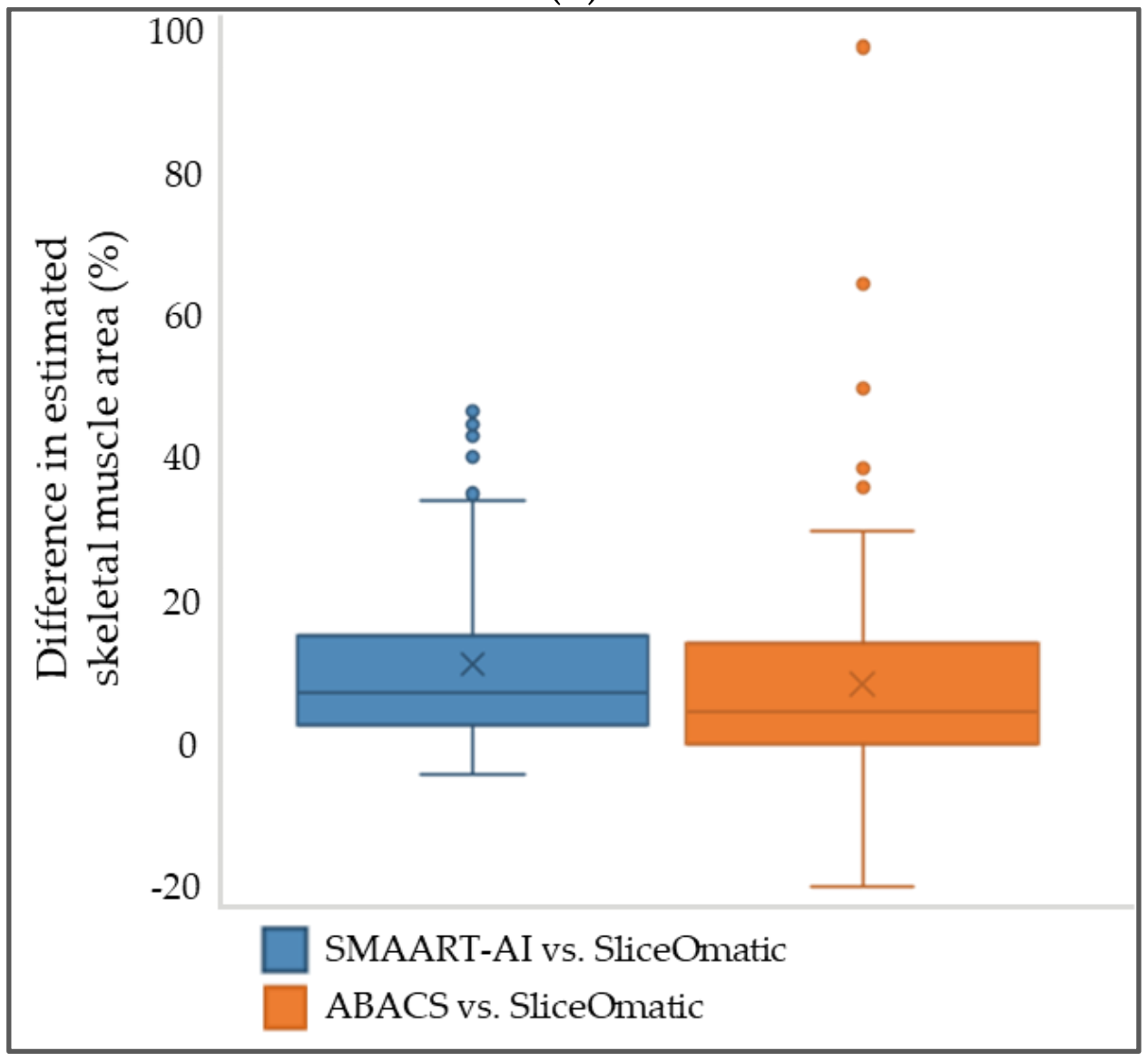

(**b**)

**Figure 7. Benchmarking SMA estimation by SMAART-AI and ABACS versus SliceOmatic for the ovarian cancer dataset**. (**a**) Comparison of SMA estimates from 154 patients' CT scans using different methods: SMAART-AI at mid-L3 (automated), the average of slices around and including SMAART-AI determined mid-L3, SMAART-AI at the manually determined mid-L3 slice, and SliceOmatic (at the same manually determined mid-L3 slice). In most cases, SMA estimates by SMAART-AI for the automatically selected mid-L3 and the average of the adjacent slices around and including mid-L3 match closely. However, there are a greater number of mismatches between the estimations made at the automatically selected mid-L3 slice and the manually selected mid-L3 slice compared to the automatically selected mid-L3 and the average of estimations at slices adjacent to the automatically selected mid-L3. The number of SMA estimations by SMAART-AI (at the automatic and manually selected mid-L3) that closely match the estimations made by SliceOmatic is less than the number of mismatches. (**b**) The box plots of the distribution of differences between SMA estimated by ABACS and SliceOmatic and SMAART-AI versus SliceOmatic, at the manually determined mid-L3 show similar performance in most cases. ABACS

versus SliceOmatic box plot shows some considerably high overestimations and some underestimations. SMAART-AI can be seen to be skewed towards overestimations. The box plot of the distribution of differences between SMA estimated by SMAART-AI versus SliceOmatic (at the same manually determined mid-L3 slice) has a slightly less widespread compared to ABACS vs. SliceOmatic differences. However, the differences between the first quartile and the median are closer to zero for ABACS versus SliceOmatic compared to SMAART-AI versus SliceOmatic. Overestimations by both ABACS and SMAART-AI are primarily attributed to out-of-distribution or noisy images, which are prevalent in this dataset.

#### 3.1.5 Sample Images, Masks, and Uncertainty Maps

Figure 8 shows some samples of noisy or out-of-distribution CT images with the corresponding skeletal muscle mask and uncertainty map produced by SMAART-AI from all datasets used in our work.

Figure 9 compares the skeletal muscle masks generated automatically by TotalSegmentator, SMAART-AI, and manually by SliceOmatic. These samples illustrate the underestimated skeletal muscle masks generated by TotalSegmentator compared to SMAART-AI and SliceOmatic at the same mid-L3 slice.

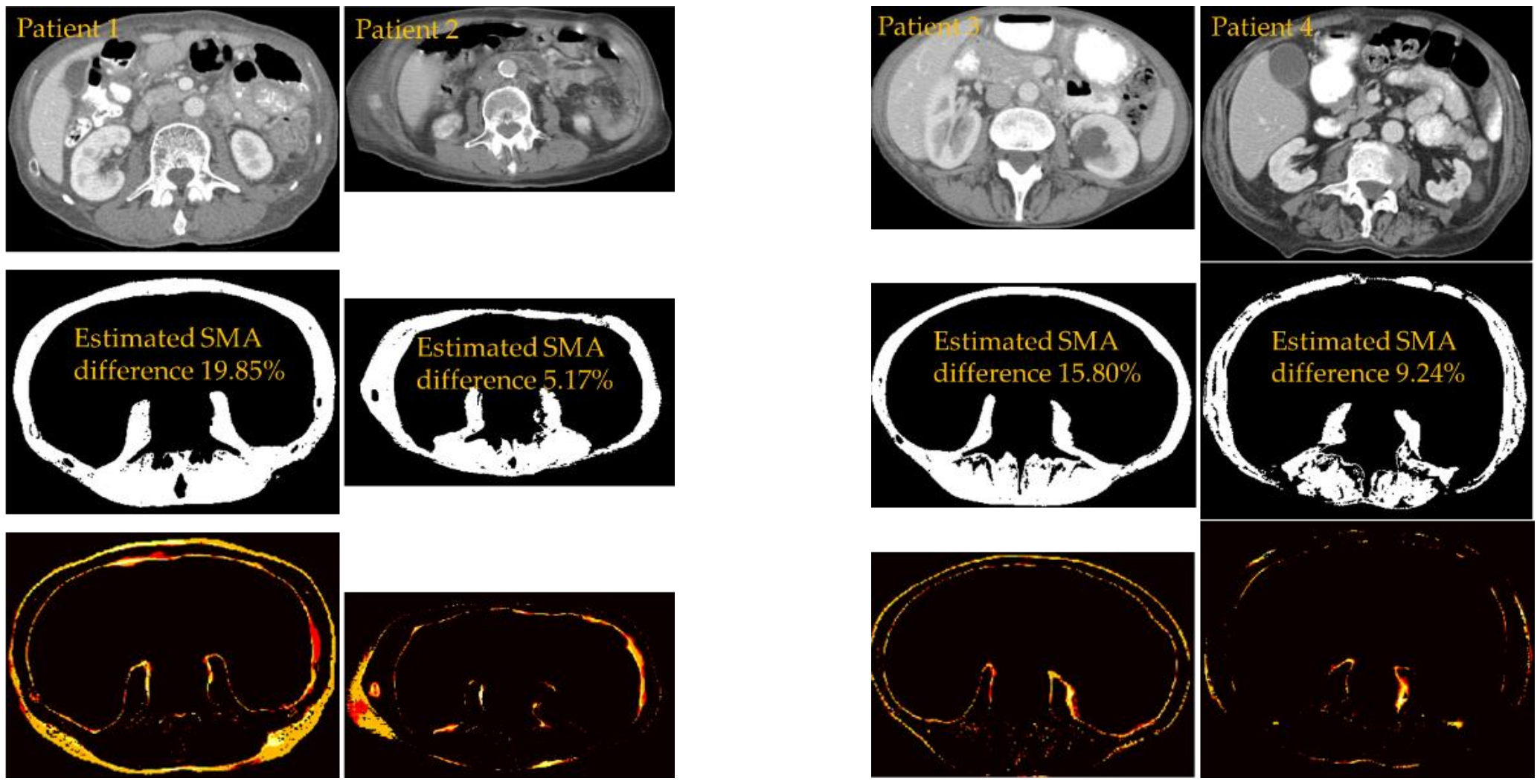

(**a**) (**b**)

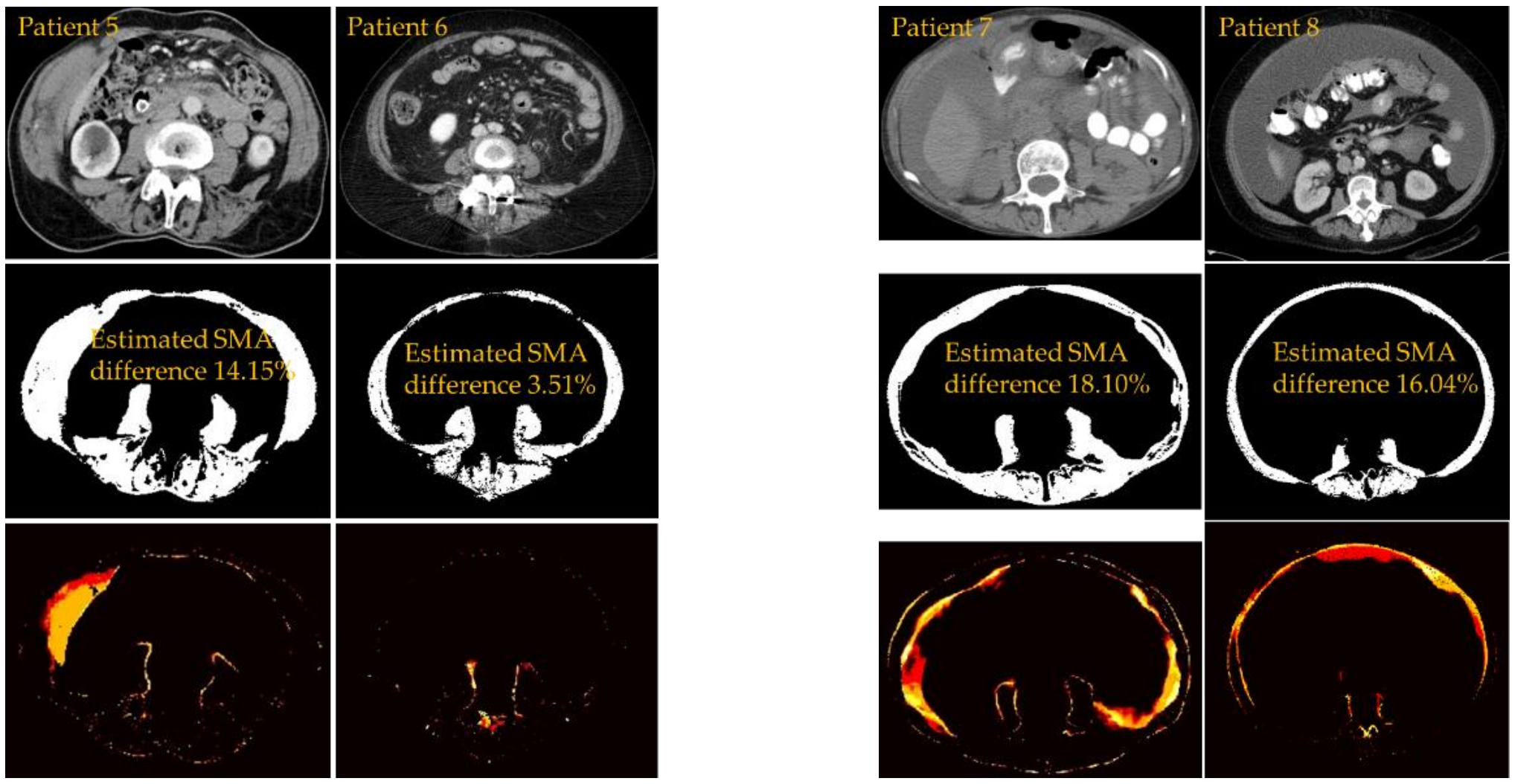

(c) (d)

**Figure 8. Samples of noisy and out-of-distribution images on which SMAART-AI did not perform well in producing the correct skeletal muscle mask**. (**a**) Gastroesophageal dataset samples. (**b**) Colorectal dataset samples for which SMAART-AI estimated SMA were 79.65/117.76 vs. DAFS estimation of 49.93/79.85 and SliceOmatic estimated 63.48/102.12. (**c**) Pancreatic dataset samples for which SMAART-AI estimated SMA were 151.4/108.4 versus AW server estimation of 111.8/100.4 and SliceOmatic estimated 132.4/104.7. (**d**) Ovarian dataset samples for which SMAART-AI estimated SMA were 98.91/138.09 versus ABACS estimation of 101.0/178.3 and SliceOmatic estimated 83.75/119.0. The percentage difference in SMA is SMAART-AI's estimation benchmarked against manual SMA segmentation using SliceOmatic.

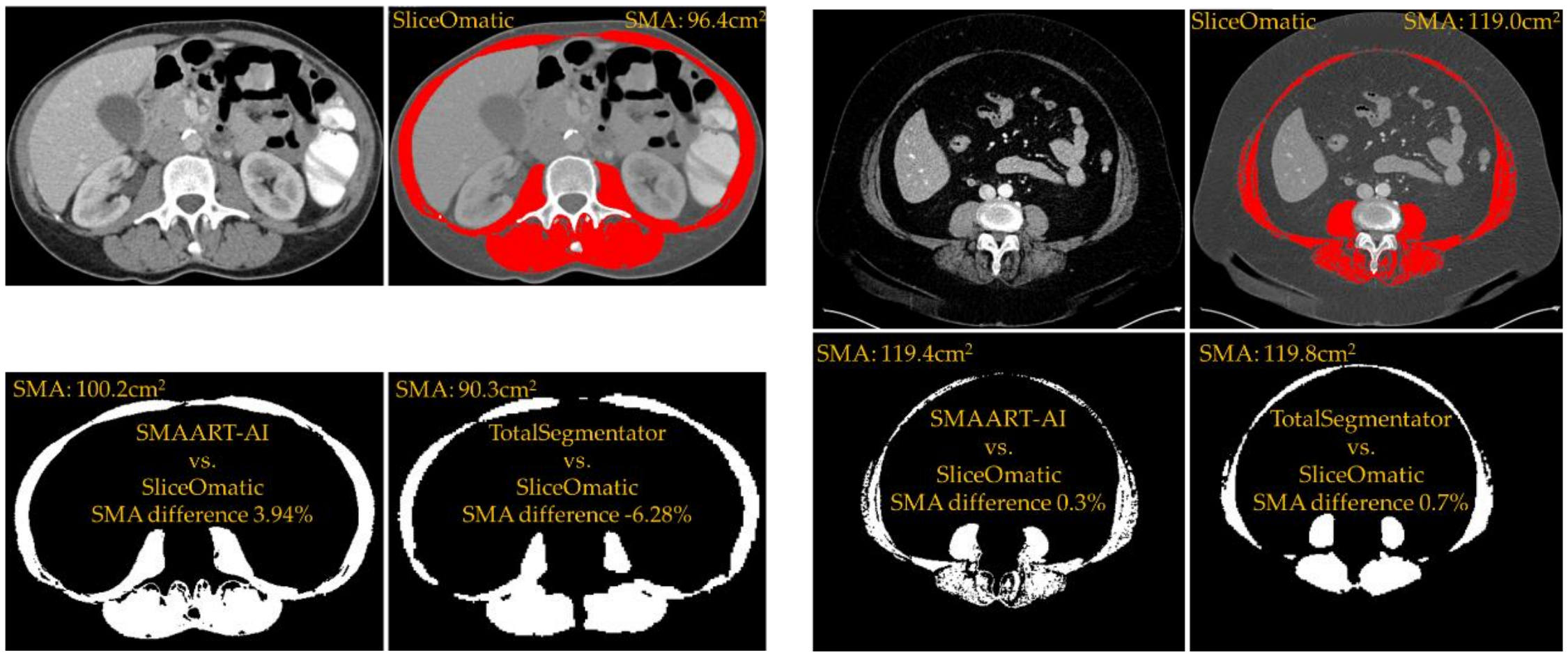

(**a**) (**b**)

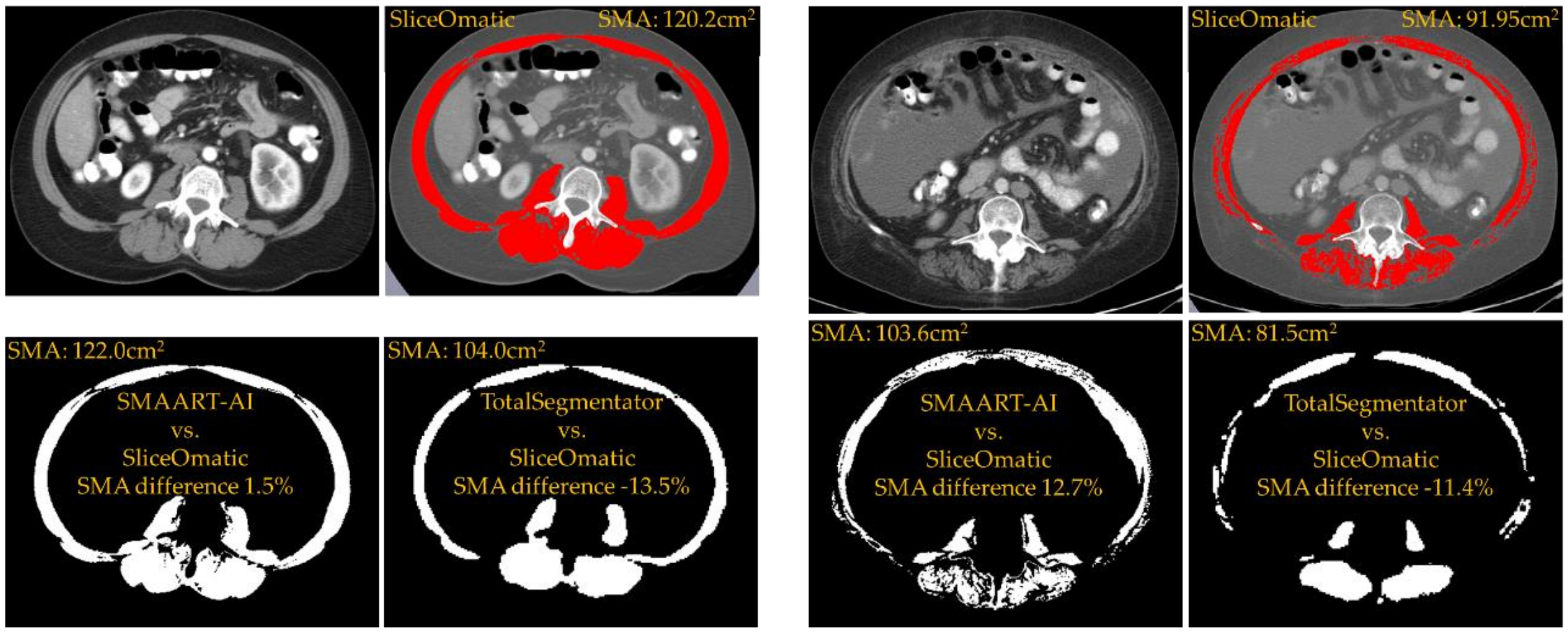

(c) (d)

**Figure 9. Comparison of the skeletal muscle mask generated using SMAART-AI, TotalSegmentator, and SliceOmatic at the same manually identified mid-L3 slice**. (**a**) Colorectal dataset sample image showing overestimation by SMART-AI and underestimation by TotalSegmentator compared to manual segmentation using SliceOmatic. SMA estimated by DAFS was 80.7cm$^2$ at the mid-L3 slice, automatically determined by its own internal mechanism. The SMA estimated by SMAART-AI, TotalSegmentator, and SliceOmatic was 100.2 cm$^2$, 90.3 cm$^2$, and 96.4 cm$^2$. (**b**) Pancreatic dataset sample image shows close SMA estimation by SMAART-AI (119.4 cm$^2$) and TotalSegmentator (119.8 cm$^2$) compared to manual segmentation using SliceOmatic (119.0 cm$^2$). The skeletal muscle mask generated by TotalSegmentator is not complete, but some pixels that do not belong to the skeletal muscle have been marked. Hence, the SMA estimated is close to that of SMAART-AI and SliceOmatic (manual segmentation). AW Server used manual estimation based on Hounsfield windowing, at the same mid-L3 slice, SMA of 112.3 cm$^2$, which is an underestimation compared to the manual segmentation using SliceOmatic. (**c**) Ovarian dataset image sample shows a close estimation of SMA by SMAART-AI (122.0 cm$^2$) and an underestimation by TotalSegmentator (104.0 cm$^2$) compared to manual segmentation using SliceOmatic (120.2 cm$^2$). ABACS underestimated the SMA to be 114.6 cm$^2$ compared to manual segmentation using SliceOmatic at the same mid-L3 slice. (**d**) Ovarian dataset sample image showing overestimation by SMAART-AI (103.6 cm$^2$) and underestimation by TotalSegmentator (81.5 cm$^2$) compared to manual segmentation using SliceOmatic (91.95 cm$^2$). ABACS overestimated the SMA (114.8 cm$^2$) compared to both SMAART-AI and manual segmentation using SliceOmatic at the same mid-L3 slice.

### *3.2. Comparison of the Uncertainty Methods and Metrics*

#### 3.2.1. Dropout Method

Table 7 presents the correlation coefficients between the uncertainty metrics calculated using the dropout method and the difference between SMAART-AI estimated SMA and the manually measured SMA from SliceOmatic. The results show a very strong correlation (r >0.7) for all metrics, except for the coefficient of variation (SMA), which exhibits a weak correlation (r = 0.296), and the average variance (SM), which maintains a strong relationship (r = 0.664).

**Table 7.** Pearson Correlation Coefficient between Skeletal Muscle Area Differences and Uncertainty Metrics using the Dropout Method on Gastroesophageal Dataset.

| Uncertainty Metrics for Dropout | Corr Coefficient |
| --- | --- |
| Average Probability* | -0.863 |
| Average Probability (SM)* | -0.823 |
| Coefficient of Variation (pixel-wise)* | 0.739 |
| Coefficient of Variation (SMA) | 0.296 |
| Average Variance* | 0.720 |
| Average Variance (SM)* | 0.664 |
| Average Entropy* | 0.867 |
| Expected Entropy of the Ensemble* | 0.869 |

*significant (p-value < 0.05), SM = Skeletal Muscle, SMA = Skeletal Muscle Area, Corr = correlation

#### 3.2.2. Ensemble Method

Table 8 presents the correlation coefficients between the differences in SMA estimates (from SliceOmatic (manual) and SMAART-AI) and various uncertainty metrics across four datasets. In the gastroesophageal cancer dataset, all methods and metrics demonstrate very strong correlations, with coefficients exceeding 0.7. The colorectal cancer dataset also shows very strong correlations for the coefficient of variation in SMA estimates from the ensemble, and strong correlations (above 0.5) for average variance (overall and for skeletal muscle pixels) as well as the coefficient of variation. The pancreatic cancer dataset exhibits strong correlations (above 0.5) for most metrics, except for the average calibrated probability and expected entropy of the ensemble, which show moderate correlations below 0.4. In the ovarian cancer dataset, correlations are strong across all metrics, with values above 0.6, and particularly very strong correlations (above 0.7) for the coefficient of variation, average entropy, and average calibrated probability.

**Table 8.** Pearson Correlation Coefficient between Skeletal Muscle Area Differences and Uncertainty Metrics using the Ensemble and Calibration Methods.

| Uncertainty Methods and Metrics | GE | CRC | Pan | Ova |
| --- | --- | --- | --- | --- |
| Average Probability* | -0.842 | -0.487 | -0.503 | -0.763 |
| Average Calibrated Probability* | -0.813 | -0.442 | -0.316 | -0.782 |
| Coefficient of Variation (pixel-wise)* | 0.852 | 0.529 | 0.526 | 0.756 |
| Coefficient of Variation (SMA)* | 0.910 | 0.759 | 0.522 | 0.660 |
| Average Variance* | 0.866 | 0.571 | 0.546 | 0.755 |
| Average Variance (SM)* | 0.723 | 0.647 | 0.523 | 0.798 |
| Average Entropy* | 0.843 | 0.474 | 0.516 | 0.749 |
| Expected Entropy of the Ensemble* | 0.701 | -0.442 | -0.316 | 0.655 |

*significant (p-value < 0.05), SM = Skeletal Muscle, SMA = Skeletal Muscle Area, CRC = Colorectal, GE = Gastroesophageal, Pan = Pancreatic, Ova = Ovarian cancer datasets.

The dropout method was tested only on the gastroesophageal dataset, as its overall performance in estimating SMA was slightly weaker than that of the ensemble method, as shown in Tables 5 and 6. For both the dropout and ensemble techniques, the average Jacquard Index—excluding the four difficult images highlighted in red—was 95.52 and 95.71, respectively, and the average Dice score was 97.70 and 97.80. The average difference in estimated SMA was 1.02% for the dropout method and 0.90% for the ensemble method.

#### 3.2.3. Uncertainty-Based Detection of Performance Degradation

From Table 8, we handpicked two representative metrics that strongly correlated with the difference in estimated SMA. Figure 10 presents scatter plots of these metrics, average variance (ensemble), and coefficient of variation (SMA-ensemble), demonstrating their utility in identifying high-error cases using the thresholding mechanism.

In Figure 10, the green and blue shaded regions represent ideal outcomes: the green region highlights cases with high differences and high uncertainty, while the blue region contains cases with low differences and low uncertainty. Both dashed lines are adjustable for analysis—the horizontal line represents the uncertainty threshold, and the vertical line separates high-difference cases from low-difference ones. The top white region shows the number of cases where uncertainty is above the threshold, yet the difference or error in SMA estimation is low, falling below the boundary that separates high and low difference cases. In contrast, the bottom white region contains cases with a high difference but low uncertainty estimates, meaning these cases are not captured by the uncertainty threshold.

For the gastroesophageal dataset (Figure 10(a) and (b)), we see that the thresholding mechanism works better with the coefficient of variation than the average variance since all five estimations with a difference higher than 2.5% are segregated with seven low difference cases having uncertainty higher than the threshold. Whereas, using the average variance if the threshold is adjusted so that all five estimations with a high difference are segregated it includes twelve low difference cases having uncertainty higher than the threshold. For the colorectal (Figure 10(c) and (d)), pancreatic (Figure 10(e) and (f)), and ovarian datasets (Figure 10(g) and (h)), both the metrics for uncertainty with the threshold work almost equally well.

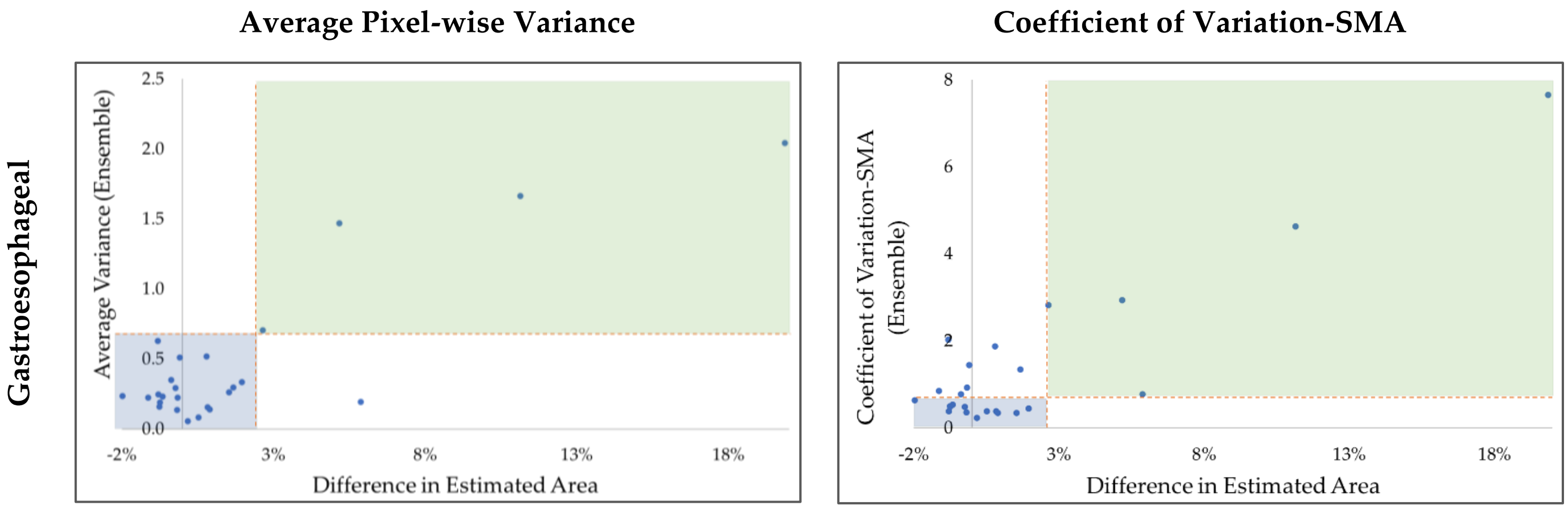

(**a**) (**b**)

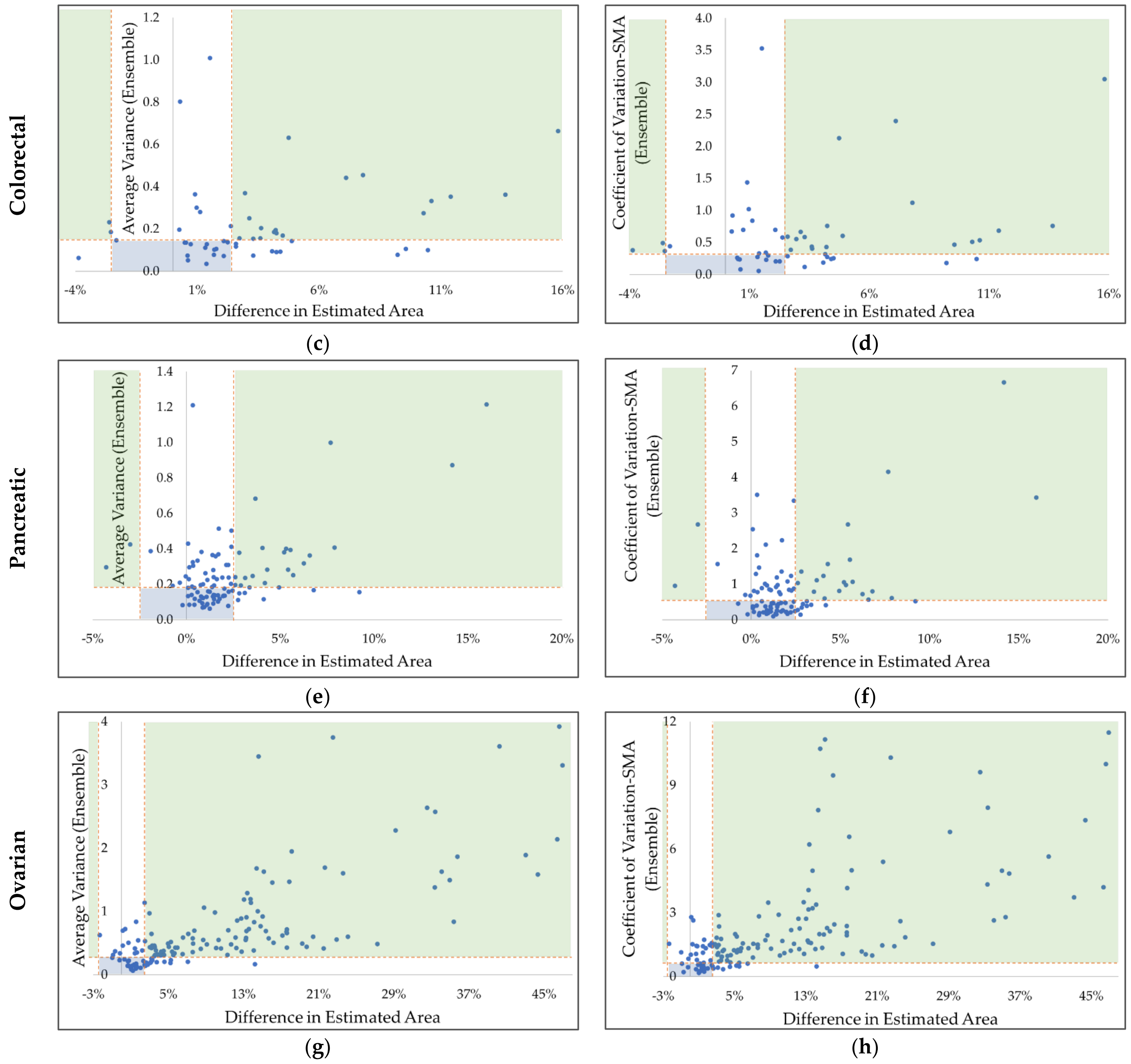

**Figure 10. Uncertainty scatter plots for the four datasets and two metrics, average variance (ensemble) and coefficient of variation (SMA)**. (**a**), (**c**), (**e**), and (**g**) display the average pixel-wise variance of the estimated SMA from the ensemble, while plots (**b**), (**d**), (**f**), and (**h**) show the coefficient of variation of the estimated SMA using the ensemble method. The horizontal dashed line represents an adjustable uncertainty threshold, which can be used to identify cases where SMAART-AI's estimated SMA may have high error. The green and blue quadrants highlight required segregation: the blue quadrant represents low-difference, low-uncertainty cases, and the green quadrant represents high-difference, high-uncertainty cases. Cases in the other two quadrants fall under either low-difference, high-variance, or high-difference, low-variance categories. Both uncertainty estimation metrics exhibit different spreads of uncertain cases, due to this, the efficiency of the thresholding mechanism for

identifying potentially high-error cases differs slightly within and between different datasets.

### *3.3. Skeletal Muscle Index (SMI) Longitudinal Tracking*

Figure 11 illustrates the longitudinal tracking of SMI during pancreatic cancer treatment of patients from the pancreatic cancer dataset. For these patients, CT scans were available at one or two additional time points after the initial cancer diagnosis, with an average interval of approximately six months. As seen in Figure 11, patients identified as cachectic at the time of diagnosis continue to lose muscle mass at varying rates. However, there are instances where an increase in SMA is also observed. Additionally, some patients initially diagnosed as non-cachectic begin to lose muscle at different rates later during their treatment.

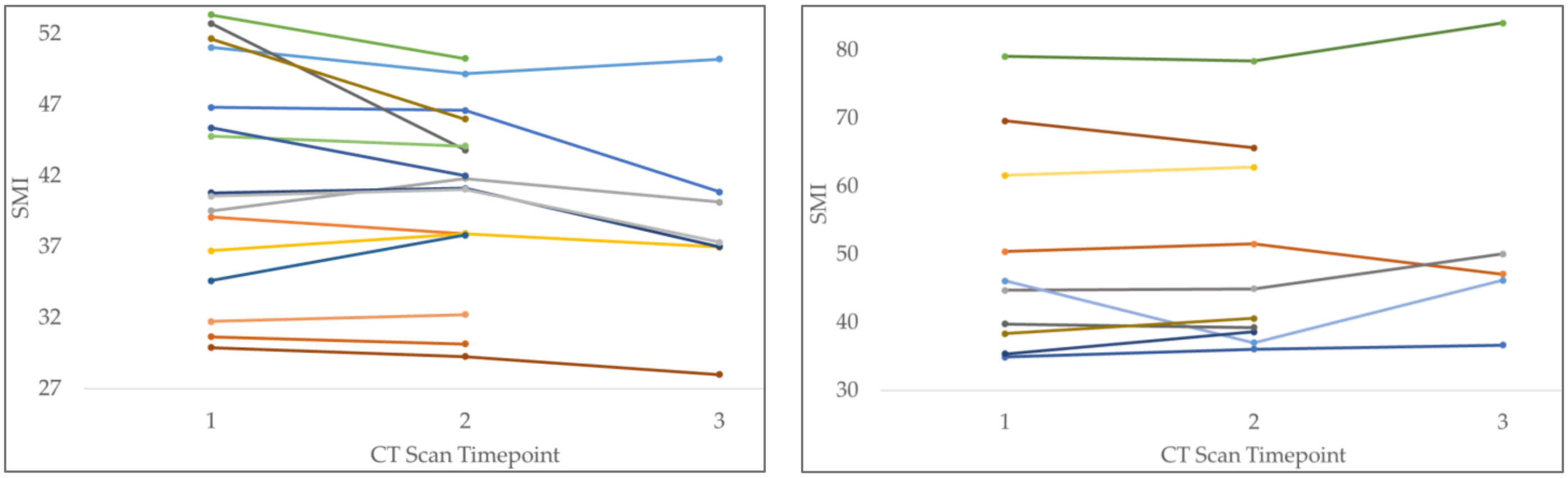

(**a**) (**b**)

**Figure 11. Longitudinal tracking of SMI to monitor (a) cachectic and (b) non-cachectic pancreatic cancer patients**. Most of the patients diagnosed as cachectic at the time of cancer diagnosis show a decrease in SMI over time (approximately six months on average between each scan). For some cachectic patients, an increase in SMI may be attributed to edema, which causes blurring between tissues and pixels that do not belong to the skeletal muscle to appear as muscle. Manual marking of skeletal muscle also becomes challenging in such cases. Some of the non-cachectic patients also show a decrease in SMA which may be indicative of the onset of cachexia.

### *3.4.Statistical Analysis*

#### 3.4.1. Overall Survival Analysis

The concordance index (Table 9) for training and test sets with different combinations of clinical data and BMI, SMI, and SMA shows that SMI, SMA, and BMI noticeably improve the concordance index, with increases of 3.0%, 6.7%, and 1.5% for the colorectal, pancreatic, and ovarian datasets, respectively compared to using only clinical data (without BMI, SMI, and SMA). Excluding BMI while retaining SMI and SMA has minimal effect, leaving the test concordance unchanged for the colorectal and ovarian datasets and resulting in only a 0.5% decrease for pancreatic cancer.

**Table 9.** Overall Survival Analysis showing Train and Test Concordance Index for Pancreatic, Colorectal, and Ovarian Cancer Dataset.

| Dataset | Penalizer | With BMI/ SMI/ SMA | With SMI/ SMA | With BMI | Without BMI/ SMI/ SMA |
|---|---|---|---|---|---|
| Colorectal | 0.5 | **0.701/0.522** | 0.705/0.507 | 0.701/0.463 | 0.705/0.463 |
| | 1.0 | 0.693/0.522 | 0.712/0.552 | 0.705/0.493 | 0.705/0.478 |
| | 1.5 | 0.693/0.522 | **0.708/0.576** | **0.701/0.493** | 0.708/0.478 |
| | 2.0 | 0.693/0.507 | 0.708/0.567 | 0.700/0.493 | **0.708/0.507** |
| Pancreatic | 0.1 | 0.670/ 0.601 | 0.671/ 0.610 | **0.650/ 0.607** | **0.649/ 0.601** |
| | 0.5 | 0.664/ 0.632 | **0.666/ 0.638** | 0.646/ 0.607 | 0.648/ 0.583 |
| | 1.0 | **0.662/ 0.641** | 0.665/ 0.626 | 0.645/ 0.604 | 0.647/ 0.571 |
| | 1.5 | 0.658/ 0.632 | 0.664/ 0.610 | 0.645/ 0.607 | 0.645/ 0.574 |
| Ovarian | 0.1 | 0.625/0.649 | 0.625/0.649 | 0.618/0.656 | 0.618/0.656 |
| | 0.5 | **0.615/0.667** | **0.615/0.667** | 0.614/0.651 | 0.614/0.651 |
| | 1.0 | 0.609/0.662 | 0.609/0.662 | 0.610/0.653 | 0.610/0.653 |
| | 1.5 | 0.606/0.664 | 0.606/0.664 | **0.608/0.657** | **0.608/0.657** |

3.4.2. Predictive Modeling (Cachexia and Recurrence)

Figure 12(a) presents the results for an MLP (multi-layer perceptron) trained to predict cachexia in pancreatic cancer patients. The model achieved an overall accuracy of 70% on a test set of 30 PDAC patients only. The model predicts whether a patient is cachectic with a precision of 79% and whether they are not with a precision of 55% at the time of cancer diagnosis. The F1 score for this task is 76.9%.

Figure 12(b) presents the results for an MLP trained to predict recurrence in ovarian cancer patients, with an overall accuracy of 62%. Notably, the MLP attained a higher precision of 78% in predicting recurrence at the time of diagnosis but a lower precision of 33% in predicting non-recurrence. The F1 score for this task is 72.5%.

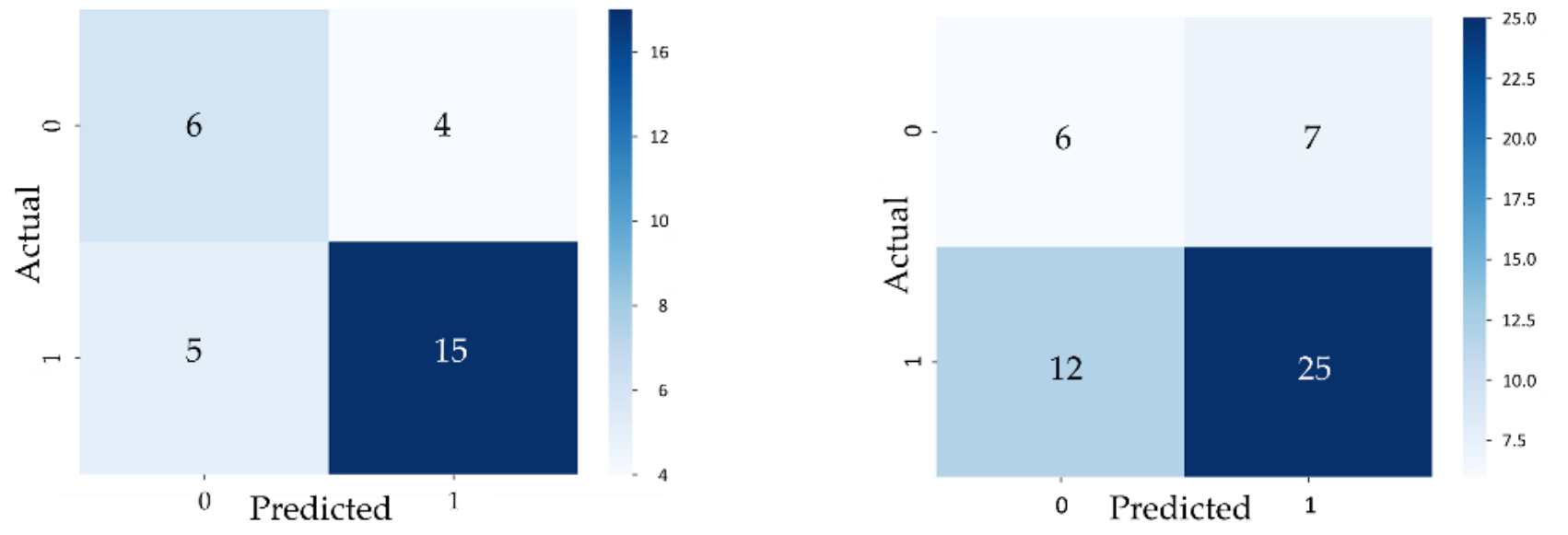

(**a**) (**b**)

**Figure 12. Confusion matrix showing prediction results for MLPs trained on multi-modal data**. (**a**) Cachexia prediction results on the pancreatic cancer test set, which included PDAC patients only, indicate an overall accuracy of 70%, with a 79% precision in predicting whether a PDAC patient is cachectic at the time of cancer diagnosis. (**b**) Recurrence prediction results on the ovarian cancer test set show an overall accuracy of 62%, with a 78% precision in predicting, at the time of diagnosis, whether a patient will experience recurrence.

### 3.5. Anectodal Evidence of SMAART-AI Tool's Utility

Figure 13 presents samples from colorectal, pancreatic, and ovarian datasets, featuring patients with nearly identical BMI but differing SMI. In each pair, one patient's SMI exceeds the literature-defined cut points for diagnosing sarcopenia/cachexia, while the other's SMI falls below these thresholds [10, 57, 58].

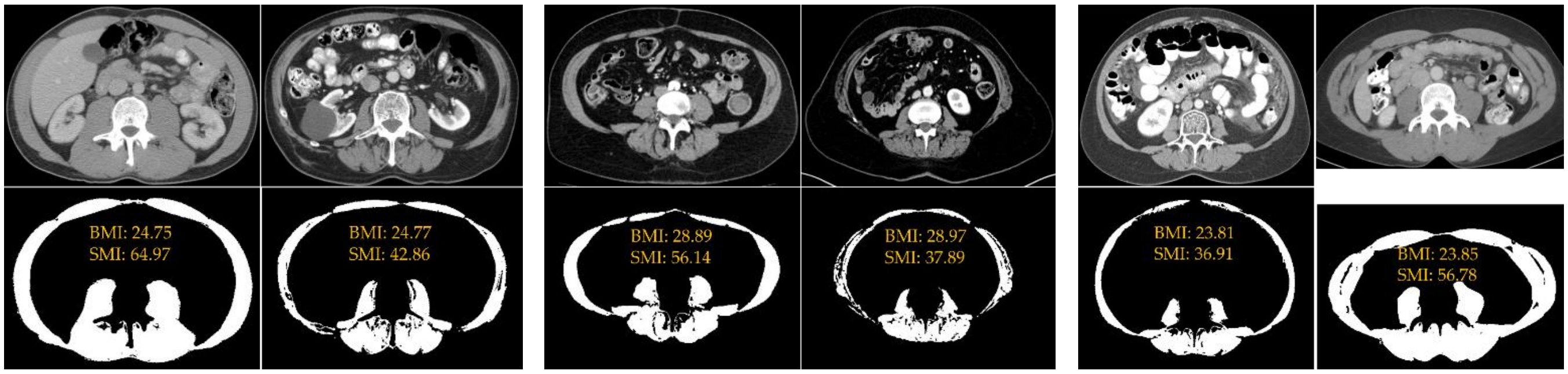

(**a**) (**b**) (**c**)

**Figure 13. CT images of patients with similar BMI but different SMI**. (a) Colorectal cancer, male patients with BMI 24.75 and 24.77, with SMI 64.97 and 42.86. (b) Pancreatic cancer, female patients with BMI 28.89 and 28.97, and SMI 56.14 and 37.89. (c) Ovarian cancer patients with BMI of 23.81 and 23.85 and SMI of 36.91 and 56.78. In all these examples, at the same BMI, one patient has SMI that is above the cut point defined in literature for diagnosing sarcopenia/cachexia, and the other patient has SMI below the cut point [2, 3, 10, 57, 58].

## 4. Discussion

Cancer cachexia is a multifactorial condition with skeletal muscle loss being one of its defining characteristics, making SMA and SMI crucial radiographic biomarkers for cachexia diagnosis and management. Our study underscores the variability in SMI among patients with similar BMIs, demonstrating the importance of longitudinal tracking of SMI during cancer treatment to provide personalized insights into disease progression and intervention strategies. Our proposed automated pipeline, SMAART-AI, integrates seamlessly into clinical workflows to monitor skeletal muscle changes with high accuracy and reliability.

SMAART-AI was mainly trained on the gastroesophageal dataset, which is reflected in its accuracy in estimating SMA for images in the gastroesophageal test set with some exceptions involving noisy or out-of-distribution images. One area of expected variation was in the colorectal cancer dataset, where differences in slice selection between pipelines naturally cause discrepancies in SMA estimates. However, this expected variation would be minimal if the slices selected as mid-L3 were within one or two slices of each other. The number of slices that can be used around the mid-L3 depends on slice thickness—thicker slices reduce the number of slices around the mid-L3 slice having almost similar SMA as the mid-slice. Comparative analysis revealed that tools like DAFS and TotalSegmentator tend to underestimate SMA considerably when manual radiologist segmentation is considered the gold standard, and such variations may not be solely due to slight differences in L3 slice selection, as observed in the skeletal muscle masks generated by TotalSegmentator. Since we did not have the skeletal muscle mask or information about the mid-L3 slice determined by DAFS, we could not further analyze the source of this difference. On the other hand, SMAART-AI did show a slight overestimation (by around 0.5-2%), for all datasets used in this study, largely due to marking connective tissues as skeletal muscle. In addition, some images offer a difference of 0.5% to 1% between different experts marking the skeletal muscle manually. Therefore, we used a margin of 2.5% from the manual annotations as a good performance in estimating SMA. Performance accuracy is particularly important when the actual SMI is close to cut points proposed in the literature to diagnose sarcopenia or cachexia and less important when the SMI is well below or above these cut points.

Comparative analysis of the different tools used to estimate SMA in the ovarian cancer dataset reveals that many images were out-of-distribution or noisy. The results highlight the sensitivity of these tools to shifts in the input data, whether due to image artifacts, scanner differences, changes in scanner settings, or unclear images with varying levels of noise. These shifts can lead to performance lapses in otherwise highly accurate AI-based models, emphasizing the need for a mechanism to alert users when the SMA estimates are likely to have considerable errors.

Among the various uncertainty estimation techniques and metrics used to assess SMAART-AI's performance, certain metrics consistently exhibited strong correlations with potential errors in SMA estimations across datasets, enabling the development of a standardized thresholding mechanism to flag cases with a high probability of error. However, some exceptions were observed, where high uncertainty coincided with low estimation errors and vice versa. High uncertainty with low errors often occurred when false positive and false negative counts were balanced, resulting in a small difference in estimated SMA—indicating model confusion despite accurate SMA predictions. Conversely, the model occasionally made confident yet incorrect predictions, leading to low uncertainty but large SMA estimation errors. Additionally, some datasets exhibited weaker correlations overall, suggesting that the effectiveness of uncertainty estimation methods depends on dataset characteristics.

SMAART-AI's ability to provide longitudinal tracking of SMA and SMI offers substantial clinical value. For patients classified as cachectic or non-cachectic at cancer onset based on established criteria defined in the literature, monitoring SMI changes during treatment revealed important trends. Some cachectic patients appeared to gain muscle mass, likely due to edema, which can obscure tissue boundaries and lead to overestimation. Conversely, some non-cachectic patients exhibited varying degrees of muscle loss, suggesting the potential onset of cachexia during treatment. These findings underscore the importance of continuous SMI monitoring for timely and precise clinical intervention.

Our survival analysis demonstrated that integrating SMA, SMI derived from radiology images, and BMI with clinical data in multimodal models improved the concordance index for predicting survival outcomes in pancreatic, colorectal, and ovarian cancers. Removing SMA and SMI from the models led to a marked drop in predictive performance, underscoring the superiority of these biomarkers over BMI alone in predicting overall survival in cancer patients.

Finally, in a binary classification task to predict cachexia in pancreatic cancer patients, our multimodal machine learning model incorporating clinical data, BMI, and SMA/SMI achieved a precision of 79%. This means that if the model predicts a patient to be cachectic, there is a 79% chance that the patient will be cachectic. Similarly, recurrence prediction in ovarian cancer patients yielded a precision of 62%. These results suggest that incorporating additional data modalities, such as laboratory reports, could further improve the predictive performance by capturing the multifactorial nature of cachexia.

Overall SMAART-AI demonstrates strong potential for clinical integration, providing accurate and reliable SMA and SMI estimates and actionable insights for cachexia diagnosis and management. By addressing the challenges of noisy data and the use of multimodal predictive models, our approach advances the field of cancer cachexia research and offers new avenues for improving patient care.

## 5. Conclusion

Cancer cachexia is a severe condition that greatly impacts patient outcomes, underscoring the need for early detection and continuous monitoring. This study presents SMAART-AI, an automated and reliable pipeline for skeletal muscle area and index estimation, leveraging the nnU-Net framework. By benchmarking SMAART-AI against existing tools such as ABACS, AW Server, DAFS, TotalSegmentator, and SliceOmatic, we demonstrate its efficacy and potential for clinical integration. Incorporating uncertainty estimation and thresholding mechanism, our approach enhances model reliability by identifying cases where SMAART-AI may perform poorly. Furthermore, we establish the clinical relevance of SMA/SMI by conducting overall survival analyses and predicting cachexia and disease recurrence at the time of cancer diagnosis. Our findings suggest that integrating SMI tracking into clinical workflows provides valuable insights beyond BMI, aiding in cachexia diagnosis and patient monitoring. This study moves us closer to practical, reliable, and robust machine learning solutions that can be deployed in clinics to enhance patient care. In future work, we aim to enhance cancer cachexia prediction by integrating additional data modalities beyond those explored in this study.

## CrediT authorship contribution statement

**Sabeen Ahmed:** Writing – original draft, Visualization, Validation, Software, Methodology, Investigation, Formal analysis, Data curation, Conceptualization. **Nathan Parker:** Writing – review and editing, Resources, Supervision, Validation, Conceptualization. **Margaret Park:** Writing – review and editing, Resources, Validation. **Daniel Jeong:** Writing – review and editing, Resources, Validation. **Lauren Peres:** Resources. **Evan W. Davis:** Writing – review and editing, Validation. **Jennifer B. Permuth:** Writing – review and editing, Resources, Validation, Funding. **Erin Siegel:** Resources. **Matthew B. Schabath:** Writing – review and editing, Resources, Validation. **Yasin Yilmaz:** Writing – review and editing, Supervision, Validation. **Ghulam Rasool:** Writing – review and editing, Resources, Supervision, Validation, Conceptualization, Funding.

## Declaration of competing interest

The authors declare that they have no known competing financial interests or personal relationships that could have appeared to influence the work reported in this paper.

Funding: This work was supported by the National Science Foundation (NSF) [grant numbers 2234468 and 2234836]; the National Institutes of Health (NIH) [grant number U01CA200464]; the James and Esther King Foundation [grant number 8JK02]; and the Department of Defense [grant number PA210192].

## Declaration of Generative AI and AI-assisted technologies in the writing process

During the preparation of this work, the author(s) used ChatGPT to improve readability and language. After using this tool/service, the author(s) reviewed and edited the content as needed and take(s) full responsibility for the content of the published article.

## Appendix A

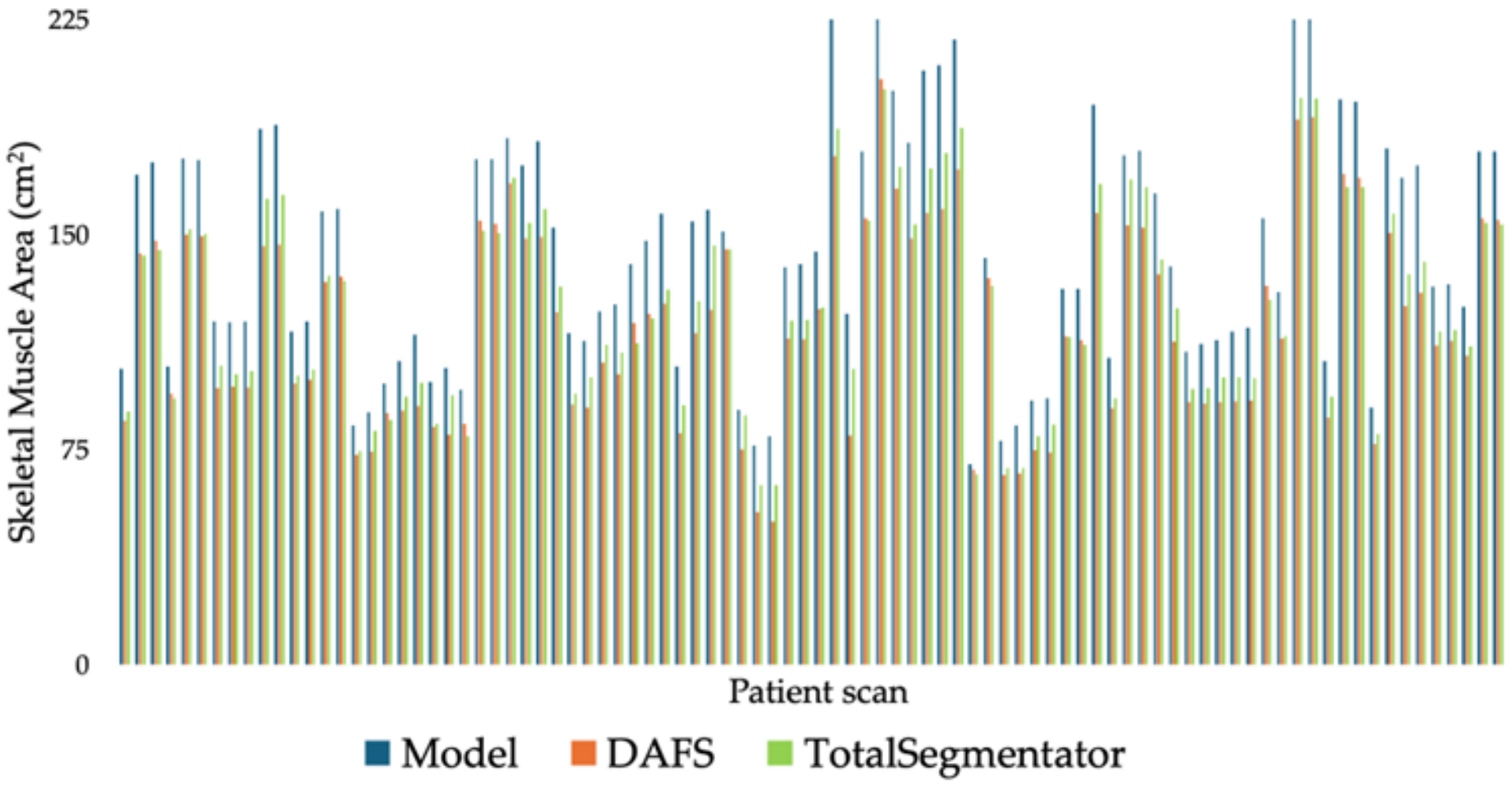

(**a**)

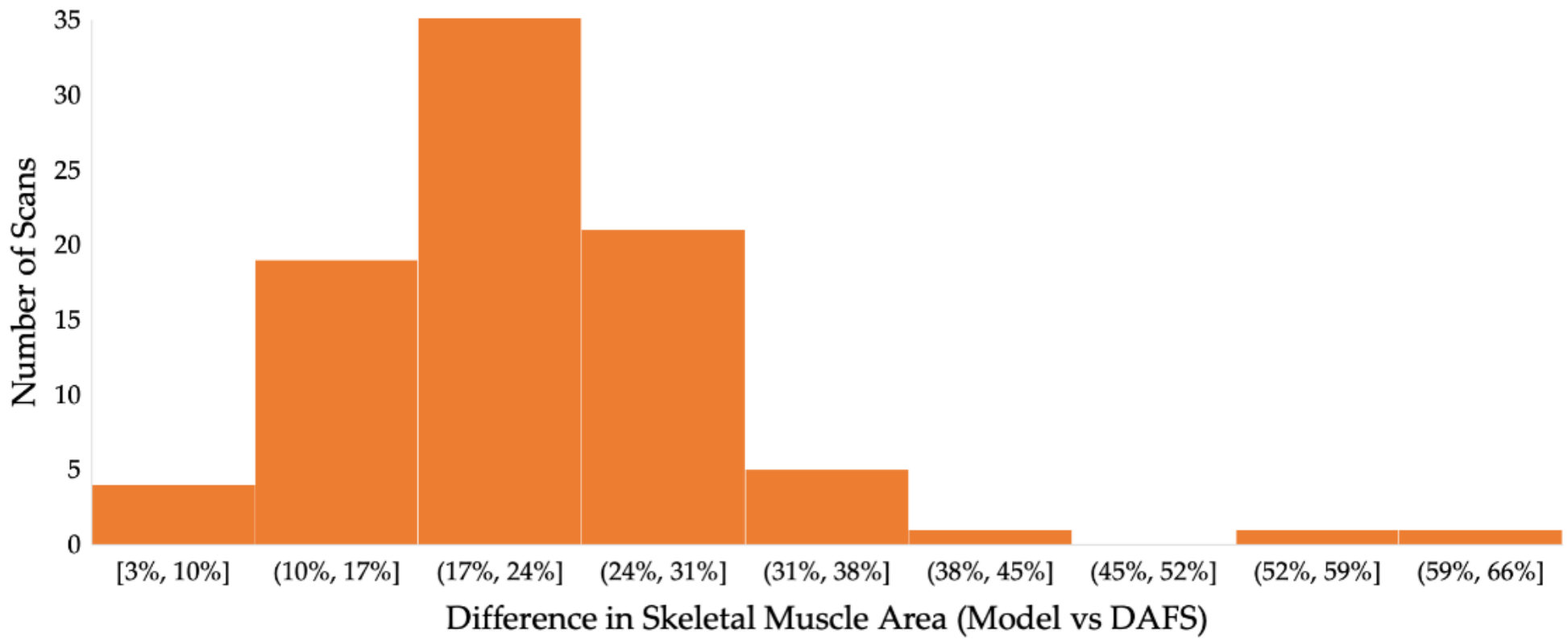

(**b**)

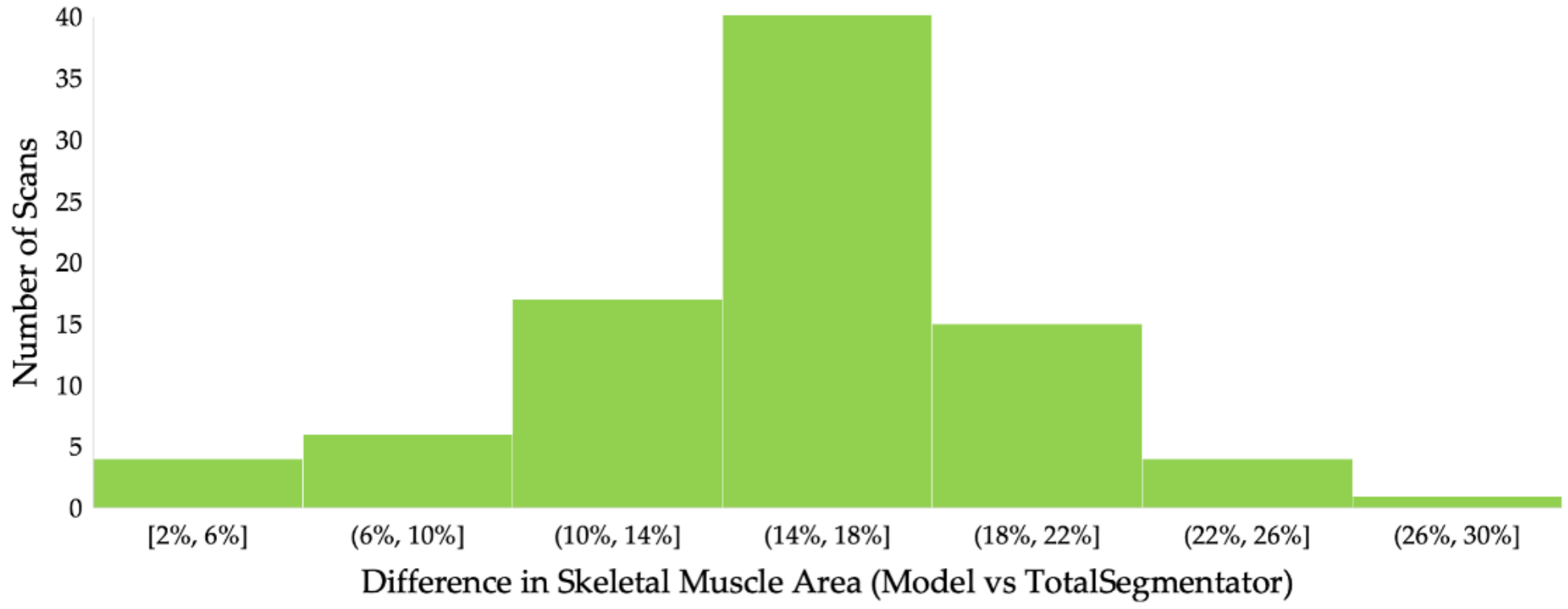

(**c**)

**Figure A1. Comparative analysis of SMA estimation using different tools for colorectal cancer**. (**a**) Comparison of SMA estimated for 60 patients (90 scans, including multiple axial series per patient) at the mid-L3 level by SMAART-AI, DAFS, and TotalSegmentator. Both DAFS and TotalSegmentator consistently estimate lower SMA values compared to SMAART-AI, with DAFS generally estimating lower values than TotalSegmentator. The mid-L3 slice used by SMAART-AI and DAFS is determined automatically by their respective pipelines, while TotalSegmentator uses the mid-L3 slice determined by our proposed pipeline. (**b**) Distribution of differences between SMA predictions by SMAART-AI and DAFS indicates a large discrepancy, potentially due to variation in the selected mid-L3 slice or poor DAFS performance on this dataset. (**c**) Distribution of differences between SMA predictions by SMAART-AI and TotalSegmentator suggests that TotalSegmentator consistently underestimates SMA in most cases.

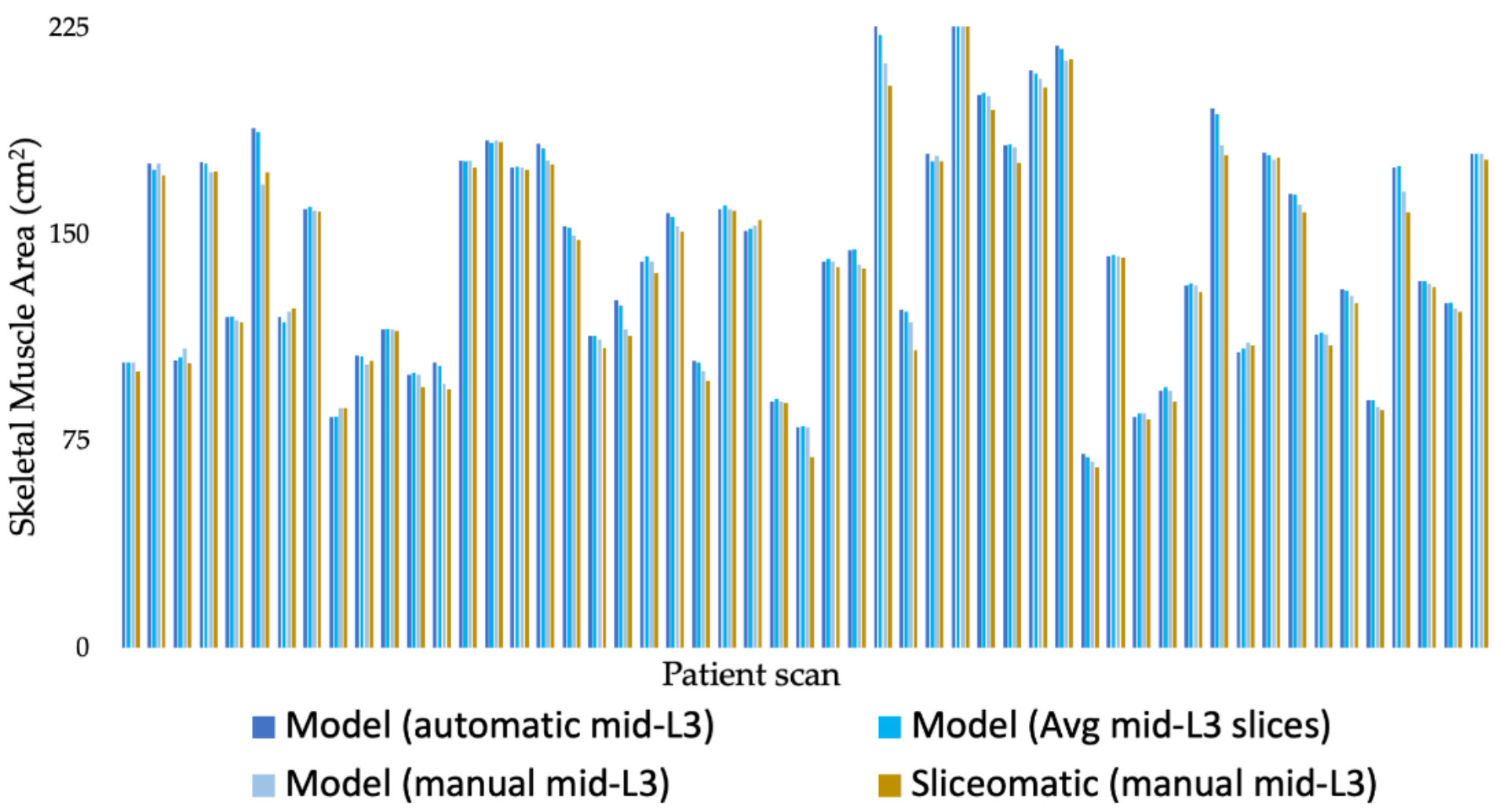

(**a**)

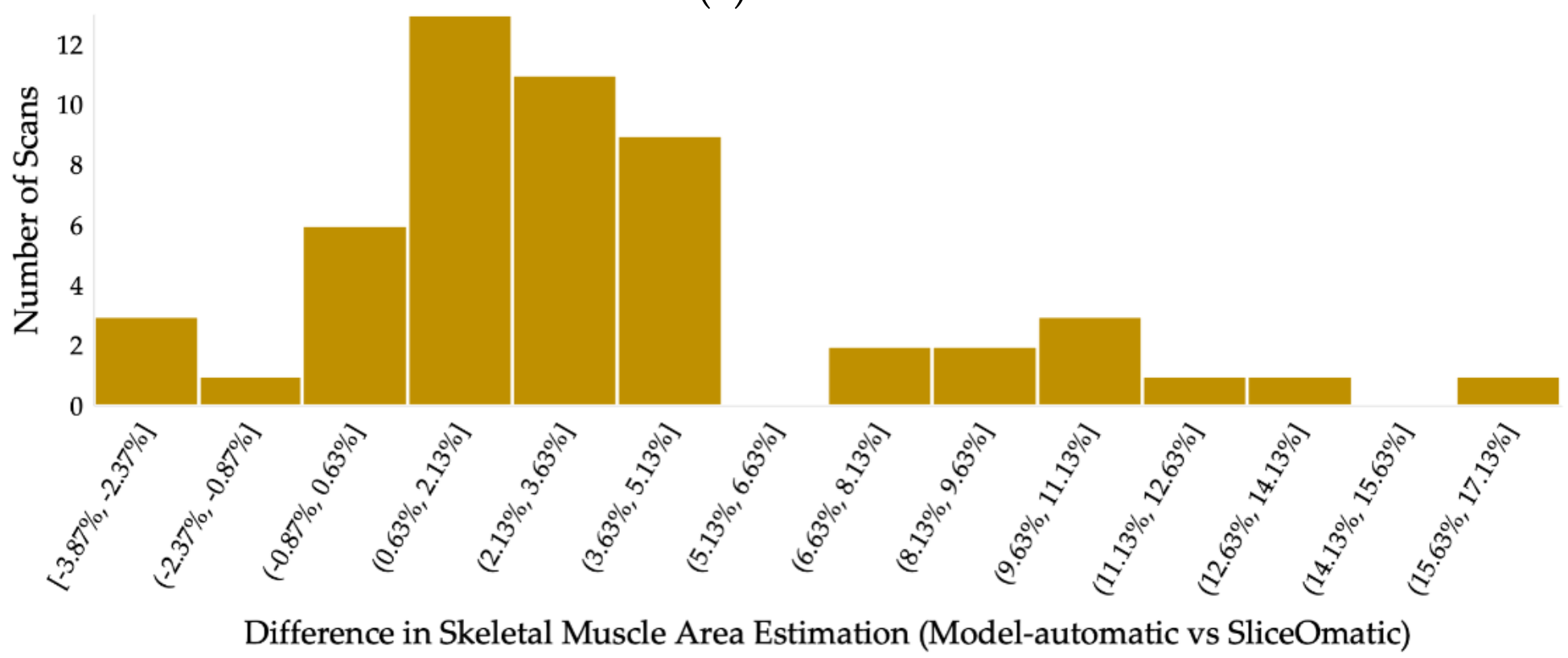

(**b**)

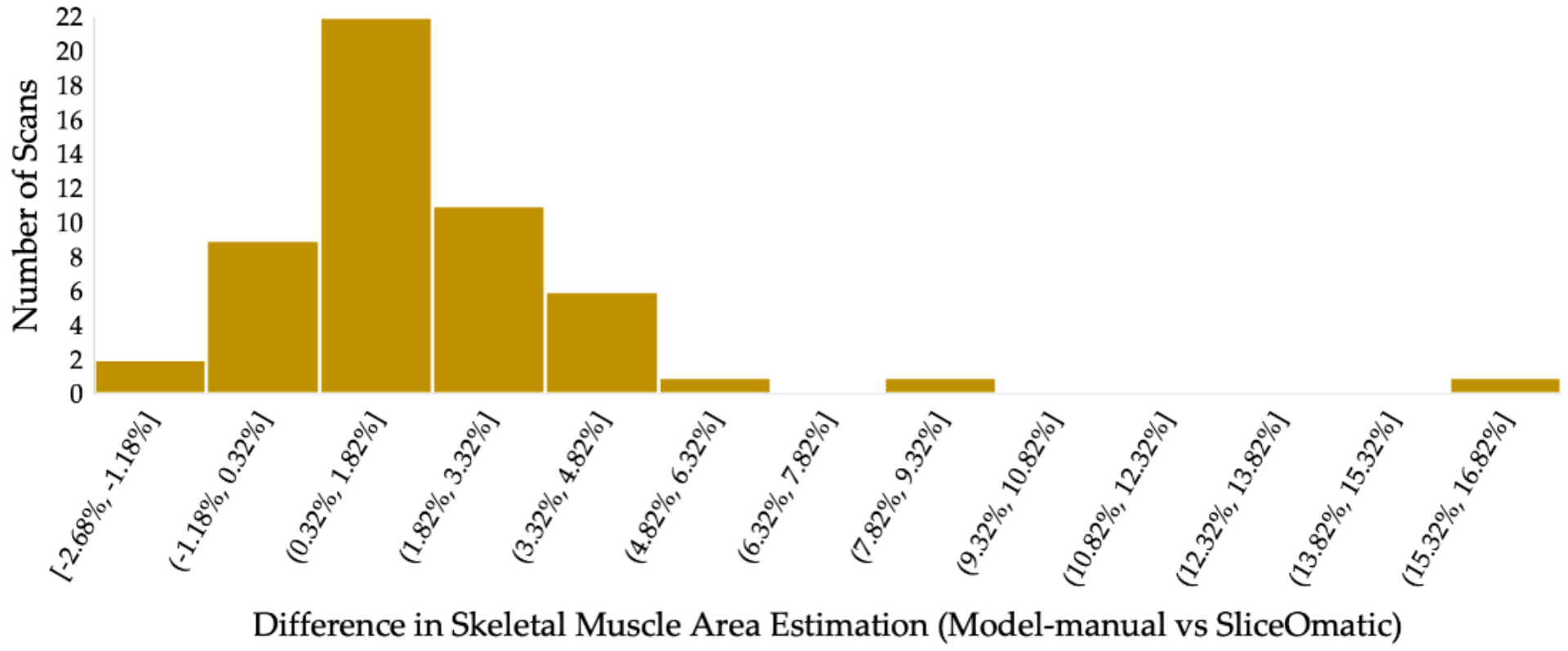

(**c**)

**Figure A2. Benchmarking SMA estimation by SMAART-AI versus SliceOmatic for colorectal cancer**. (**a**) Comparison of SMA estimates for 53 patients by different methods: SMAART-AI at mid-L3 (automated pipeline), the average of slices around the pipeline determined mid-L3, the model's prediction at the manually determined mid-L3 slice, and SliceOmatic (manual). The SMA values for the manual and automated mid-L3 slices, as well as the average of surrounding slices, are nearly identical in most cases, closely aligning with the estimations from SliceOmatic. (**b**) The distribution of differences between SMAART_AI's SMA estimates (mid-L3 determined by the pipeline) and SliceOmatic (manual) shows strong agreement. Differences greater than 3% are mainly observed when the automated and manually selected mid-L3 slices differ significantly or when the CT image is noisy or out-of-distribution. (**c**) The difference between SMAART-AI's estimation and SliceOmatic at the manually determined mid-L3 indicates strong model performance overall. Larger discrepancies are mainly observed in low-quality images, while differences up to 2% may be attributed to the fact that the model may include connective tissues as part of skeletal muscle. The comparison of the average area of slices near mid-L3 with a single mid-L3 slice has a median difference of 0.53% and a mean difference of 0.66%, indicating that adjacent slices provide similar area estimates. Overall, SMAART-AI outperforms both DAFS and TotalSegmentator.

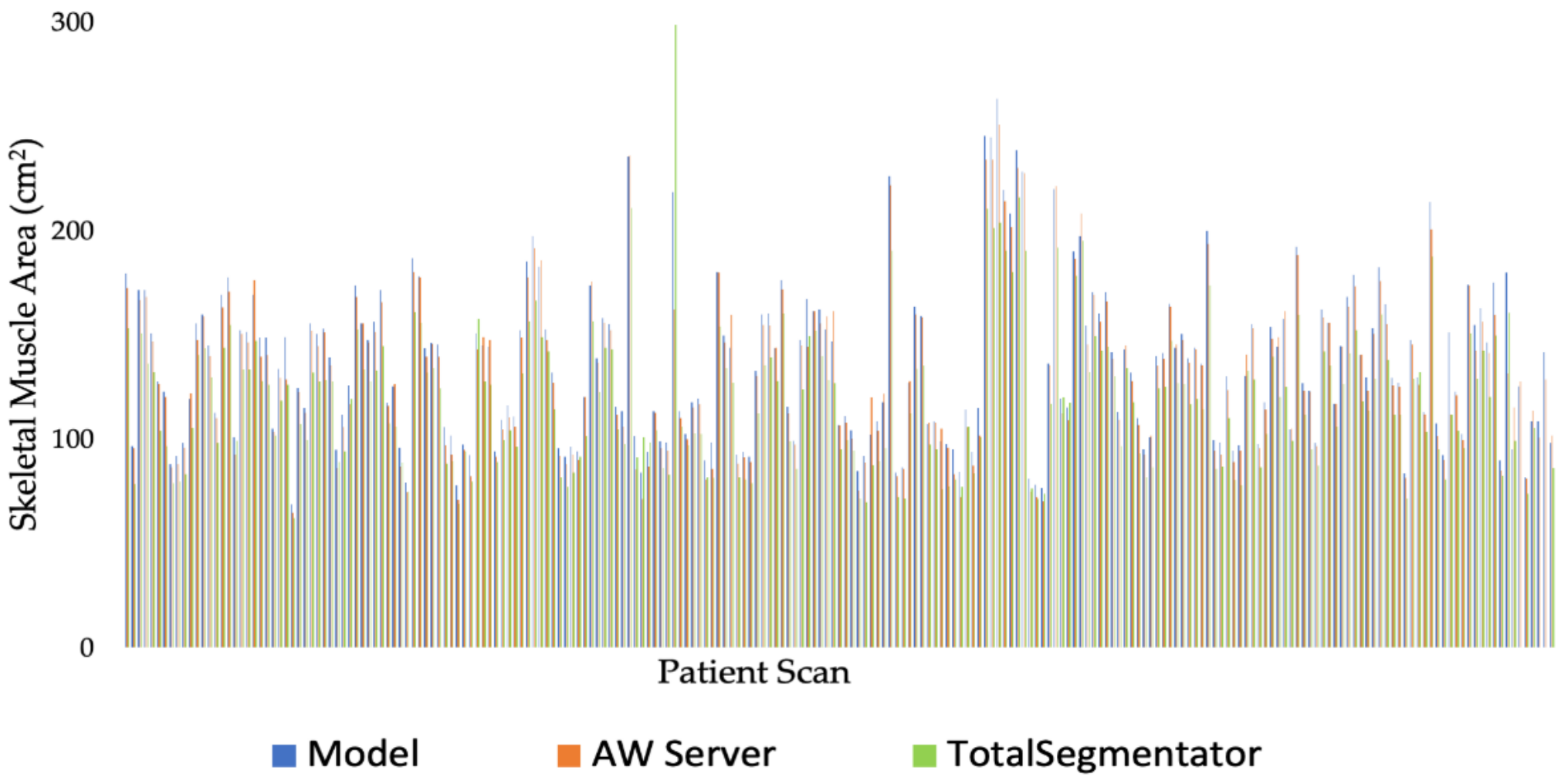

(**a**)

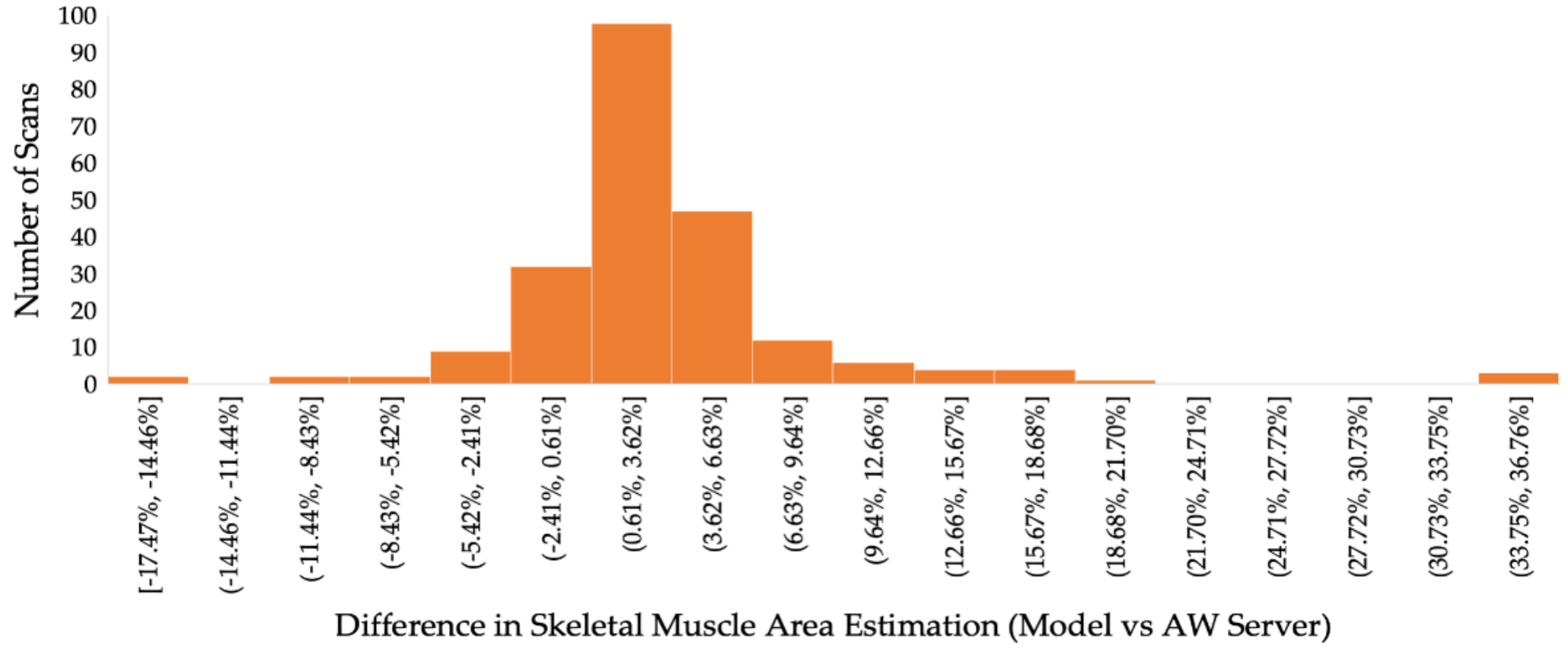

(**b**)

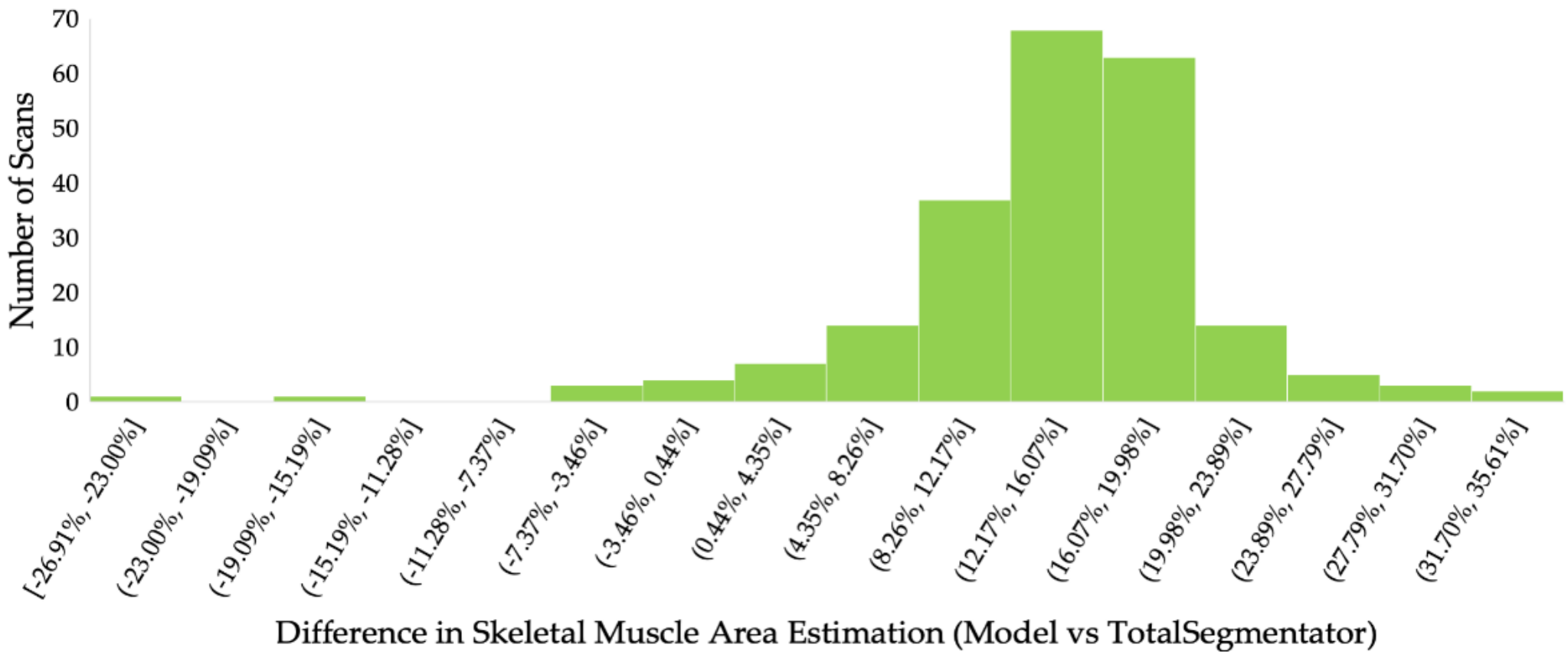

(**c**)

**Figure A3. Comparative analysis of SMA estimation using different tools for pancreatic cancer dataset**. (**a**) Comparison of skeletal muscle area (SMA) estimates from 222 patient scans at the manually determined end-L3 level by SMAART-AI, AW Server, and TotalSegmentator shows that TotalSegmentator consistently estimates lower values in most cases compared to both SMAART-AI and AW Server. (**b**) The distribution of differences between the SMA estimated by SMAART-AI and AW Server shows an approximately 3% absolute difference in around 60% of cases. Overall, SMAART-AI tends to slightly underestimate in some cases but generally overestimates compared to AW Server. (**c**) The distribution of differences between SMAART-AI and TotalSegmentator indicates significant disagreement in most cases.

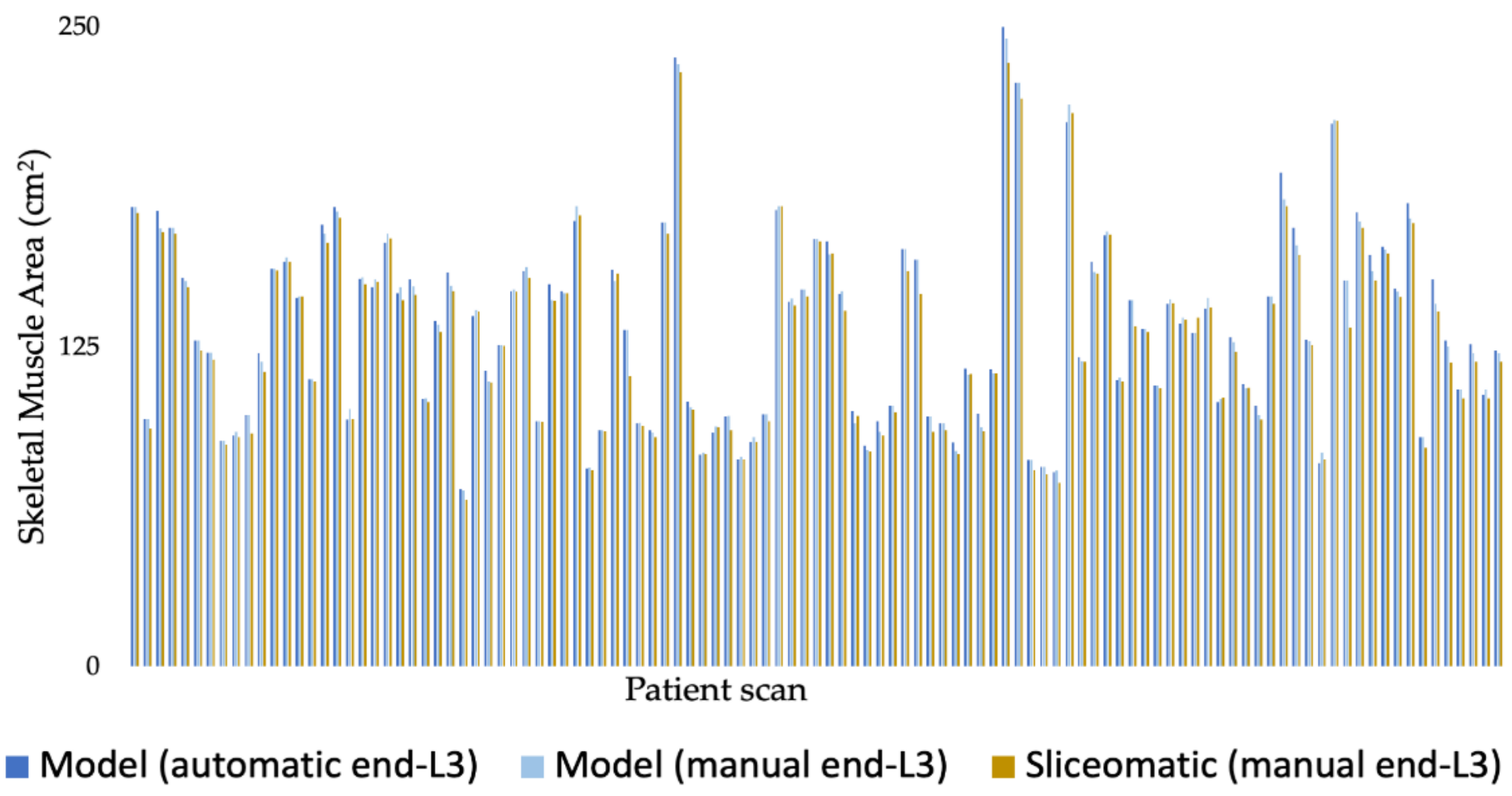

(**a**)

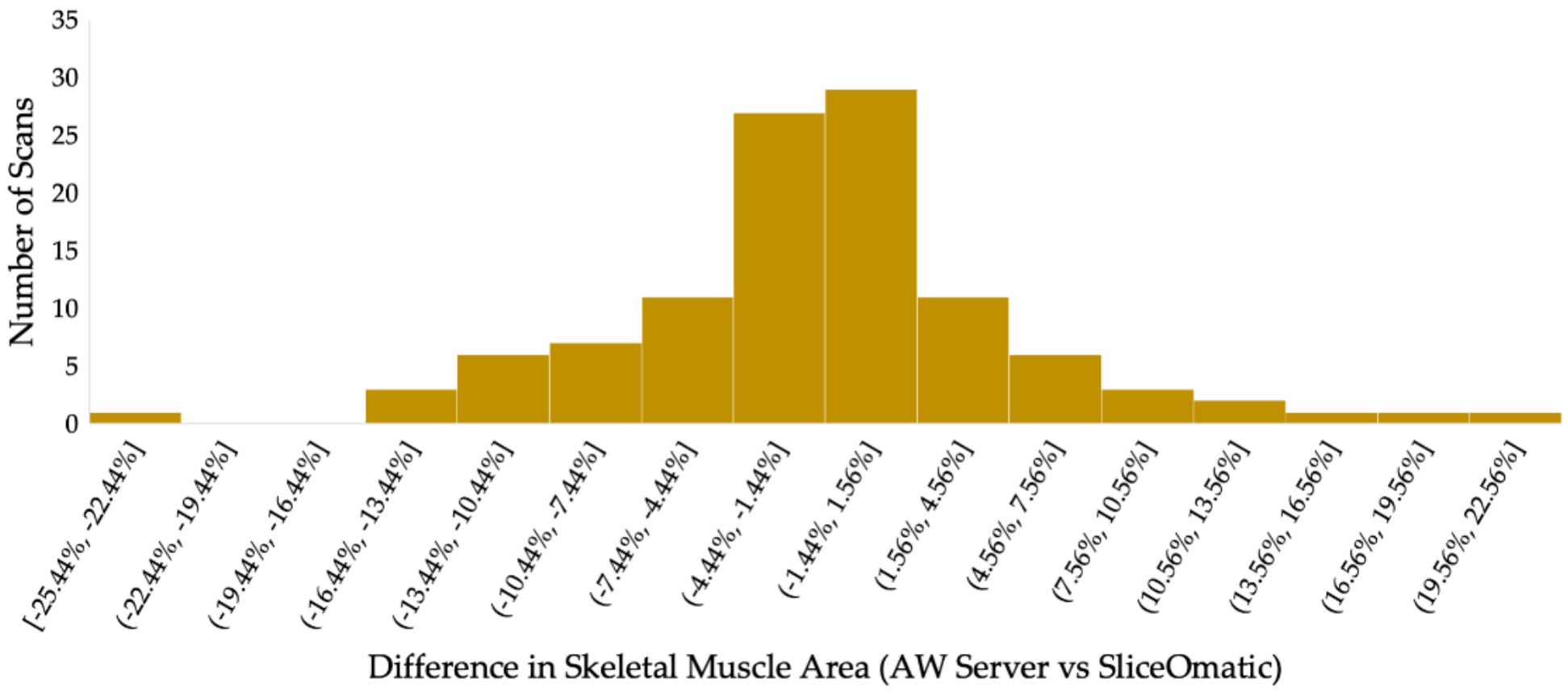

(**b**)

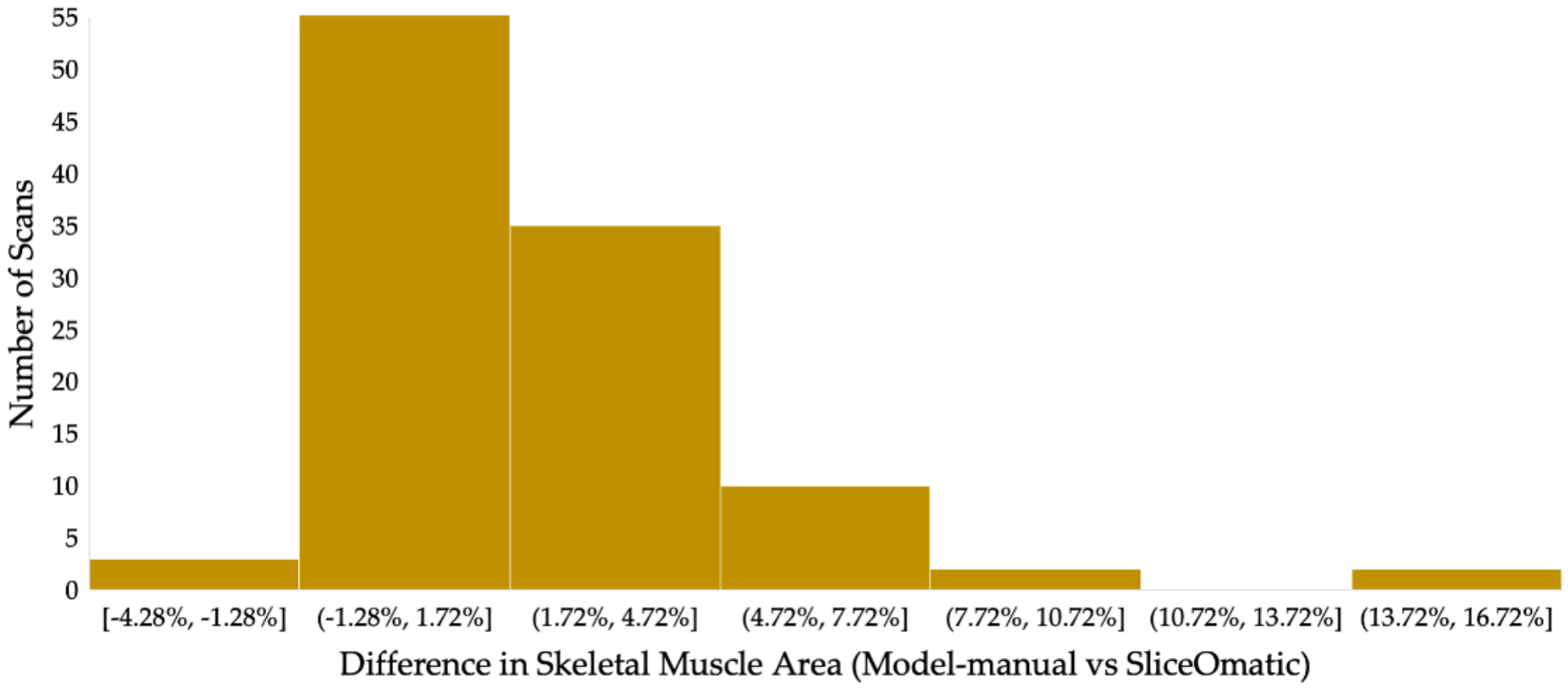

(**c**)

**Figure A4. Benchmarking SMA estimation by SMAART-AI and AW server versus SliceOmatic for pancreatic cancer dataset**. (**a**) Comparison of SMA estimates for 109 patient scans, using SMAART-AI at both automatically and manually identified end-L3 slices, along with SliceOmatic (at the same manually selected end-L3 slices), shows that the SMA values from both automated and manually selected slices are nearly identical in most cases, closely matching the reference end-L3 SMA determined by SliceOmatic. (**b**) The distribution of differences in SMA between estimates from AW Server and SliceOmatic at the manually determined end-L3 slice shows good agreement in approximately 61% of cases. (**c**) The distribution of differences in SMA between SMAART-AI's estimate at the manually determined end-L3 and SliceOmatic reveals good agreement in around 87% of cases. SMAART-AI tends to overestimate the SMA in some cases, whereas AW Server shows a slight bias towards underestimation. Overall, SMAART-AI performs well.

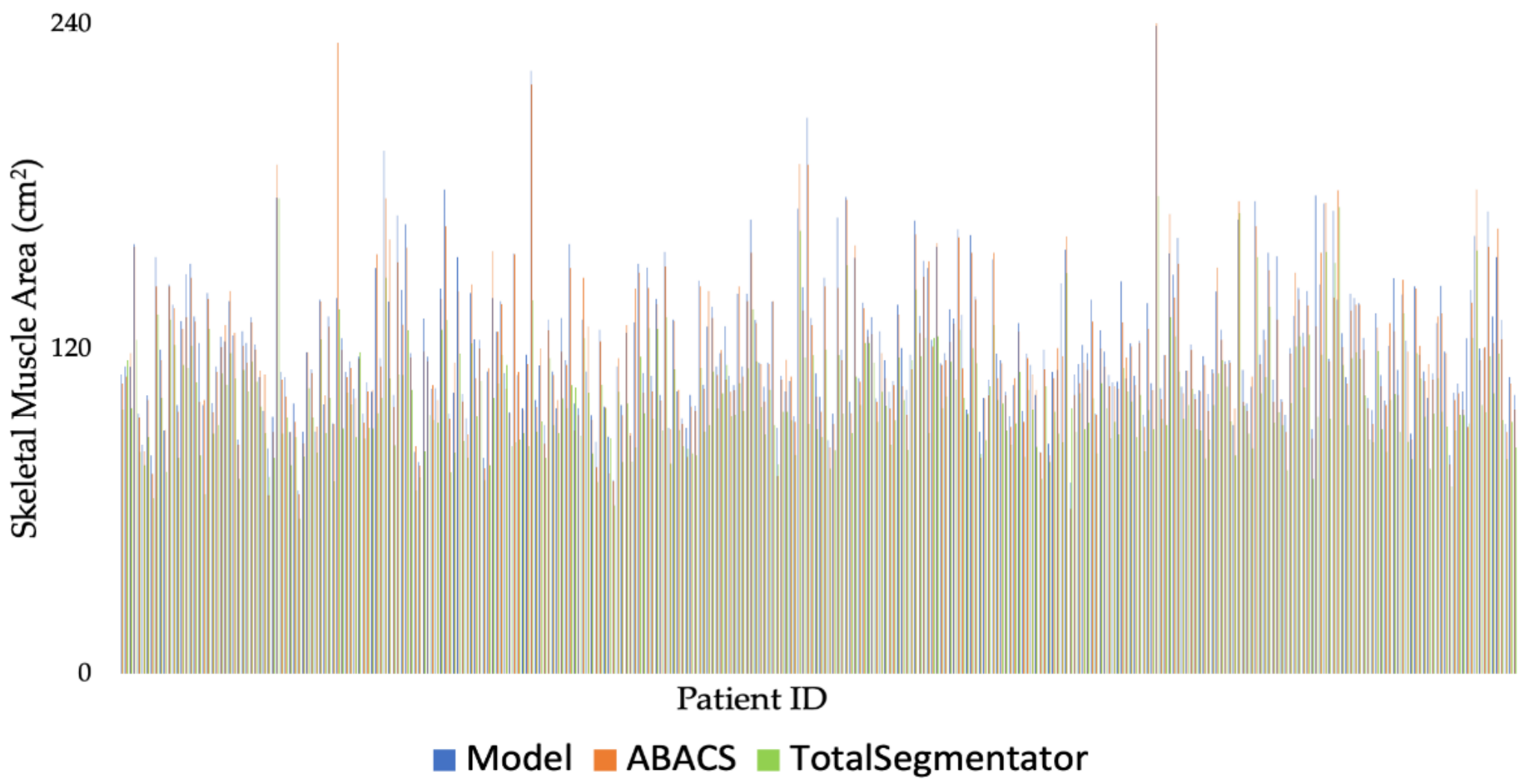

(**a**)

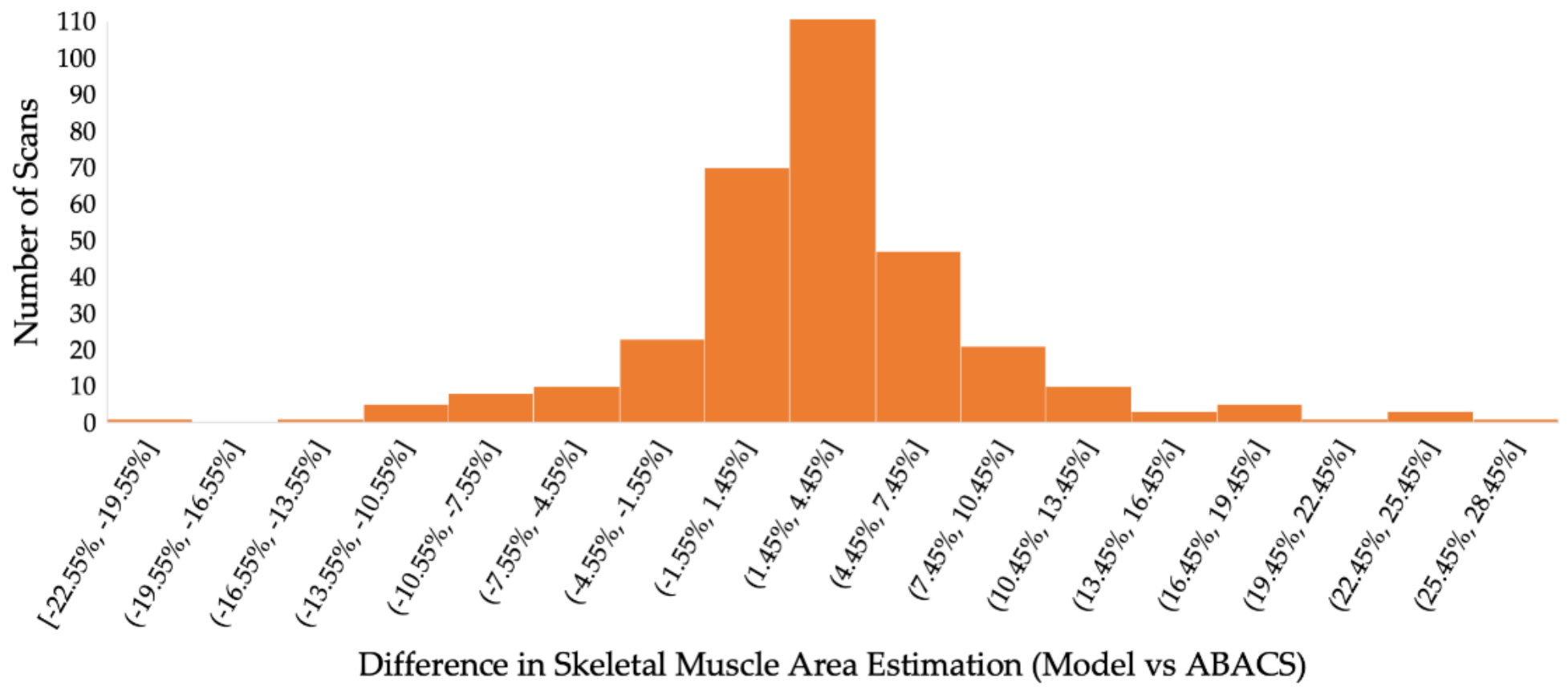

(**b**)

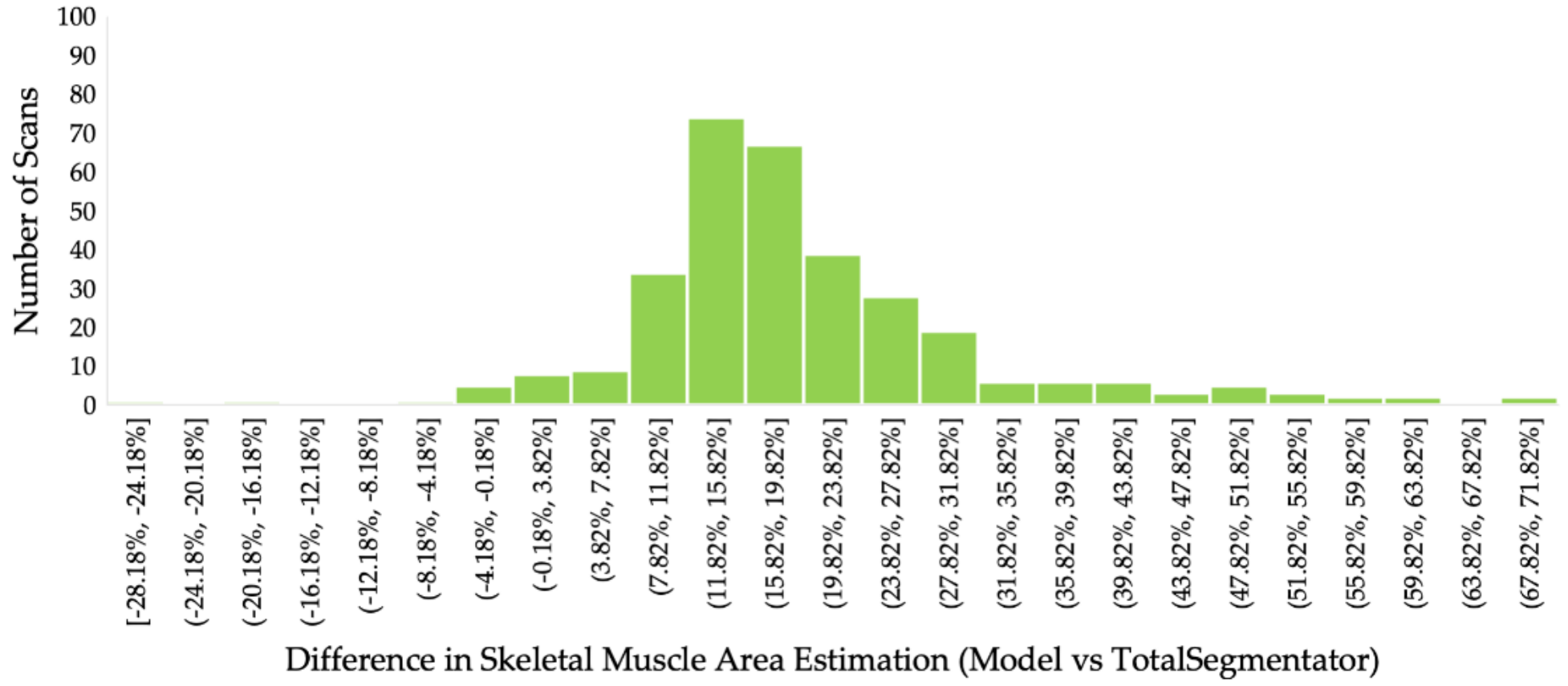

(**c**)

**Figure A5. Comparative analysis of SMA estimation using different tools for ovarian cancer**. (**a**) A comparison of skeletal muscle area (SMA) from 324 patient scans at the manually determined mid-L3 level, estimated by SMAART-AI, ABACS, and TotalSegmentator, shows that TotalSegmentator estimates lower values compared to the trained model and ABACS with some exceptions. There are a few odd cases where ABACS is estimating very large values for the area compared to SMAART-AI and TotalSegmentator. (**b**) The distribution of the difference between the SMA estimated

by the SMAART-AI and ABACS indicates close agreement in around 63% of cases. (**c**) The distribution of the difference between SMAART-AI and TotalSegmentator shows high disagreement in most cases and a close estimation in only around 4% of the cases.

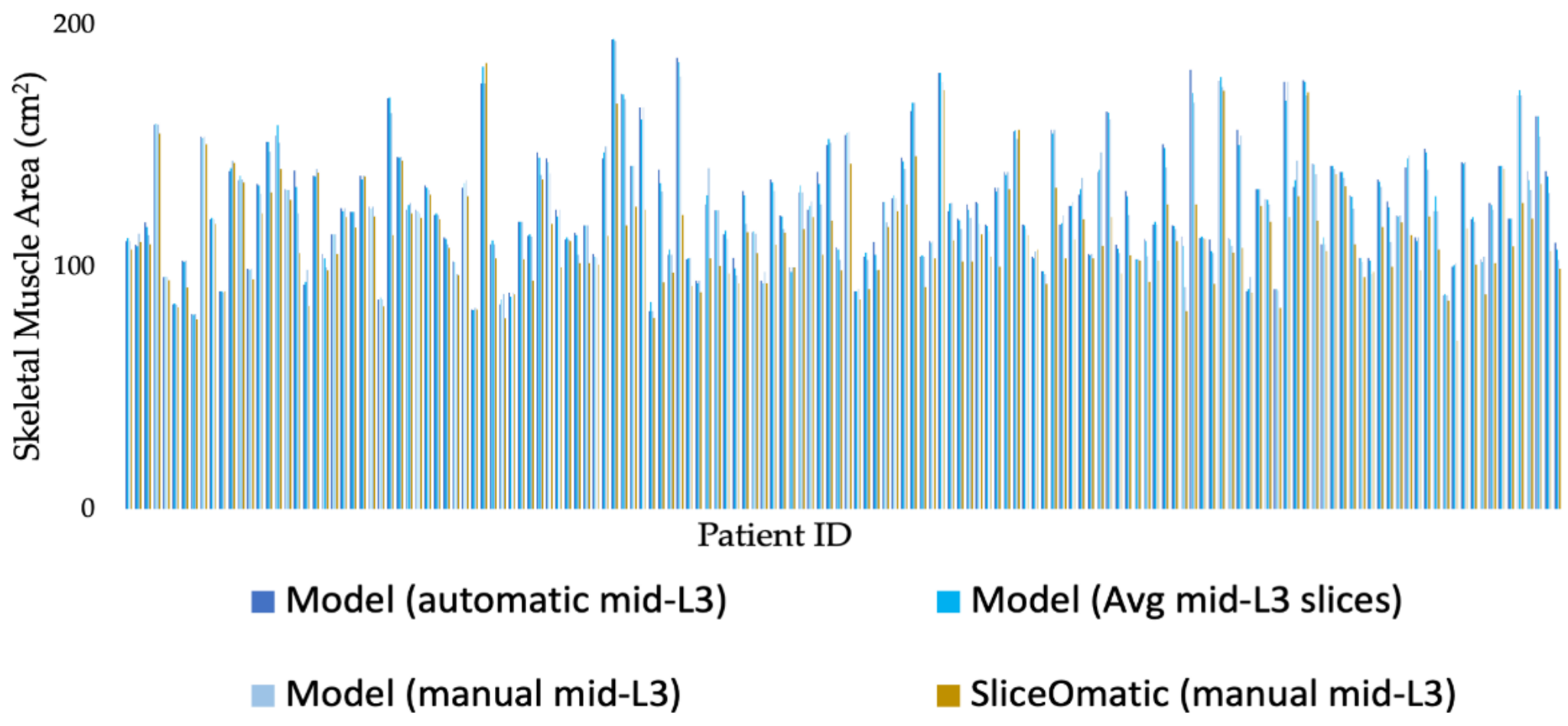

(**a**)

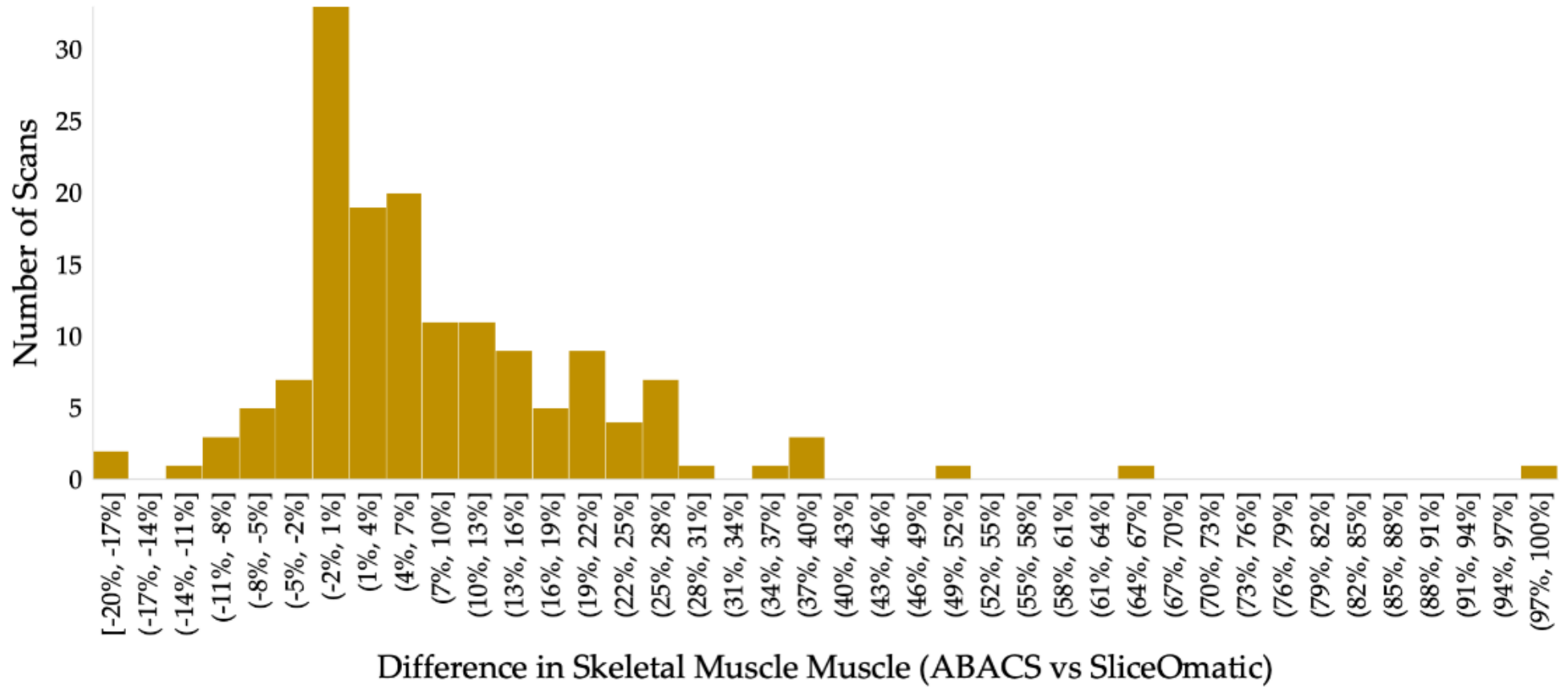

(**b**)

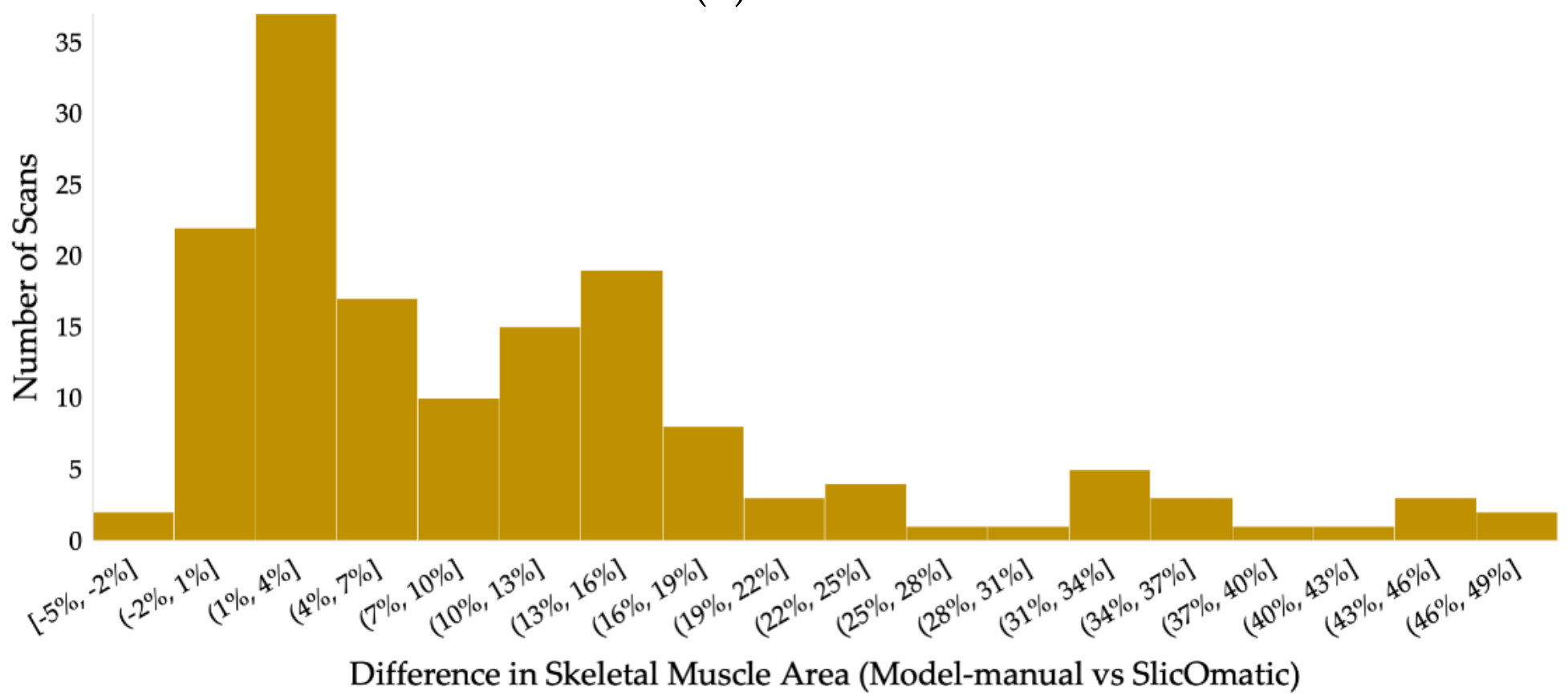

(**c**)

**Figure A6. Benchmarking SMA estimation by SMAART-AI and ABACS versus SliceOmatic for ovarian cancer**. (**a**) Comparison of SMA estimates from 154 patients' CT scans using different methods: SMAART-AI at mid-L3 (automated), the average of slices around SMAART-AI determined mid-L3, SMAART-AI at the manually

determined mid-L3, and SliceOmatic (manual). In most cases, SMA estimates for mid-L3 (both manual and automatic), and the average of adjacent slices closely match the reference SMA determined by SliceOmatic. (**b**) The distribution of differences between SMA estimated by ABACS and SliceOmatic at the manually determined mid-L3 shows that ABACS provides estimates with low differences in approximately 32% of the cases, with a trend of overestimations and occasional underestimations. (**c**) The distribution of differences between SMA estimated by SMAART-AI at the manually determined mid-L3 and SliceOmatic shows estimation with low difference in around 26% of cases but tends to overestimate in most cases. Overestimations by both ABACS and SMAART-AI are primarily attributed to out-of-distribution or noisy images, which are prevalent in this dataset.